\documentclass[iop]{emulateapj}

\usepackage{natbib}

\newcommand{\lsim}{\,\rlap{\raise 0.35ex\hbox{$<$}}{\lower 0.7ex\hbox{$\sim$}}\,}
\newcommand{\gsim}{\,\rlap{\raise 0.35ex\hbox{$>$}}{\lower 0.7ex\hbox{$\sim$}}\,}

\def \msun{{\,M_\odot}}

\def \dg{^{\circ}}

\def \8{SN~1987A}
\def \kms{\rm{km\ s^{-1}}}

\shorttitle{The ejecta of Supernova 1987A at 10000 days}
\shortauthors{Larsson et al.}

\begin{document}

\title{Three-dimensional distribution of ejecta in Supernova 1987A at 10 000 days}


\author{J.~Larsson\altaffilmark{1}, C.~Fransson\altaffilmark{2}, J.~Spyromilio\altaffilmark{3}, B.~Leibundgut\altaffilmark{3}, P.~Challis\altaffilmark{4}, R.~A.~Chevalier\altaffilmark{5}, K.~France\altaffilmark{6,7}, A.~Jerkstrand\altaffilmark{8}, R.~P.~Kirshner\altaffilmark{4}, P.~Lundqvist\altaffilmark{2}, M.~Matsuura\altaffilmark{9}, R.~McCray\altaffilmark{10}, N.~Smith\altaffilmark{11}, J.~Sollerman\altaffilmark{2}, P.~Garnavich\altaffilmark{12}, K.~Heng\altaffilmark{13},  S.~Lawrence\altaffilmark{14}, S.~Mattila\altaffilmark{15,16}, K.~Migotto\altaffilmark{2}, G.~Sonneborn\altaffilmark{17}, F.~Taddia\altaffilmark{2}, and J.~C.~Wheeler\altaffilmark{18}}


\altaffiltext{1}{KTH, Department of Physics, and the Oskar Klein Centre, AlbaNova, SE-106 91 Stockholm, Sweden}
\altaffiltext{2}{Department of Astronomy and the Oskar Klein Centre,  Stockholm University, AlbaNova, SE-106 91 Stockholm, Sweden}
\altaffiltext{3}{ESO, Karl-Schwarzschild-Strasse 2, 85748 Garching, Germany}
\altaffiltext{4}{Harvard-Smithsonian Center for Astrophysics, 60 Garden Street, MS-78, Cambridge, MA 02138, USA}
\altaffiltext{5}{Department of Astronomy, University of Virginia, P.O. Box 400325, Charlottesville, VA 22904-4325, USA}
\altaffiltext{6}{Laboratory for Atmospheric and Space Physics, University of Colorado, 392 UCB, Boulder, CO 80309, USA}
\altaffiltext{7}{Center for Astrophysics and Space Astronomy, University of Colorado, 389 UCB, Boulder, CO 80309, USA}
\altaffiltext{8}{Astrophysics Research Centre, School of Maths and Physics, Queen's University Belfast, Belfast BT7 1NN, UK}
\altaffiltext{9}{School of Physics and Astronomy, Cardiff University, Queen's Buildings, The Parade, Cardiff, CF24 3AA, UK}
\altaffiltext{10}{Department of Astronomy, University of California, Berke- ley, CA 94720-3411, USA}
\altaffiltext{11}{Steward Observatory, University of Arizona, 933 North Cherry Avenue, Tucson, AZ 85721, USA}
\altaffiltext{12} {Nieuwland Science, University of Notre Dame, Notre Dame, IN 46556-5670, USA}
\altaffiltext{13}{University of Bern, Center for Space and Habitability, Si- dlerstrasse 5, CH-3012 Bern, Switzerland}
\altaffiltext{14}{Department of Physics and Astronomy, Hofstra University, Hempstead, NY 11549, USA}
\altaffiltext{15}{Tuorla Observatory, Department of Physics and Astron- omy, University of Turku, Vislntie 20, FI-21500 Piikki, Finland}
\altaffiltext{16}{Institute of Astronomy, University of Cambridge, Madingley Road, Cambridge CB3 0HA, UK}
\altaffiltext{17}{Observational Cosmology Laboratory Code 665, NASA Goddard Space Flight Center, Greenbelt, MD 20771, USA}
\altaffiltext{18}{Department of Astronomy, University of Texas, Austin, TX 78712-0259, USA}

\begin{abstract}

Due to its proximity, SN~1987A offers a unique opportunity to directly observe the geometry of a stellar explosion as it unfolds. Here we present spectral and imaging observations of SN~1987A obtained \hbox{$\sim\ 10,000$} days after the explosion with HST/STIS and VLT/SINFONI at optical and near-infrared wavelengths. These observations allow us to produce the most detailed 3D map of H$\alpha$ to date, the first 3D maps for [Ca~II]$\ \lambda \lambda 7292,\ 7324$, [O~I]$\ \lambda \lambda 6300,\ 6364$ and Mg~II$\ \lambda \lambda 9218,\ 9244$, as well as new maps for [Si~I]$+$[Fe~II]$\ 1.644\ \mu$m and He I $2.058~\mu$m. A comparison with previous observations shows that the [Si~I]$+$[Fe~II] flux and morphology have not changed significantly during the past ten years, providing evidence that it is powered by  $^{44}$Ti.  The time-evolution of H$\alpha$ shows that it is predominantly powered by X-rays from the ring, in agreement with previous findings.  All lines that have sufficient signal show a similar large-scale 3D structure, with a north-south asymmetry that resembles a broken dipole. This structure correlates with early observations of asymmetries, showing that there is a global asymmetry that extends from the inner core to the outer envelope. On smaller scales, the two brightest lines, H$\alpha$ and  [Si~I]$+$[Fe~II]$\ 1.644\ \mu$m, show substructures at the level of $\sim 200 - 1000\ \kms$ and clear differences in their 3D geometries. We discuss these results in the context of explosion models and the properties of dust in the ejecta.

\end{abstract}

\keywords{Supernovae:general \--- Supernovae: individual: SN 1987A}

\section{Introduction}

Supernova (SN) 1987A, located in the Large Magellanic Cloud, is the only modern SN where the ejecta are spatially resolved in optical/NIR imaging observations. This makes it a unique target for studying ejecta composition, energy sources and asymmetries. In this work, we focus on asymmetries by studying the spatial distribution of the inner ejecta. These ejecta have not been affected by the reverse shock and are therefore still in the homologous expansion phase, which was reached about one week after the explosion (\citealt{Gawryszczak2010}). The spatial distribution reflects the conditions at the time of explosion and hence carries information about the progenitor and explosion mechanism (e.g.,  \citealt{Wongwathanarat2015}).   

The ejecta of \8 are surrounded by a triple ring system that was created approximately $20,000$ years before the explosion, possibly as a result of a binary merger (\citealt{Morris2007}). The outermost ejecta have been interacting with the inner, equatorial ring since day $\sim 3000$, resulting in a number of hotspots appearing in the optical images as well as a sharp increase in flux across the electromagnetic spectrum (e.g, \citealt{Groningsson2008a,Helder2013,Ng2013}). The optical emission from the ring peaked around $8000$ days and is now decaying as the ring is being destroyed by the shocks (\citealt{Fransson2015}). The same behavior is seen in the infrared (\citealt{Arendt2016}), while the soft X-ray light curve leveled off at $\sim 9500$ days (\citealt{Frank2016}). The X-ray emission from the ring is also affecting the inner ejecta. In particular, it is most likely responsible for the increase in optical emission seen after day $\sim 5000$ (\citealt{Larsson2011}). 

In the freely-expanding ejecta the observed Doppler shifts are directly proportional to the distance along the line of sight to the centre of the explosion ($v_{\rm{obs}} = z/t$, where $t$ is the time since the explosion on 1987 February 23 and $z$ is the distance). This means that the combination of imaging and spectroscopy makes it possible to infer the three-dimensional (3D) distribution of ejecta. This technique has previously been used for \8 in \cite{Kjaer2010} and \cite{Larsson2013} (L13 from here on). 

\cite{Kjaer2010} used observations obtained in 2005 with the integral field spectrograph SINFONI at the Very Large Telescope (VLT) to study the  [Si~I]$+$[Fe~II]$\ 1.644\ \mu$m and He I $2.058~\mu$m lines. A clearly asymmetric distribution was found, with the ejecta being predominantly blueshifted in the north and redshifted in the south. In L13 we confirmed these results for the $1.644\ \mu$m line with more recent SINFONI data, and also used Hubble Space Telescope (HST) spectra and imaging to compare with the distribution of H$\alpha$. The large-scale 3D emissivity of H$\alpha$ was found to be similar to the $1.644\ \mu$m line, except for H$\alpha$ extending to higher blueshifted velocities, although it should be noted that the 3D information for  H$\alpha$ had significantly lower resolution and suffered from strong contamination by scattered light from the ring. In addition, we studied the temporal evolution of the ejecta morphology in the HST images, finding that it changes from an approximately elliptical shape before $\sim 5000$ days, to an edge-brightened more irregular morphology thereafter. This transition coincides with and can be explained by the change in the dominant energy source powering the ejecta (from radioactive decay of $^{44}$Ti to X-rays from the ring, see also \citealt{Fransson2013}).  

Here we extend the work on the 3D distribution of ejecta using new observations obtained in 2014, approximately 10,000 days after the explosion, with HST/STIS and SINFONI. 
The STIS observations cover the whole ejecta with narrow (0$\farcs{1}$) slits. This represents a major improvement compared to our previous observations. The last STIS observations covering the whole ejecta were obtained using 0$\farcs{2}$ slits in 2004, when the ejecta were $\sim 2/3$ of the current size. The new observations allow us to create the most detailed map of H$\alpha$ to date, as well as the first spatially-resolved maps of the ejecta in [Ca II]~$\lambda \lambda 7292,\ 7324$,  [O~I]~$\lambda \lambda 6300,\ 6364$ and Mg~II~$\lambda \lambda 9218,\ 9244$. From the SINFONI observations, we present new maps of the [Si~I]$+$[Fe~II]$\ 1.644\ \mu$m and He I $2.058~\mu$m lines, which we compare to the STIS data. We also compare with our previous SINFONI observations in order to assess the time evolution of these two lines.  

This paper is organized as follows: we describe the observations and data reduction in section \ref{obs}, present the analysis and results in section \ref{analysis}, and discuss our findings in section \ref{discussion}. We summarize our conclusions and discuss future prospects in section \ref{conclusions}.

\section{Observations and data reduction}
\label{obs}

\subsection{HST/STIS observations}
\label{stisdata}

\begin{figure}
\begin{center}
\resizebox{70mm}{!}{\includegraphics{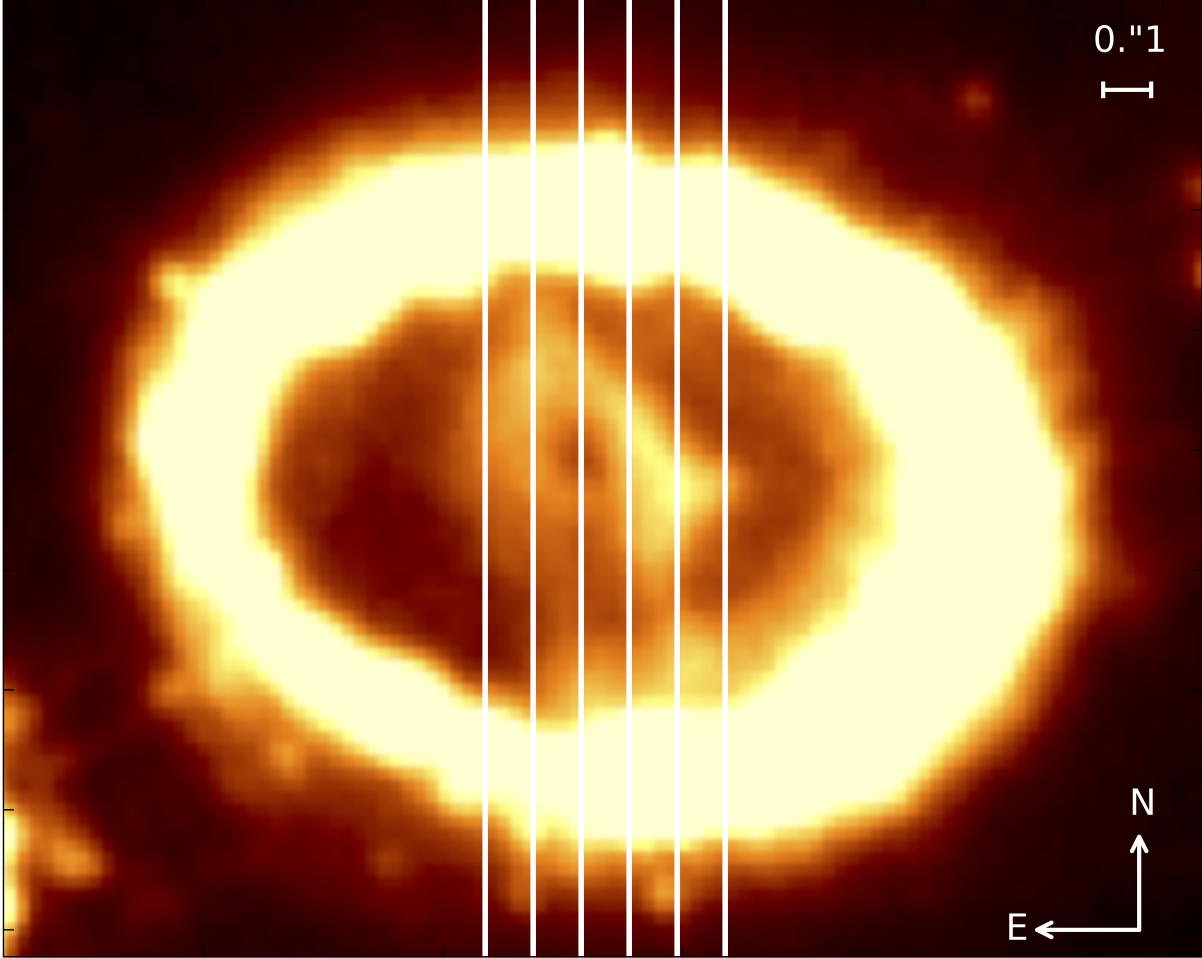}}
\caption{Slit positions for the STIS observations from day 10,035 shown superposed on the WFC3/F625W image from day 9973. The slit width is 0$\farcs{1}$. The ring has a radius of approximately $0\farcs{8}$ ($\sim 0.6$~ly) and an inclination of $44\dg$ \citep{Plait1995}.}
\label{slitpos}
\end{center}
\end{figure}

\begin{deluxetable}{ccccl}
\tabletypesize{\scriptsize}
\tablecaption{SINFONI observations \label{sinftable}}
\tablewidth{0pt}
\tablehead{
\colhead{Date\tablenotemark{1}} & \colhead{Filter} & \colhead{Exposure.\tablenotemark{2}} & \colhead{Seeing} & \colhead{Standard stars}\\
 & &  \colhead{(s)} &  \colhead{(")} & 
}
\startdata
2014-10-10 & H & 3600  & 0.81-1.20 & Hip025889, \\ 
& & & &Hip037007\\
2014-10-12 & K & 2400 & 0.59-0.80 & Hip040908\\
2014-11-12 & K & 1200 & 0.54-0.57 & Hip043987 \\
2014-11-13 & K & 1200 & 1.05-0.89 & Hip037597 \\
2014-12-01 & K & 2400  & 0.65-0.73 & Hip044317, \\
& & & &Hip032080  
\enddata
\tablenotetext{1}{The dates correspond to $10,090 - 10,152$ days after explosion}
\tablenotetext{2}{Integrated from single exposures of 600~s}
\end{deluxetable}

\begin{figure*}
\begin{center}
\resizebox{\hsize}{!}{\includegraphics{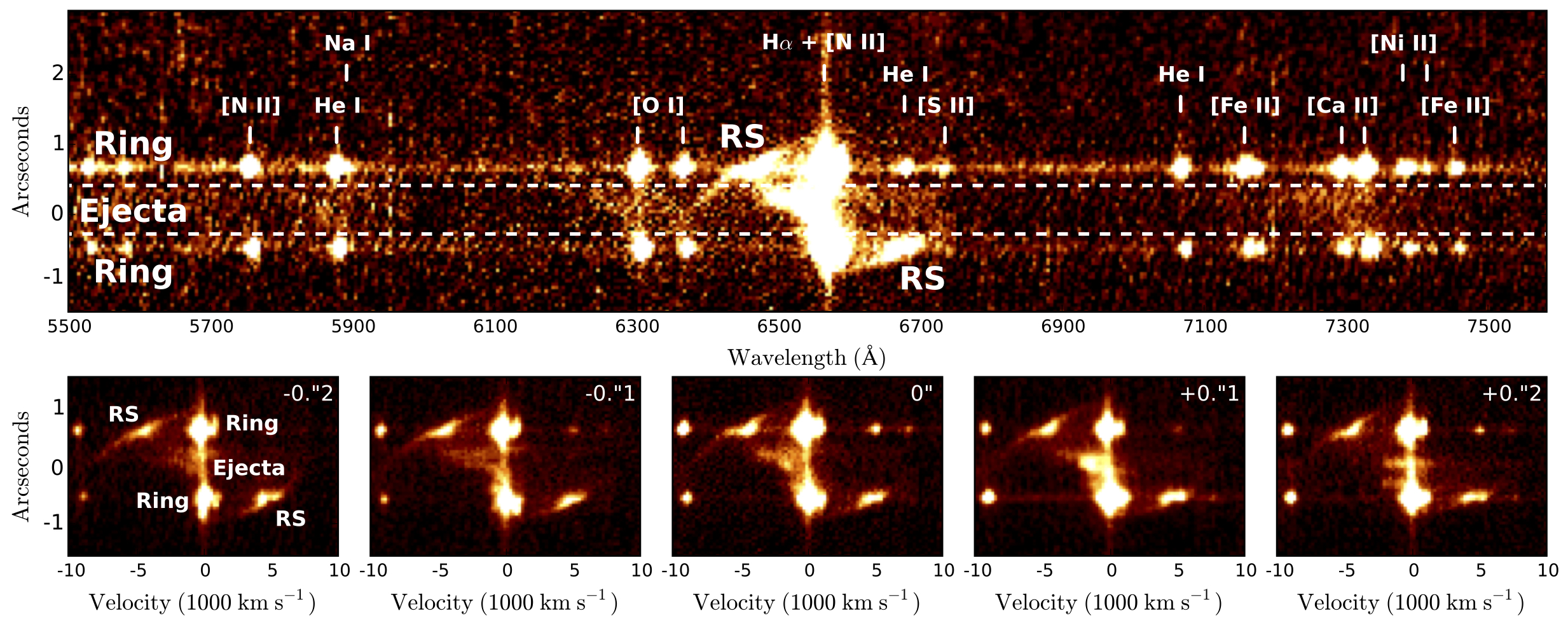}}
\caption{Two-dimensional STIS G750L spectra. Upper panel: spectrum from the central slit position in Fig.~\ref{slitpos} with all emission lines identified. The main emission components are also labeled: the ring, the ejecta and the reverse shock (RS). The latter is only seen in H$\alpha$. The dashed lines show the region used to extract the ejecta spectrum in Fig.~\ref{totspec}. Significant emission from the ejecta is detected in Na~I~$\lambda \lambda 5890,\ 5896$,  [O I]~$\lambda \lambda 6300,\ 6364$, H$\alpha$, [Ca~II]~$\lambda \lambda 7292,\ 7324$,  as well as Mg~II~$\lambda \lambda 9218,\ 9244$. The latter lines are outside the plotted region but shown in  Fig.~\ref{totspec}. Bottom panels: emission from H$\alpha$ for all five slit positions in Fig.~\ref{slitpos}, with the leftmost panel corresponding to the position furthest to the east, etc. The main emission components are also indicated in the left panel.}
\label{stis2d}
\end{center}
\end{figure*}

\8 was observed by HST/STIS between 2014 Aug $16 - 20$ ($10,035-10,039$ days after the explosion) using the G750L grating. This grating covers the wavelength range $\sim 5300-10,000\ \rm{\AA}$, where emission from H$\alpha$ dominates the spectrum. Around the H$\alpha$ line, the spectral resolution is $\sim 450\ \kms$. Nearly all of the ejecta were covered using five adjacent slit positions and a slit width of 0$\farcs{1}$. The slit positions are shown in Fig.~\ref{slitpos}, superposed on the WFC3/F625W image from day 9973 (see section \ref{w3data} below). We verified that the HST pointing was accurate to within a few per cent of the slit width.\footnote{See section 5.2 of the STIS data handbook (\citealt{Boestrom2011})} 

For each slit position, the total exposure time was $8100$~s, divided into 11 separate exposures dithered along the slit. We used the STISTOOLS package to combine these exposures, thus removing cosmic rays and bad pixels, and to create flux-calibrated 2D spectra. The absolute photometry for STIS L modes should be accurate to within $5\%$. We extracted 1D spectra by summing over rows in the 2D spectra, and corrected all spectra for the \8 systematic velocity of $287\ \kms$ (e.g,~\citealt{Groningsson2008a}). For all 1D spectra a background spectrum was also subtracted. The background was created as an average of spectra extracted from source-free regions just north and south of the SN.

\subsection{HST/WFC3 observation}
\label{w3data}

The HST imaging observations that are closest in time to the STIS observations discussed above were obtained on 2014 June 15 (9973 days after the explosion) using the WFC3 instrument. The image in the broad F625W filter, shown in  Fig.~\ref{slitpos}, is completely dominated by emission from  H$\alpha$ and is thus very useful for comparison with the STIS results. The image has previously been analyzed in \cite{Fransson2015}, where further details about the observations are provided.

\subsection{VLT SINFONI observations}
\label{sinfoniobs}

Observations of \8 were carried out by the SINFONI Integral Field Spectrograph at the VLT (\citealt{Eisenhauer2003,Bonnet2004}) between 2014 Oct-Dec ($10,090 - 10,152$ days after explosion) in the H and K bands . Details of the observations are provided in Table~\ref{sinftable}. The data were processed with the standard ESO pipeline (\citealt{Modigliani2007}) with improvements in the sky subtraction according to \cite{Davies2007}. Dedicated software for combination of individual cubes was developed by us. A detailed description of the data reduction and calibration is given in \cite{Kjaer2010}. 

The flux calibration and correction for telluric absorption was performed by
observing a number of $\sim$7.5 magnitude (H and K bands) B stars selected 
from the Hipparchos catalogue with 2MASS magnitudes. We adopt the 
zero points from \cite{Cohen2003}. The stellar atmosphere spectrum was
approximated by a Planck curve of the appropriate temperature for the spectral
type of the standard. Corrections for the residual H lines were made
manually. 

The accuracy of the fluxing and telluric correction was checked by fluxing the 
standard stars against each other, which results in a maximum error of 10\% in the 
absolute flux that we determine. The list of standards for each observation is 
provided in Table~\ref{sinftable}.

The spectral resolution is  approximately $110\ \kms$ and $60\ \kms$ for the H and K bands, respectively. The point spread function in SINFONI has an enhanced core and broad wings. As a measure of the spatial resolution we calculate the radius for 50\% (80\%) encircled energy as in \cite{Kjaer2010}, which gives $0\farcs{15}\ (0\farcs{24})$ for the H band, and  $0\farcs{12}\ (0\farcs{22})$ for the K band.

\subsection{Spatial alignment of data sets}
\label{alignment}

A good spatial alignment of the data sets is needed in order to make a detailed comparison of the 3D emissivity in different lines. The equatorial ring has a very similar shape in the optical and NIR and we therefore use this as a reference. In particular, we translate the positions on the images to positions on the STIS 2D spectra using the hotspots in the ring that fall inside the slits. In the case of the WFC3/F625W image (which has a much wider field of view than SINFONI) we were able to check the accuracy of this alignment using the outer rings and stars that fall within the slits further away from the SN. We found the agreement to be good to within $0\farcs{013}$, which corresponds to about  $25 \%$ of the STIS pixel scale along the slit (i.e. in the north - south direction). Uncertainties in position from the expansion of the ejecta in the time between the first (HST) and last (SINFONI K-band) observations are negligible, with the age-difference corresponding to an increase in size by $\sim 100~\rm{days}/10,000~\rm{days} =1\%$.

\section{Analysis and results}
\label{analysis}

\subsection{Ejecta spectra} 
\label{ejspectra}

\begin{figure}
\begin{center}
\resizebox{\hsize}{!}{\includegraphics{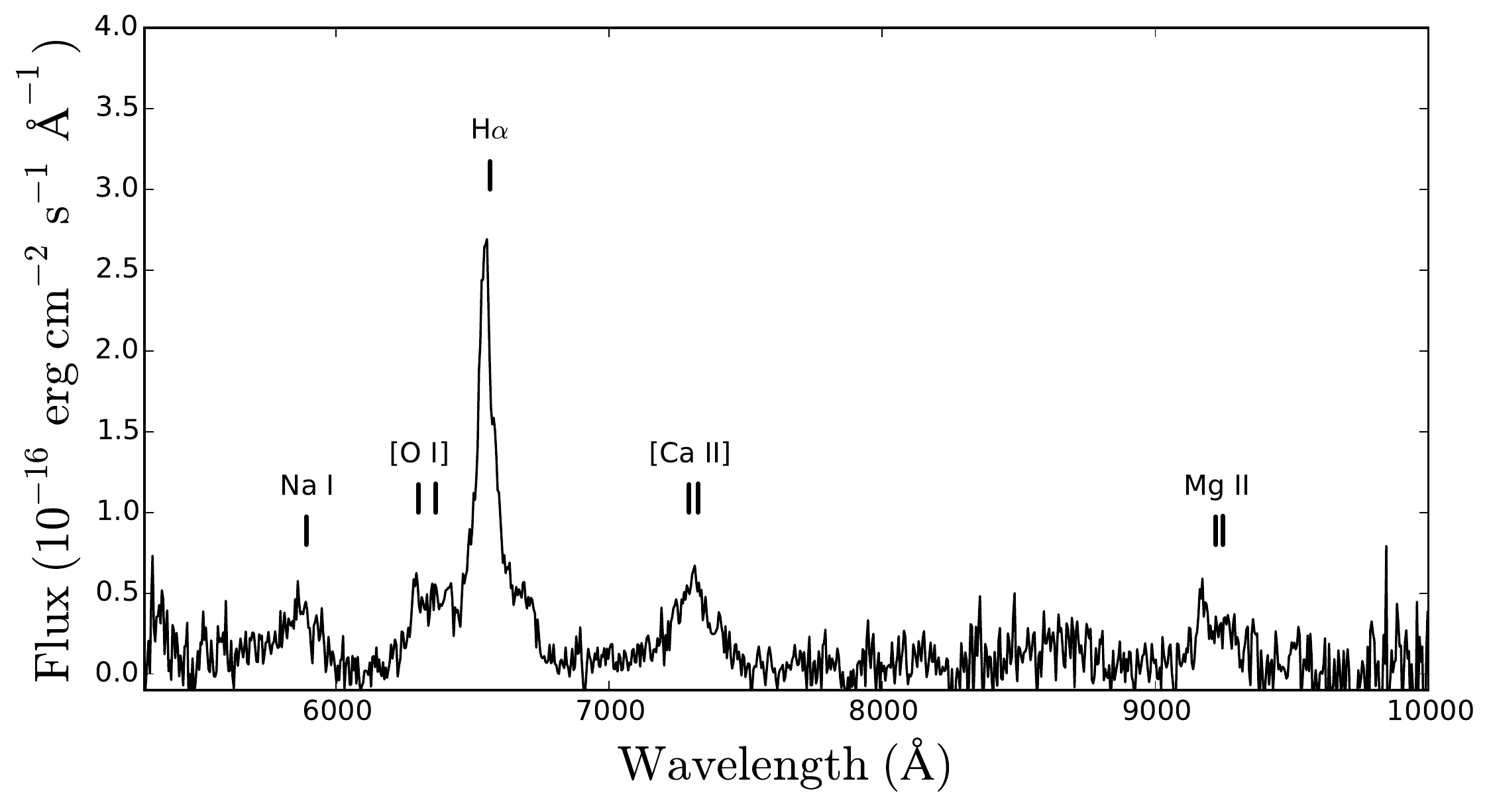}}\\
\resizebox{\hsize}{!}{\includegraphics{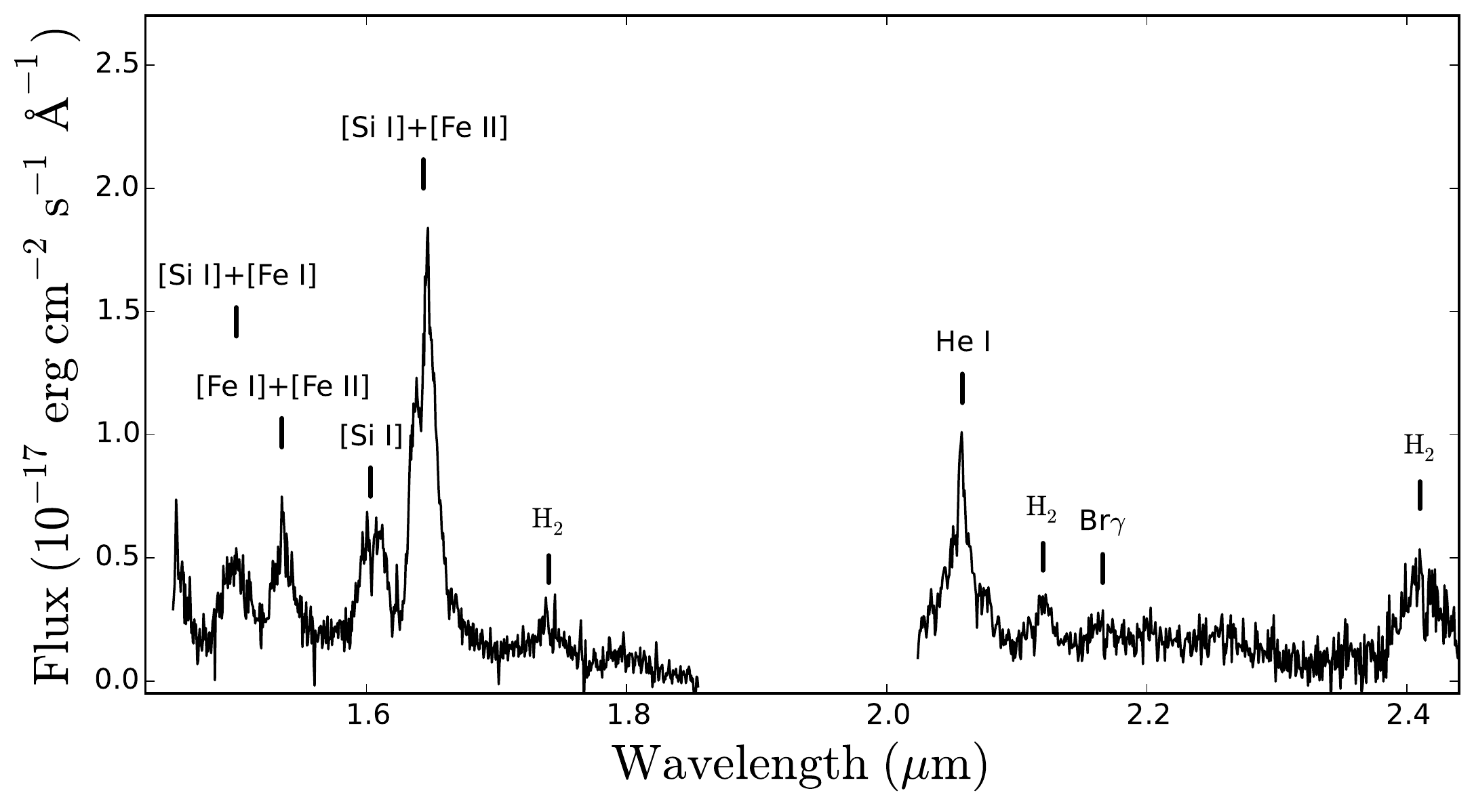}}
\caption{Upper panel: Total STIS G750L spectrum of the ejecta obtained by summing all five slits. The extraction region is indicated in Fig.~\ref{stis2d}. The spectrum has been corrected for scattered light from the ring. Lower panel: Total ejecta spectrum in the SINFONI H and K bands. The data have been binned by a factor of three and corrected for scattered light from the ring.}
\label{totspec}
\end{center}
\end{figure}

\begin{figure}
\begin{center}
\resizebox{70mm}{!}{\includegraphics{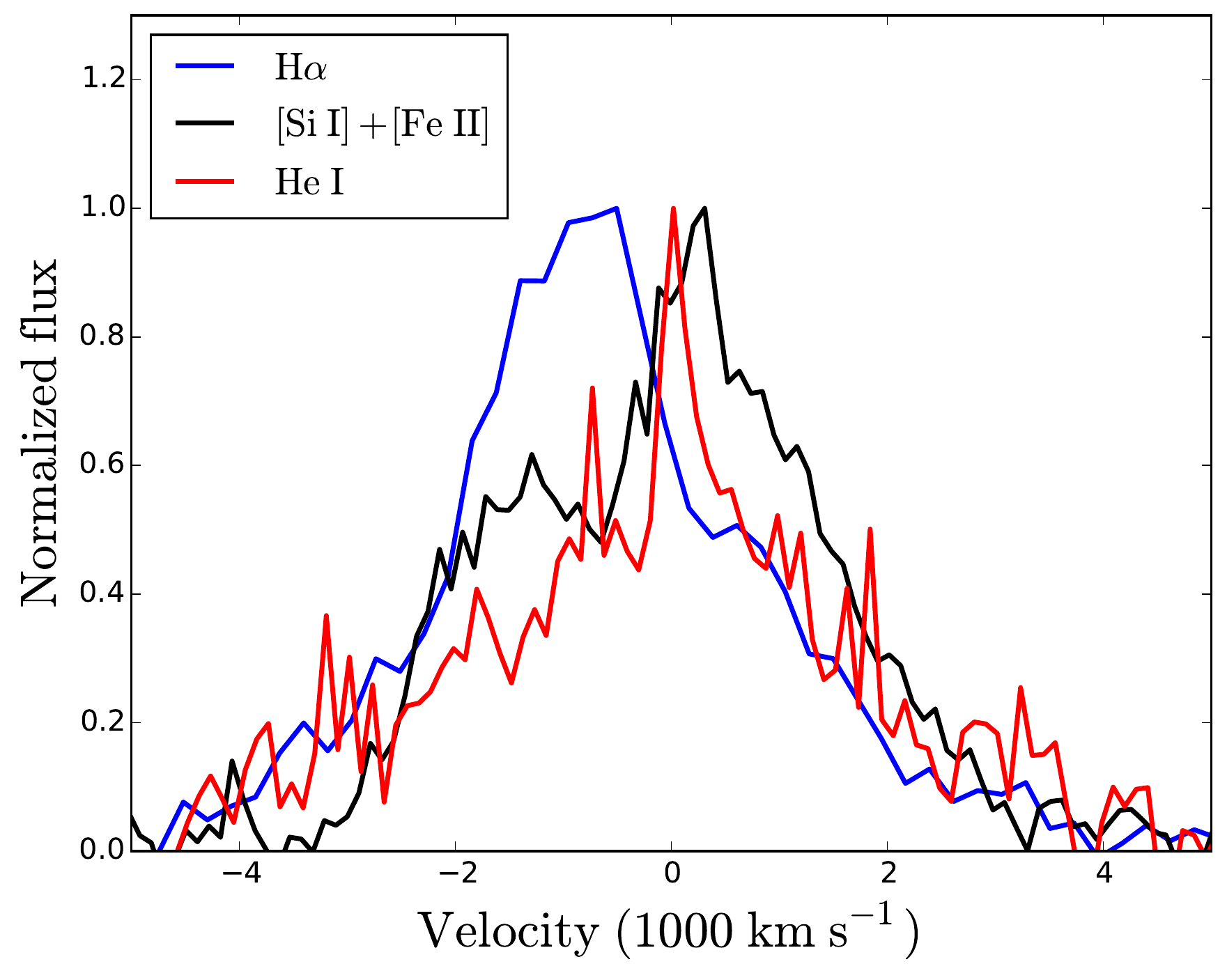}}
\caption{Comparison of the velocity profiles of the three strongest lines from the ejecta; H$\alpha$, [Si~I]$+$[Fe~II]~$1.644\ \mu$m and He~I~$2.058\ \mu$m. The lines have been normalized to the same maximum value after subtracting the continuum. The SINFONI spectra have been binned by a factor of three.}
\label{fullprofiles}
\end{center}
\end{figure}
\begin{figure*}
\begin{center}
\resizebox{38mm}{!}{\includegraphics{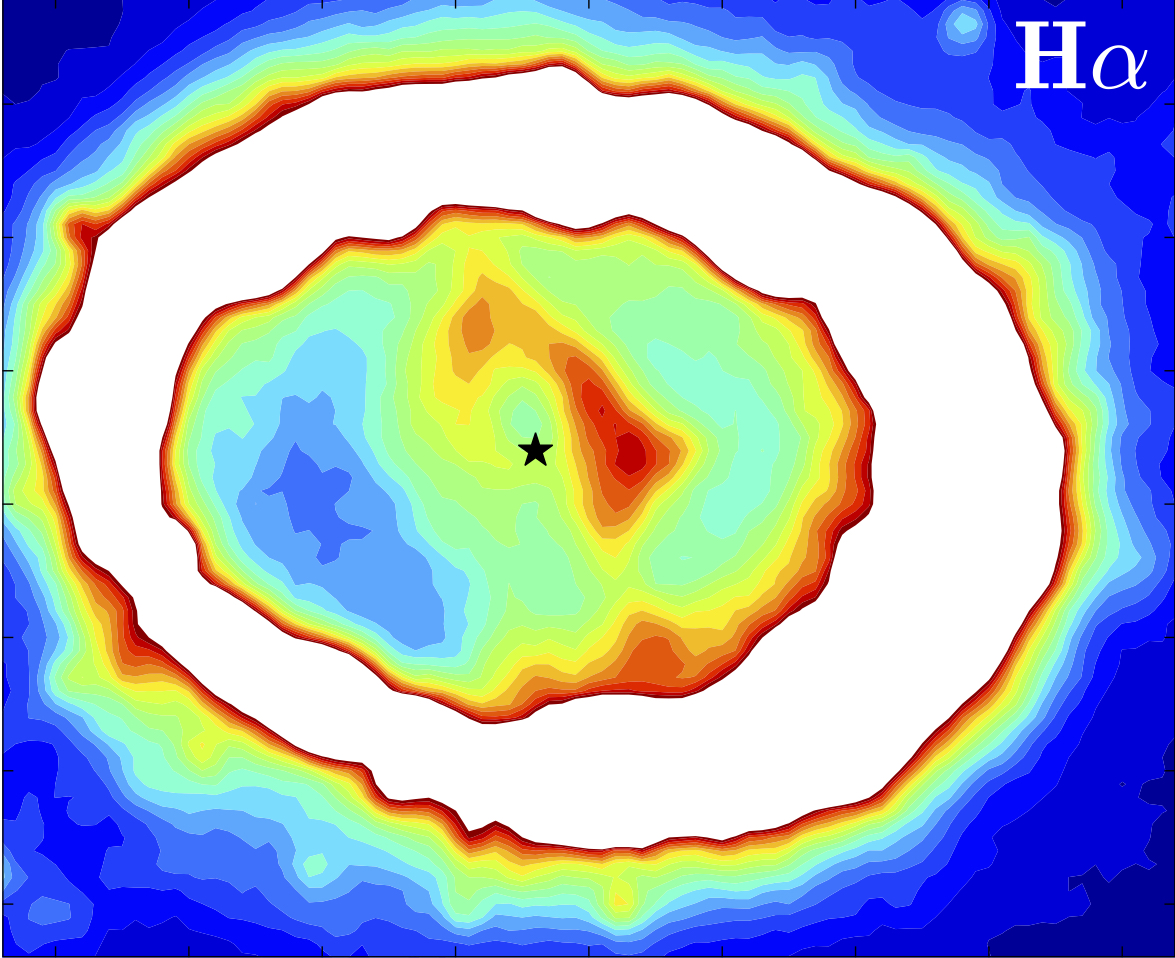}}
\resizebox{38mm}{!}{\includegraphics{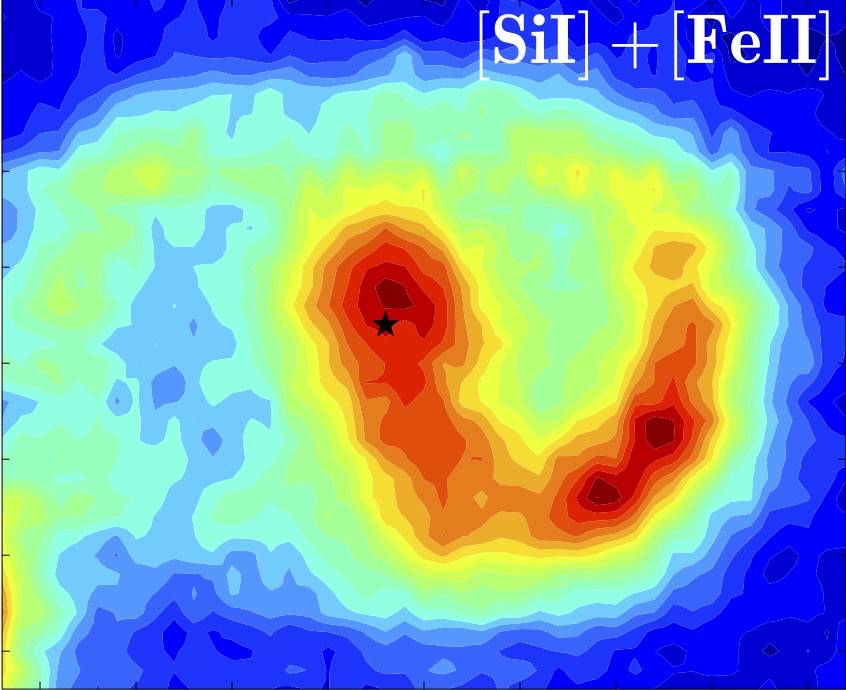}}
\resizebox{38mm}{!}{\includegraphics{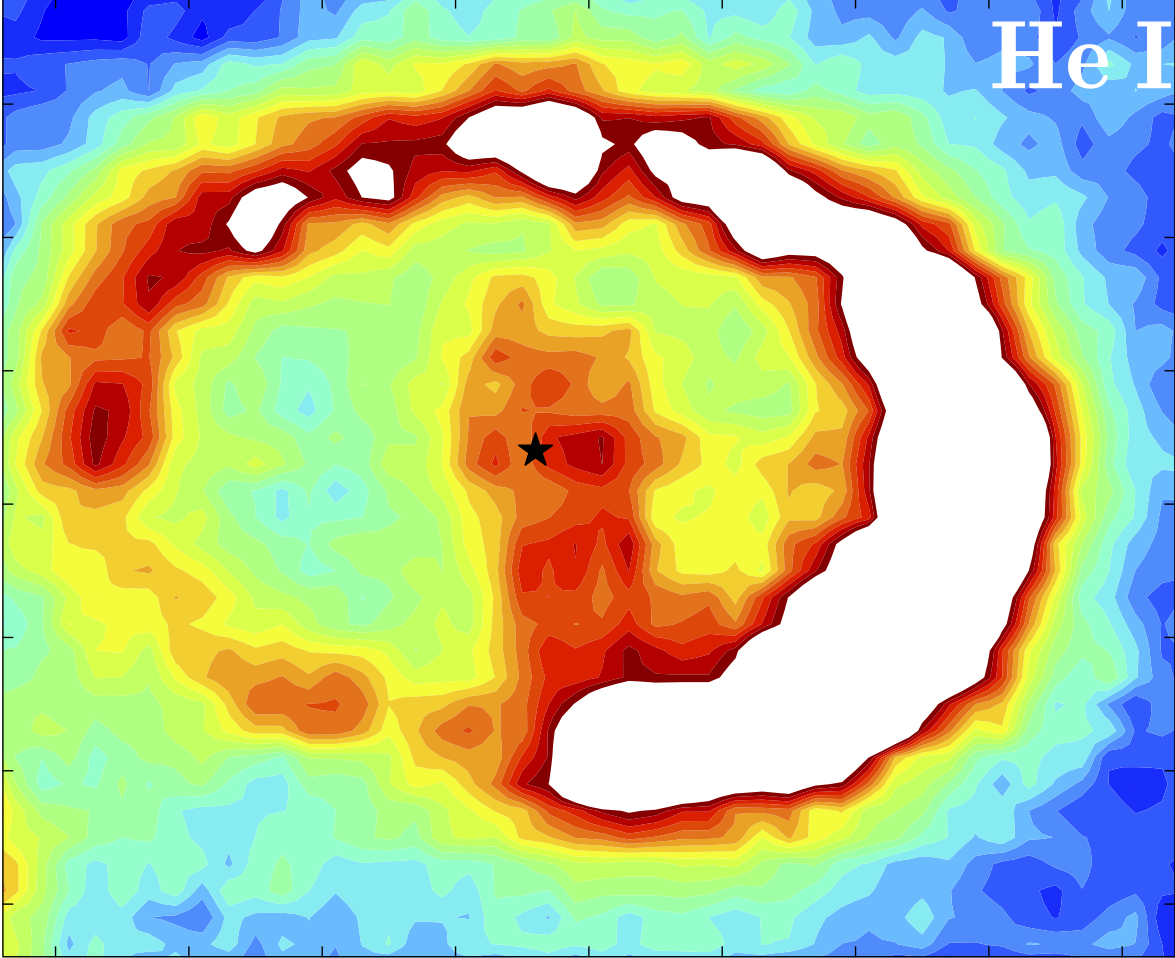}}
\resizebox{38mm}{!}{\includegraphics{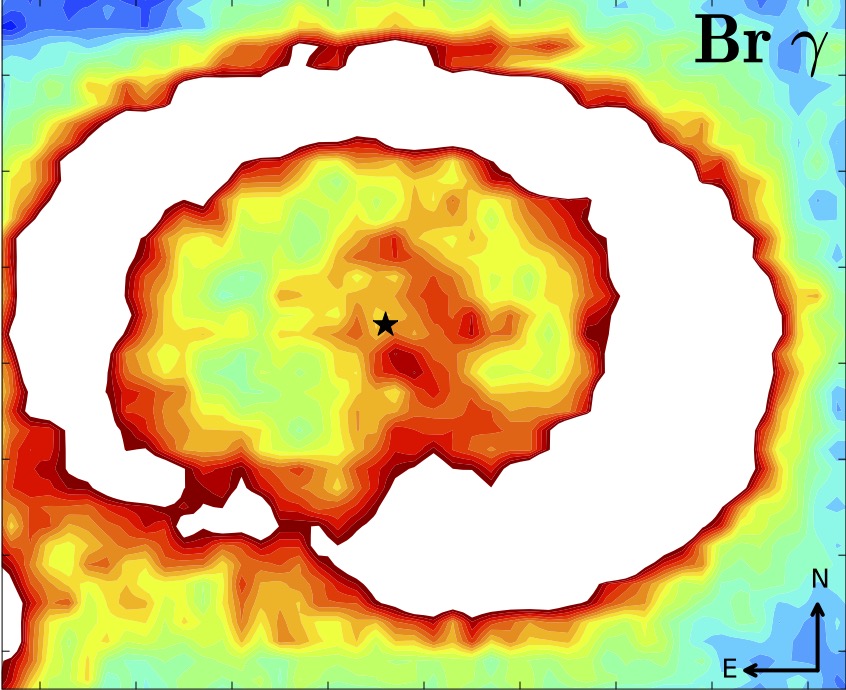}}
\caption{Contour plots of the ejecta in different lines. The left panel was produced from the WFC3 F625W image in Fig.~\ref{slitpos}, which is dominated by emission from H$\alpha$. The next three panels show emission from [Si~I]$+$[Fe~II], He~I and Br$\gamma$, respectively. These three panels were produced from SINFONI data, covering $\pm 3000~\rm{km\ s^{-1}}$ around each line with the central $\pm 450~\rm{km\ s^{-1}}$  removed. The latter cut removes most of the contribution from the narrow lines from the ring. The remaining emission from the ring seen in the images is primarily continuum emission. This is particularly strong in the south-west part of the ring.  The contours in each plot are linearly spaced in flux between the maximum (red) and minimum (blue) of the ejecta region. The ring is shown in white in those parts where it is brighter than the maximum of the ejecta region. The black star marks the geometric centre of the ring.}
\label{conts}
\end{center}
\end{figure*}

Fig.~\ref{stis2d} shows the STIS 2D spectrum in the $5600-7600\ \rm{\AA}$ range from the central slit position (see Fig.~\ref{slitpos}) together with the H$\alpha$ emission from all five slit positions. H$\alpha$ is by far the strongest line in the spectrum with clear emission components originating from the inner ejecta, the ring and the reverse shock. The latter is seen as thin streaks near the ring, extending out to $\sim \pm 10,000\ \kms$ (\citealt{Michael2003,Heng2006,France2010,France2015}). In addition to H$\alpha$, we also detect the following emission lines from the ejecta (listed in order of decreasing intensity): [Ca II]~$\lambda \lambda 7292,\ 7324$, [O~I]~$\lambda \lambda 6300,\ 6364$, Na~I~$\lambda \lambda 5890,\ 5896$ and Mg~II~$\lambda \lambda 9218,\ 9244$.  These lines have all been detected in previous observations (\citealt{Fransson2013}). In the case of the [O~I] doublet we note that the H$\alpha$ from the reverse shock overlaps with the $6364\ \rm{\AA}$ component and that there may also be blending with emission from Fe~I~$\lambda 6300$  (\citealt{Fransson2002}). It is also possible that there is a contribution from  He~I~$\lambda 5876$ blended with the Na~I doublet, although \cite{Jerkstrand2011} find that Na~I dominates. The identification of the Mg~II lines is discussed in \cite{Jerkstrand2011} and \cite{Fransson2013}. While this emission feature is outside the plotted region in Fig.~\ref{stis2d}, it is seen in the upper panel of Fig.~\ref{totspec}, which shows the full 1D ejecta spectrum obtained by summing all five slits.  

The STIS spectrum Fig.~\ref{totspec} has been corrected for scattered light from the ring using the fact that the lines from the ring are narrow (FWHM $\sim300\ \kms$, e.g. \citealt{Fransson2015}) compared to the lines from the ejecta, which have FWHM $\sim 2500\ \kms$. The narrow lines can thus be removed by subtracting a rescaled ring spectrum from the spectrum extracted from the ejecta region. At this late epoch, there is no significant contribution from the fading outer rings, which emit very narrow lines with FWHM $\sim 20\ \kms$ (e.g.~\citealt{Tziamtzis2011}). The subtraction of scattered light from the ring is further described in Section \ref{3ddist} below.  We note that the latest STIS observations discussed in L13 (from days 6360 and 8378) did not have sufficient resolution to perform a reliable correction for the ring component. 

The lower panel of Fig.~\ref{totspec} shows the SINFONI spectrum of the ejecta in the H and K bands. The spectra have been corrected for scattered light from the ring in the same way as for STIS. These spectra are consistent with those presented in \cite{Kjaer2010}. In the H-band the strongest line from the ejecta is the $1.644\ \mu$m line, which is a blend of [Si I] and [Fe II], with [Si I] likely dominating. This band also contains several other lines from [Si~I], [Fe~I] and [Fe~II] (see \citealt{Kjaer2010} for details). We further identify the feature at $1.74\ \mu$m with  H$_2$ (Fransson et al.~2016). In the K-band the strongest line is the $2.058~\mu$m line from He~I, followed by the rotational-vibrational lines from H$_2$ at $2.40~\mu$m and $2.122~\mu$m, as well as a weak signal from Br~$\gamma$ at $2.166~\mu$m. Additional weak lines from H$_2$ may also be present, making up part of the ``continuum". The H$_2$ lines and continuum are discussed in detail in \cite{Fransson2016}. 

Fig.~\ref{fullprofiles} shows a comparison of the line profiles for the three strongest lines; H$\alpha$, [Si~I]$+$[Fe~II]~$1.644\ \mu$m and He~I~$2.058\ \mu$m. The profiles are  asymmetric and clear differences can be seen between the different lines. The origin of these differences are investigated in more detail below.  

\subsection{Images} 
\label{images}
\begin{figure*}
\begin{center}
\resizebox{160mm}{!}{\includegraphics{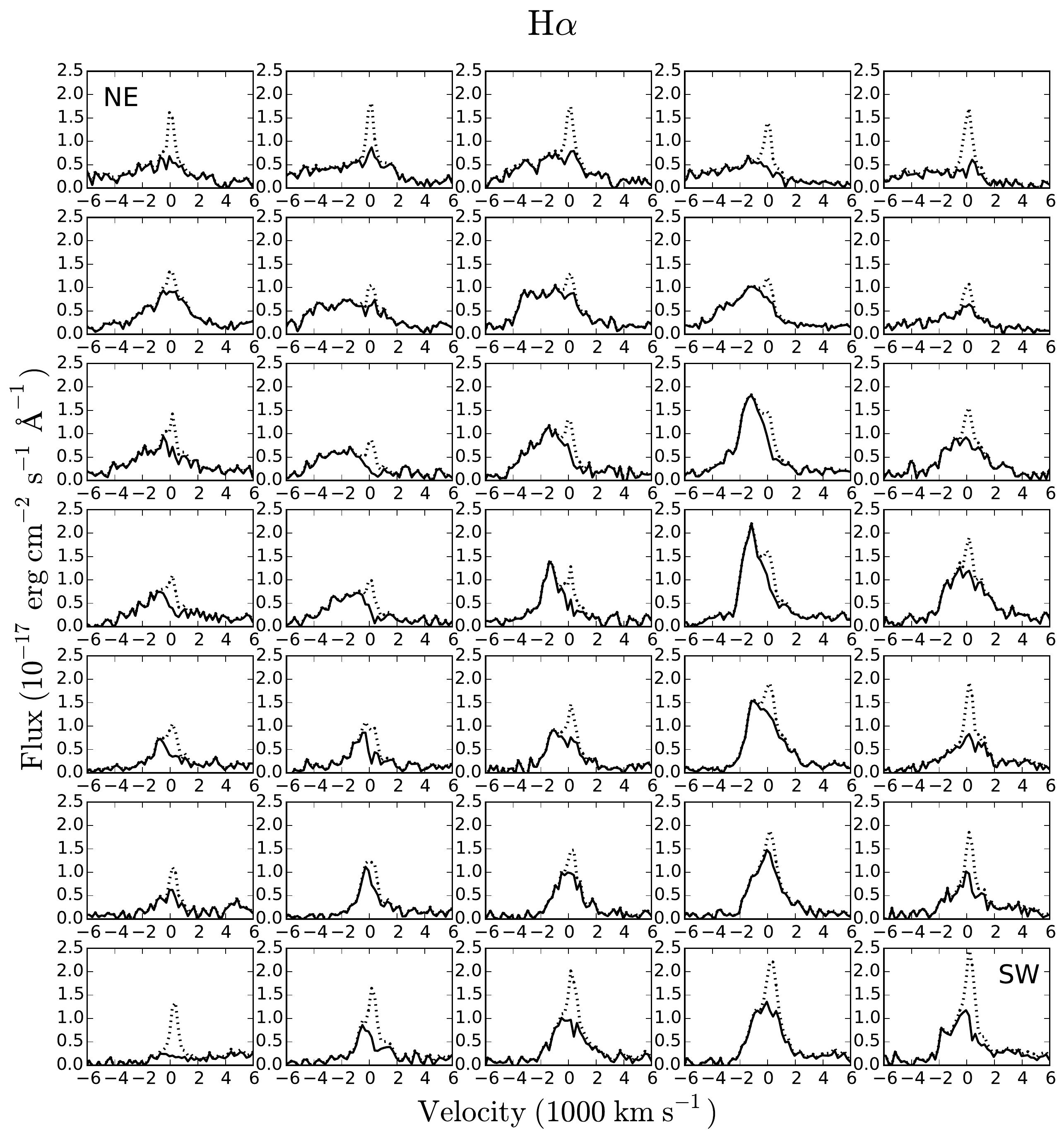}}
\resizebox{60mm}{!}{\includegraphics{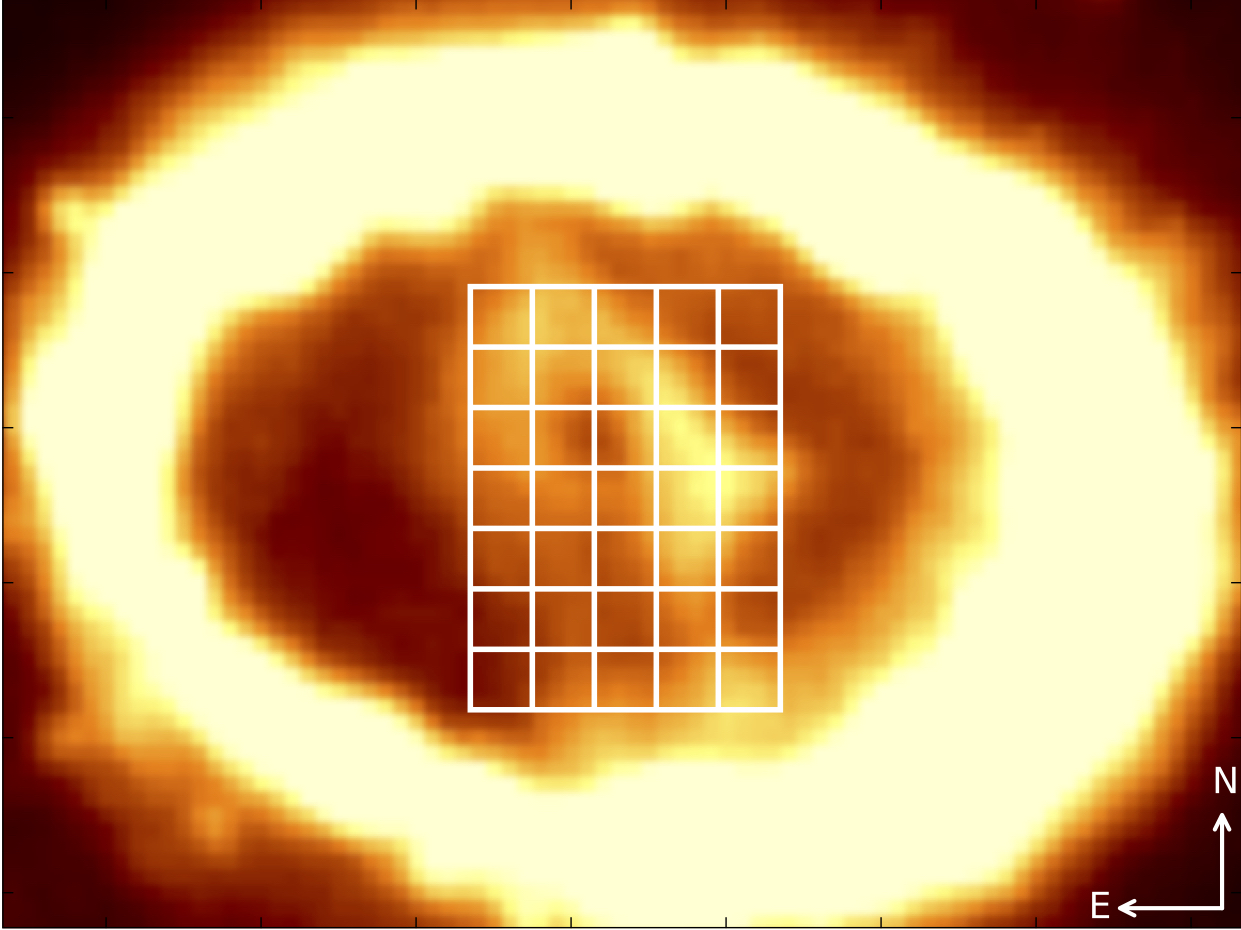}}
\caption{Spatially resolved STIS G750L spectra of H$\alpha$, extracted from $0\farcs{1} \times 0\farcs{1}$ regions of the ejecta. The regions are shown superposed on the WFC3/F625W image in the bottom panel. The panels with the line profiles are organized in the same way as the regions, with the upper, left panel corresponding to the region in the north-east corner.  At this epoch $0\farcs{1}$ corresponds to $866\ \kms$. The dashed and solid lines show the spectra before and after correcting for scattered light from the ring, respectively. }
\label{hamap}
\end{center}
\end{figure*}
\begin{figure*}
\begin{center}
\resizebox{160mm}{!}{\includegraphics{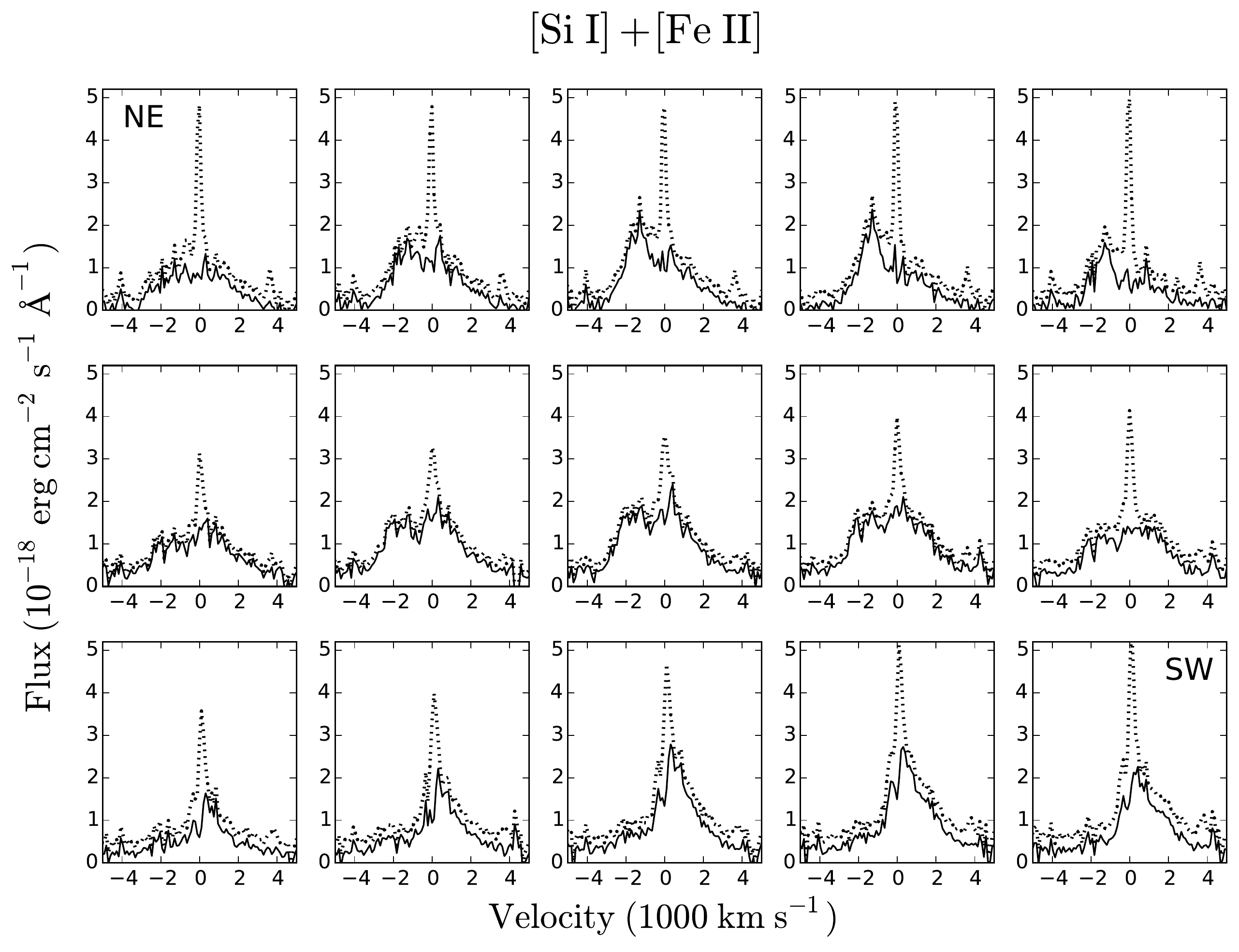}}
\resizebox{60mm}{!}{\includegraphics{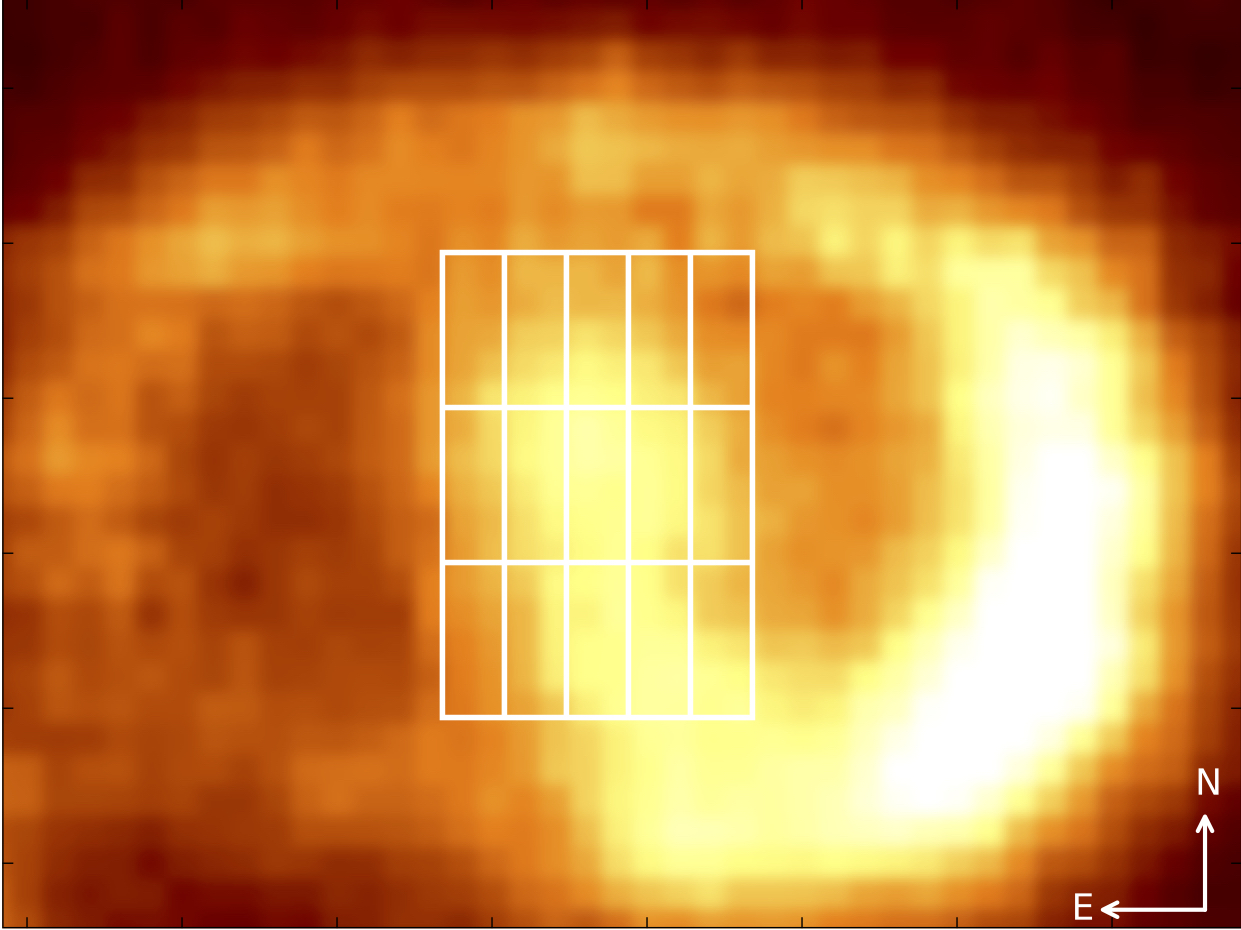}}
\caption{Spatially resolved SINFONI H-band spectra extracted from $0\farcs{1} \times 0\farcs{25}$ regions of the ejecta.  The spectra are centered on zero velocity for the  [Si~I]$+$[Fe~II]~$1.644\ \mu$m line. The regions are shown superposed on the image of the line ($\pm 3000~\rm{km\ s^{-1}}$) in the bottom panel. The panels with the line profiles are organized in the same way as the regions, with the upper, left panel corresponding to the region in the north-east corner. The dashed and solid lines show the spectra before and after correcting for scattered light from the ring, respectively.}
\label{simap}
\end{center}
\end{figure*}
\begin{figure*}
\begin{center}
\resizebox{160mm}{!}{\includegraphics{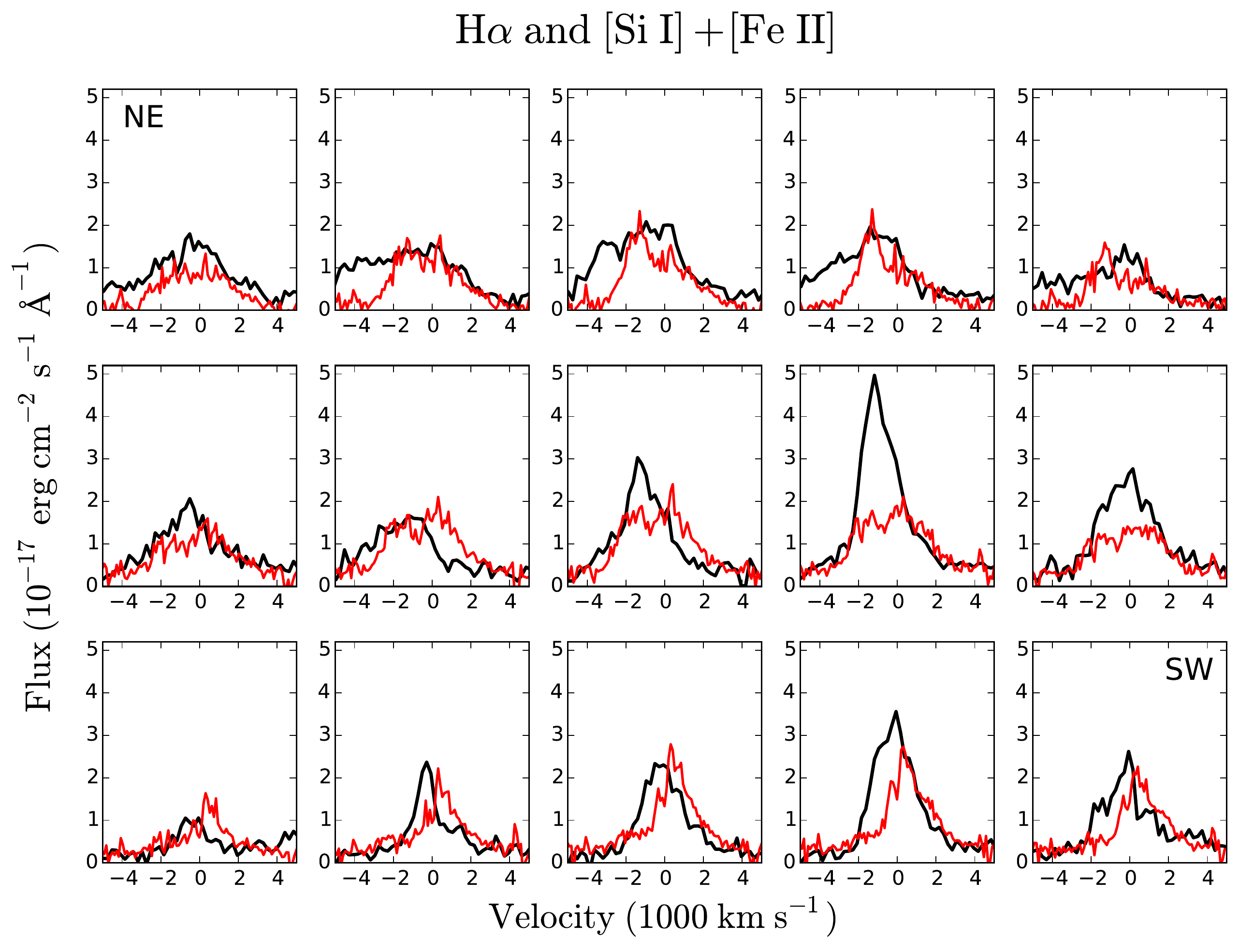}}
\resizebox{60mm}{!}{\includegraphics{regions_map_sinfoni_v3_nice_vb.jpeg}}
\resizebox{60mm}{!}{\includegraphics{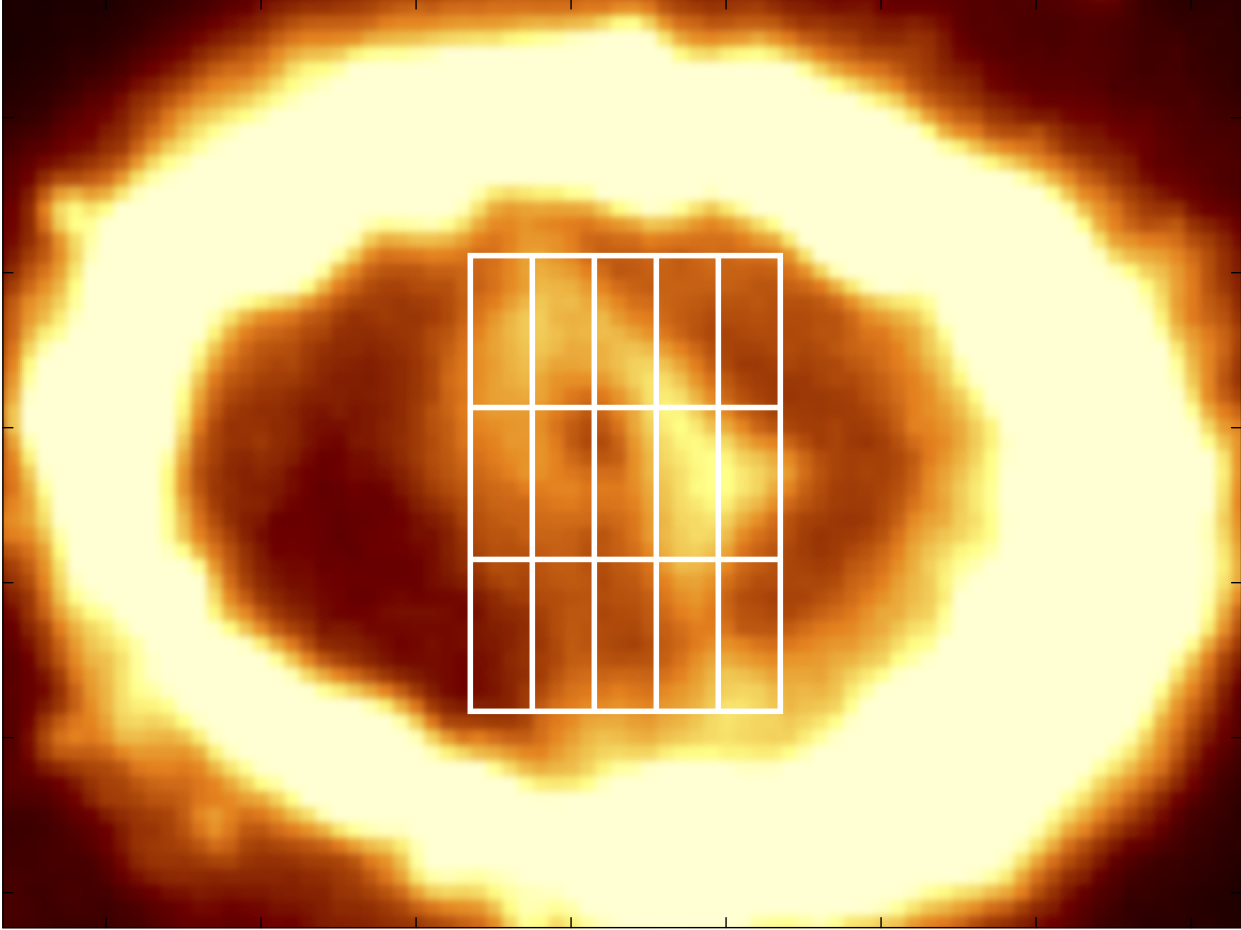}}
\caption{Comparison of the  [Si~I]$+$[Fe~II]  (red) lines and H$\alpha$ (black) extracted from $0\farcs{1} \times 0\farcs{25}$ regions of the ejecta. The regions are shown in the bottom panel, superposed on the image of the $1.644~\mu \rm{m} \pm 3000~\rm{km\ s^{-1}}$ line (left) and the WFC3/F625W image (right). The panels with the line profiles are organized in the same way as the regions, with the upper, left panel corresponding to the region in the north-east corner. The spectra have been corrected for emission from the ring as demonstrated in Figs.~\ref{hamap} and \ref{simap}. The SINFONI spectra have been multiplied by a factor of ten to match the flux level of the STIS spectra.}
\label{hsimap}
\end{center}
\end{figure*}

In Fig.~\ref{conts} we show contour plots of the ejecta for [Si~I]$+$[Fe~II] ($1.644\ \mu$m, SINFONI H-band), He~I ($2.058\ \mu$m, SINFONI K-band), Br$\gamma$ (SINFONI K-band) and H$\alpha$ (WFC3/F625W image). The plots produced from SINFONI data cover $\pm 3000\ \kms$ around each line, excluding the central $\pm 450\ \kms$  in order to remove most of the emission from the ring. The remaining emission from the ring  in these images is mainly  bound-free emission from recombination of H~II and  He~II. 

The ejecta are elongated in the north-east -- south-west direction in all images, but the detailed morphology differs significantly in the different lines. While both the [Si~I]$+$[Fe~II] and He~I emission is centrally peaked, H$\alpha$ and Br$\gamma$ show an edge-brightened morphology. The morphologies of the former three lines are consistent with previous observations in terms of being centrally peaked or edge-brightened (\citealt{Kjaer2010} and L13), although the H$\alpha$ morphology clearly continues to evolve (see section \ref{time} for more details on the time evolution). The morphology of Br$\gamma$ has not been investigated in previous observations. This image shows a low-surface brightness region in the middle, which coincides with the ``hole" in the H$\alpha$-dominated HST image. The image of the Br$\gamma$ line also has a significant contribution from the continuum emission from the ejecta (see Fig.~\ref{totspec}). 

In the freely expanding ejecta, an angle of $0\farcs{1}$ corresponds to $866\ \kms$ at the distance of 50~kpc and 10,000 days since the explosion. This means that the inner ejecta (defined by the red contours in Fig.~\ref{conts}) extend in the range $\sim [-2000,+2300]\ \kms$ along the east-west direction and  $\sim [-4500 ,+2800]\ \kms$ along the south-north direction. The extent of the ejecta to the south is more uncertain than the other directions  due to the overlap with the ring. 

\subsection{3D emissivities}
\label{3ddist}

\begin{figure*}
\begin{center}
\resizebox{\hsize}{!}{\includegraphics{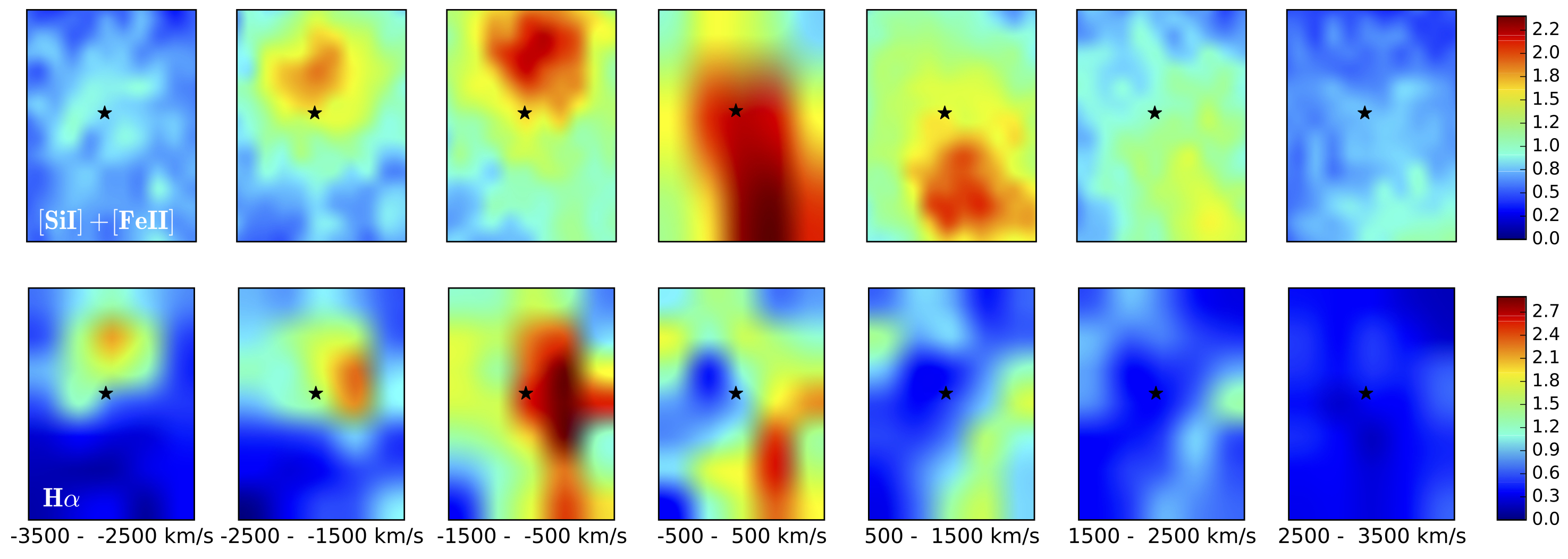}}
\vspace{0.1cm}\\
\resizebox{50mm}{!}{\includegraphics{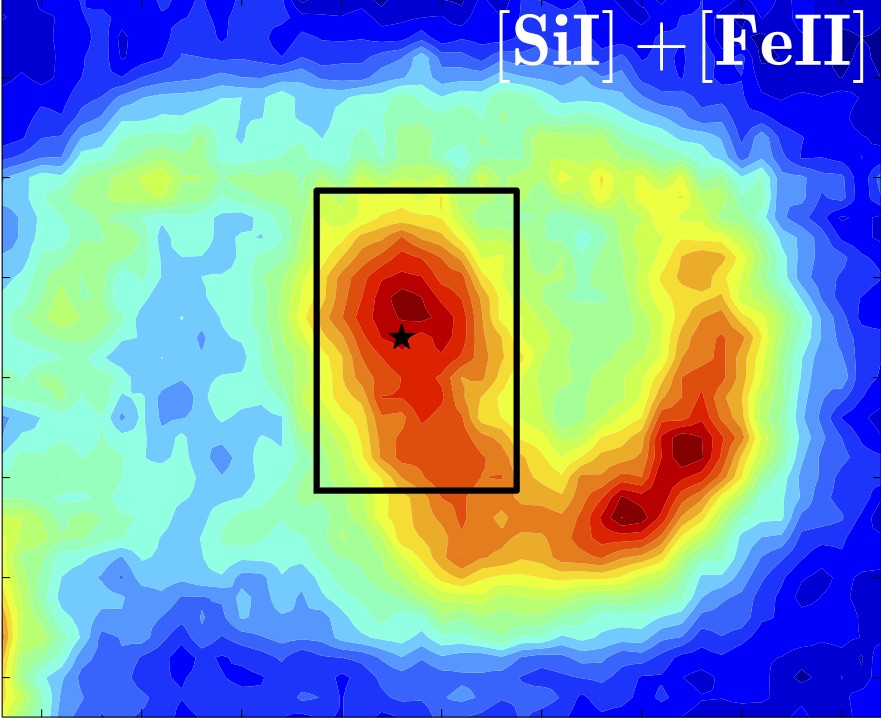}}
\hspace{0.5cm}
\resizebox{50mm}{!}{\includegraphics{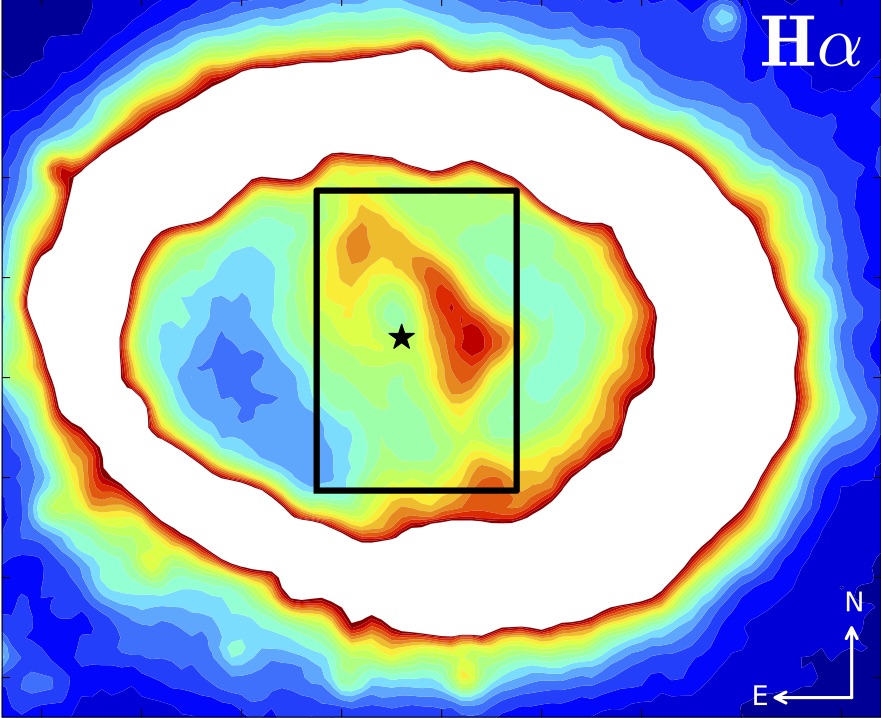}}
\caption{Images of the ejecta in velocity slices for [Si~I]$+$[Fe~II] (top row) and H$\alpha$ (middle row), smoothed with a 2D spline function. The color bars show the intensity in units of $10^{-17}\ \rm{erg\ cm^{-2}\ s^{-1}}$ and $10^{-16}\ \rm{erg\ cm^{-2}\ s^{-1}}$, respectively. The plotted regions are indicated by the black boxes in the bottom row on the contour plots for the SINFONI [Si~I]$+$[Fe~II] $1.644\ \mu \rm{m} \pm 3000\ \kms$ image (left) and the WFC3/F625W image (right). Both data sets have been corrected for emission from the ring. In the case of [Si~I]$+$[Fe~II], this results in a significantly lower spatial resolution in the central bin (see text for details).  Taking the centre of the ring (black star) as the point of zero velocity, the plotted area covers $[-2000,+2300]\ \kms$ in the x-direction and $[-3300,+2800]\ \kms$ in the y-direction.  } 
\label{slices}
\end{center}
\end{figure*}
\begin{figure*}
\begin{center}
\resizebox{80mm}{!}{\includegraphics{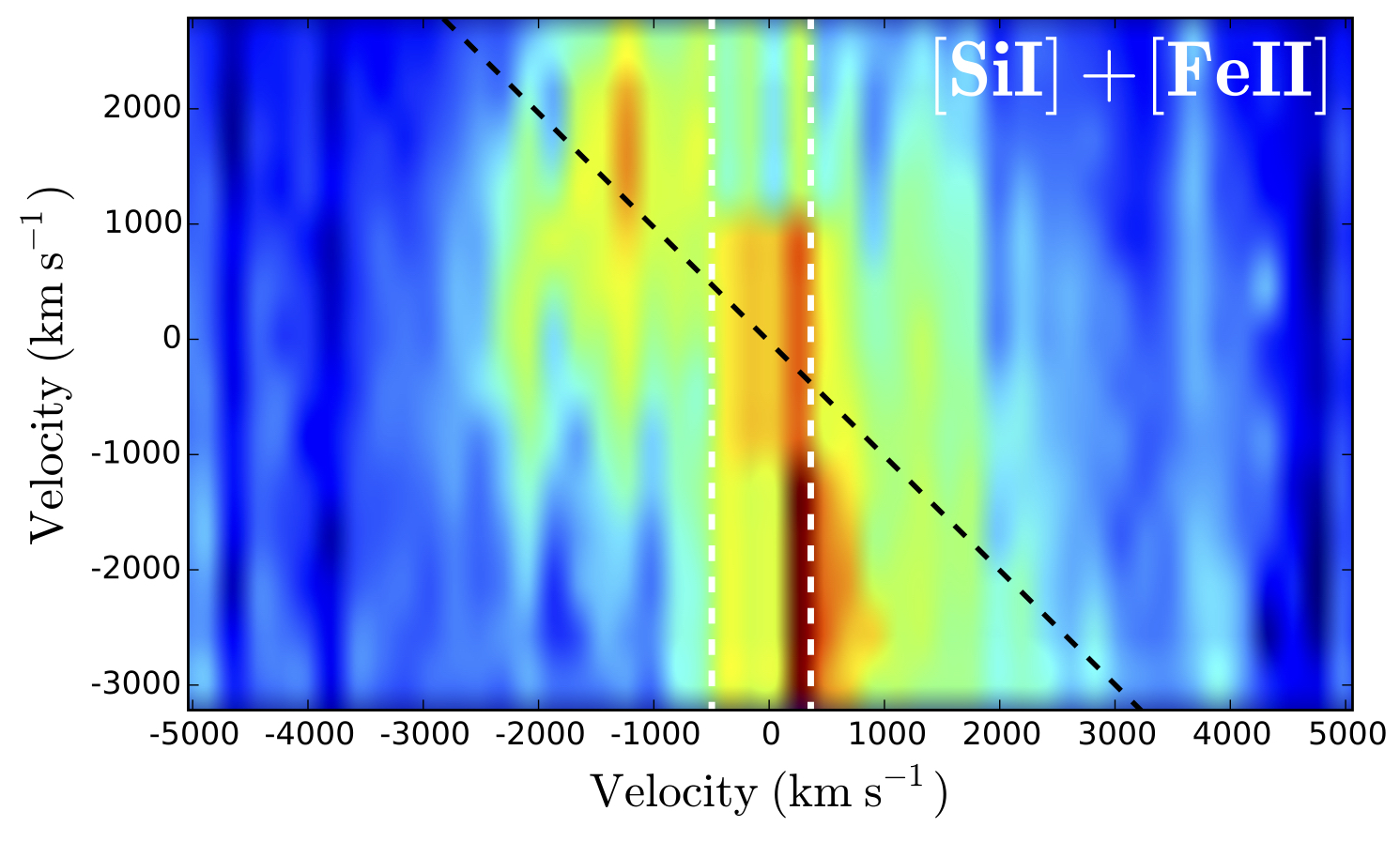}}
\resizebox{80mm}{!}{\includegraphics{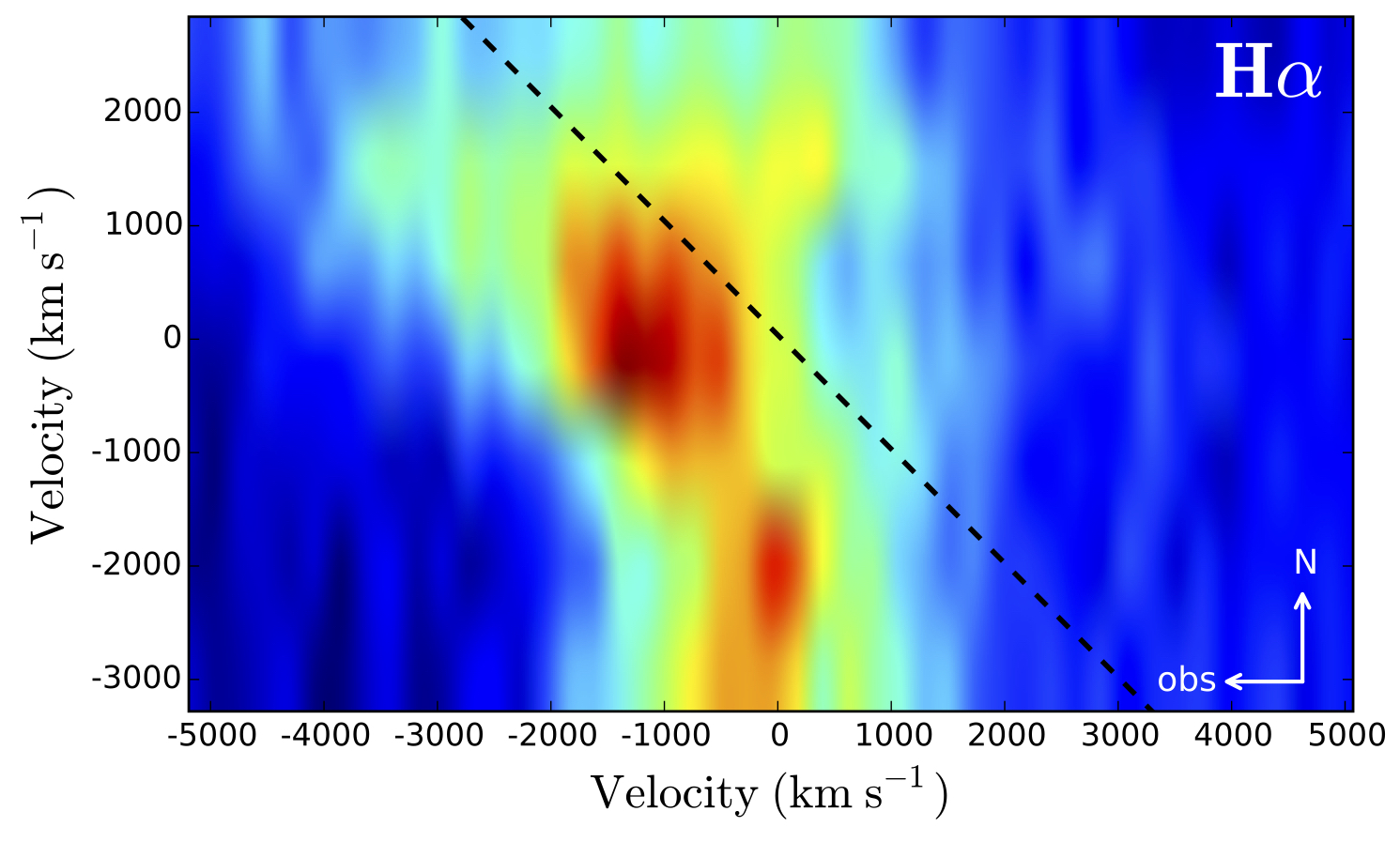}}
\caption{Side views of the ejecta for [Si~I]$+$[Fe~II] (left) and H$\alpha$ (right). The observer is located to the left and the plane of the ring is indicated by the dashed, black line. The spatial region covered is the same as in Fig.~\ref{slices} above. The spatial scale was converted to a velocity scale in the north-south direction using the relation $0\farcs{1}=866\ \kms$. The SINFONI spectra have been binned by a factor of six compared to the original resolution. Both data sets have been corrected for emission from the ring and smoothed by a 2D spline function. In the case of [Si~I]$+$[Fe~II], the ring correction results in a significantly lower resolution (by a factor of five) in the north-south direction inside the region marked by the white dashed lines. The color scale is set between the maximum (red) and minimum (blue) of each image.}
\label{sideview}
\end{center}
\end{figure*}
\begin{figure*}
\begin{center}
\resizebox{59mm}{!}{\includegraphics{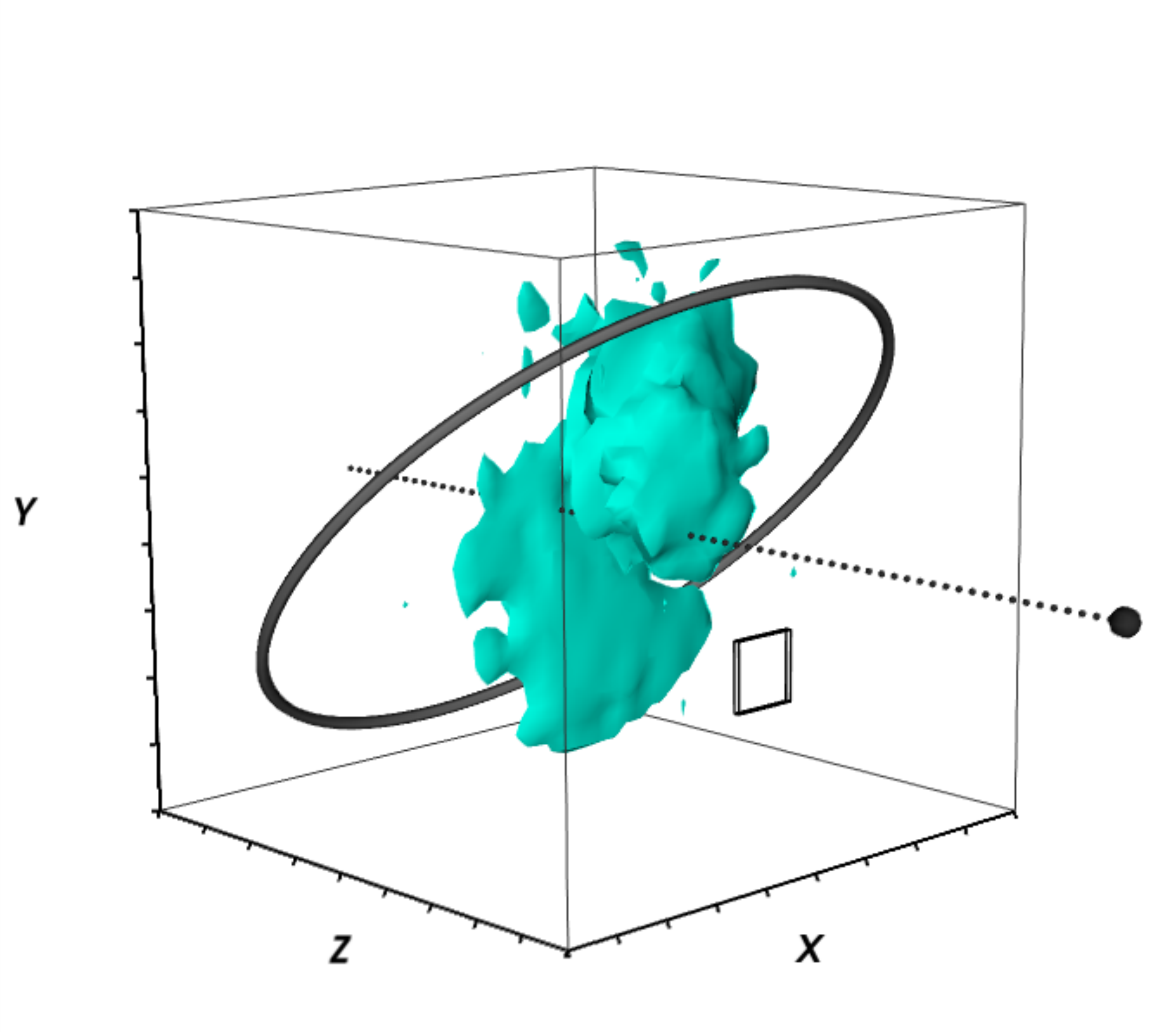}}
\resizebox{59mm}{!}{\includegraphics{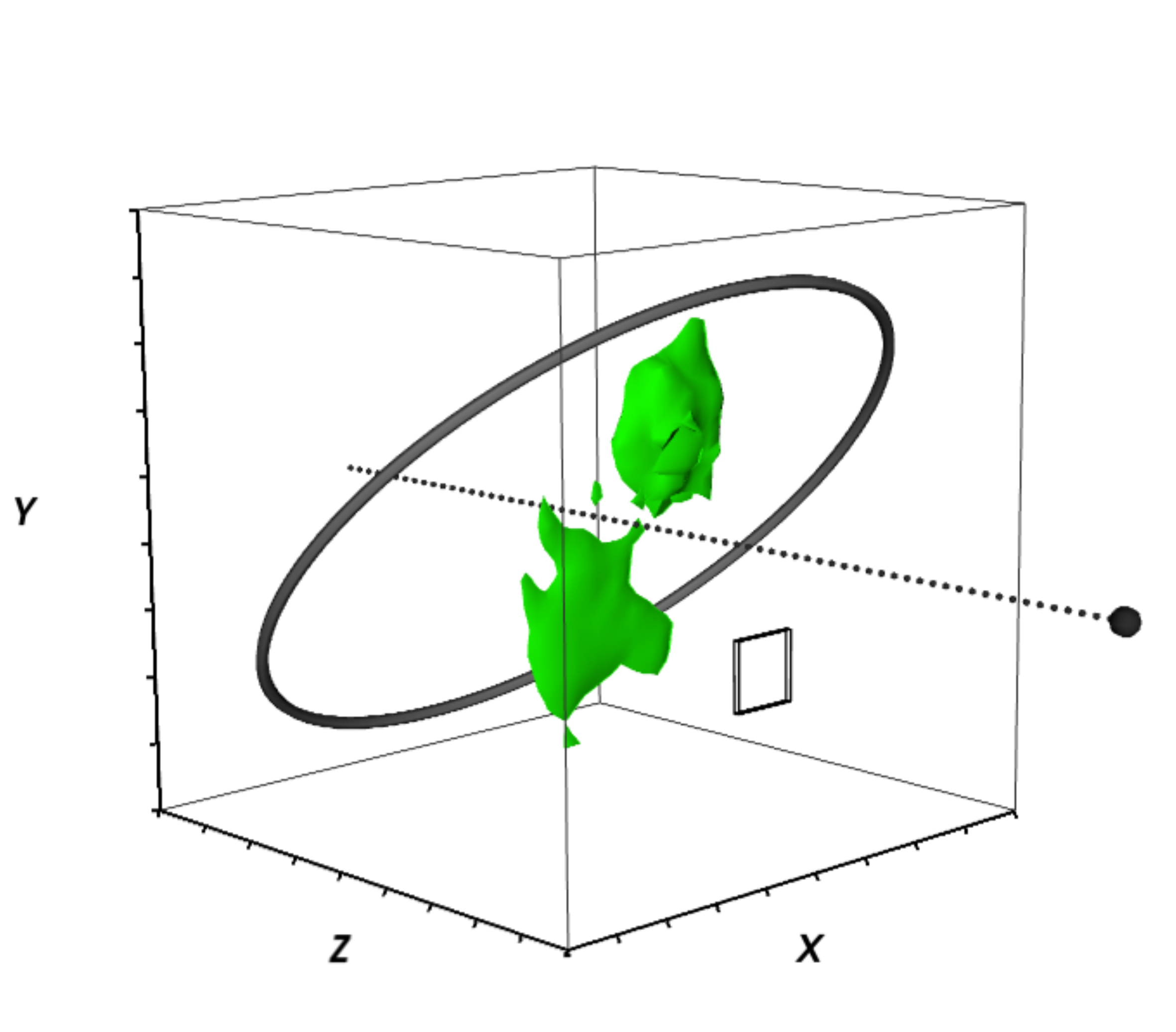}}
\resizebox{59mm}{!}{\includegraphics{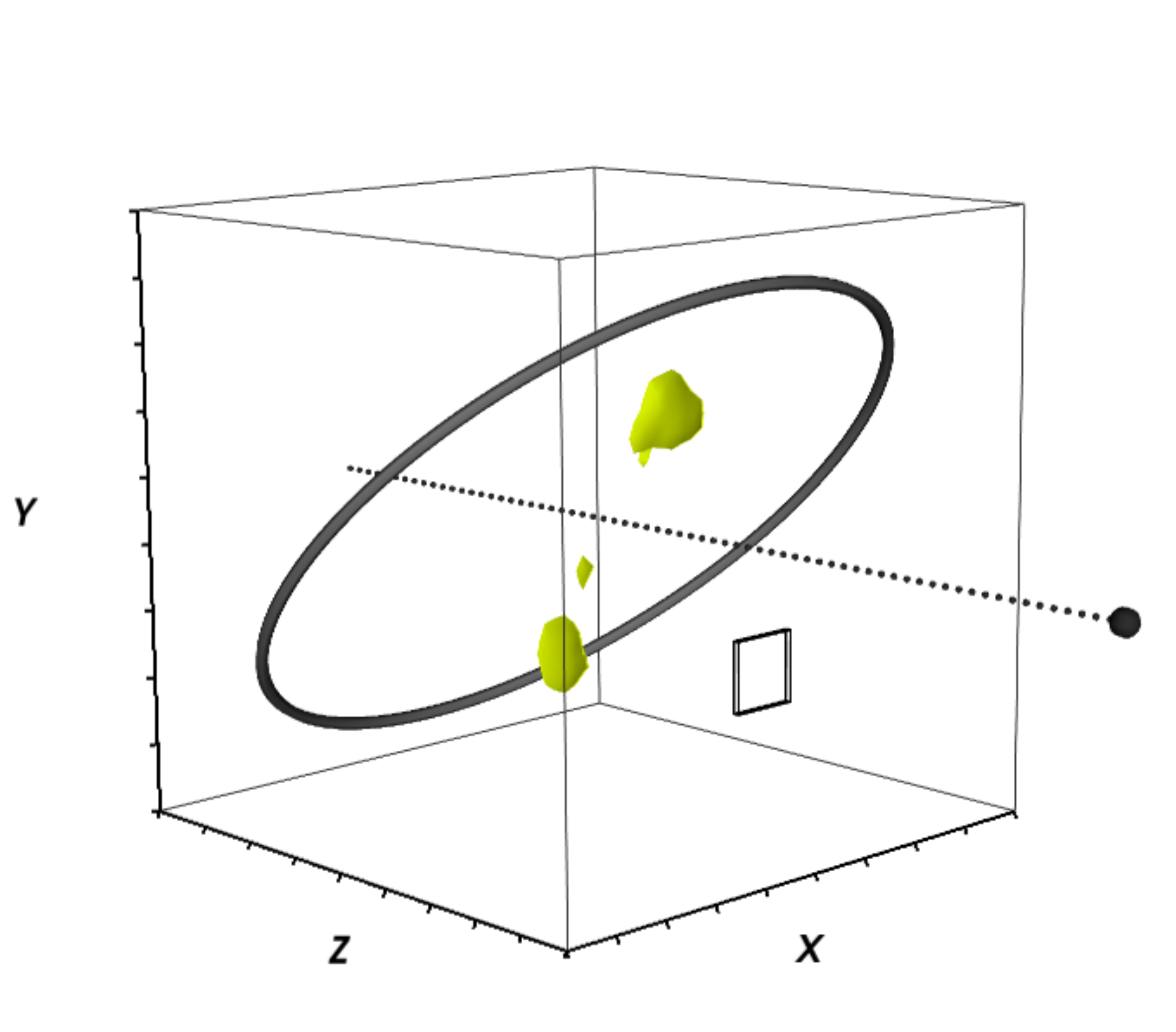}}
\resizebox{59mm}{!}{\includegraphics{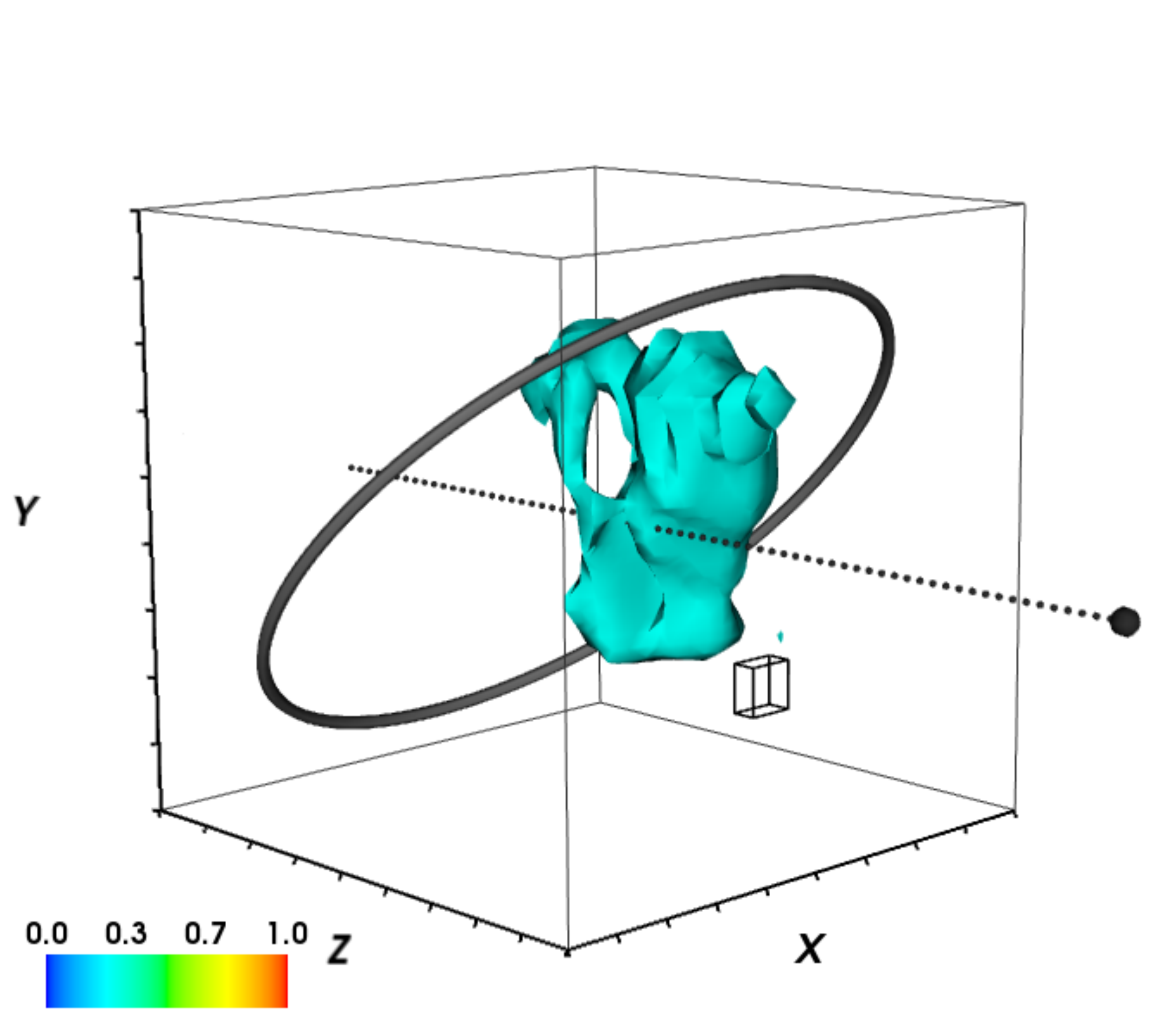}}
\resizebox{59mm}{!}{\includegraphics{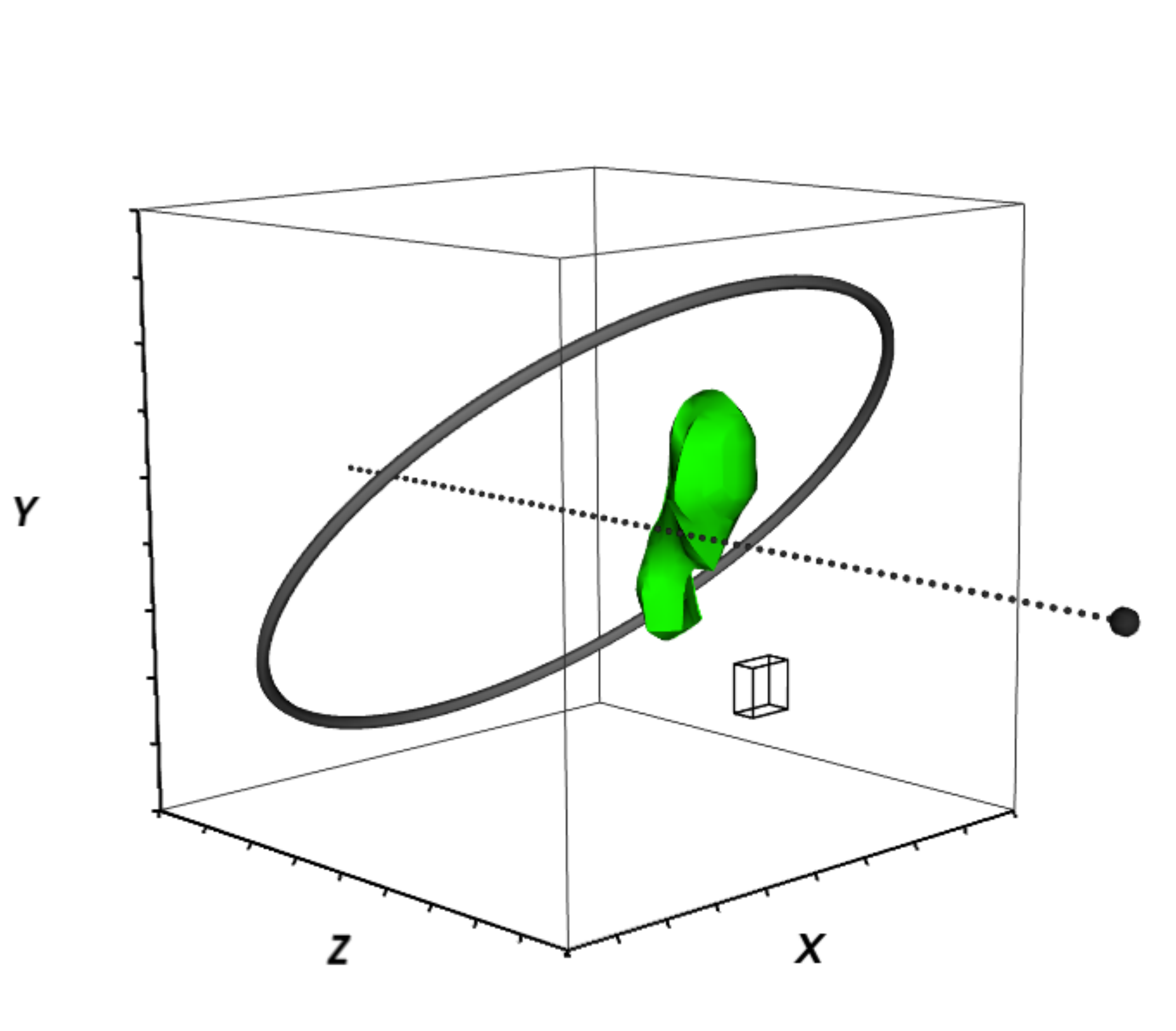}}
\resizebox{59mm}{!}{\includegraphics{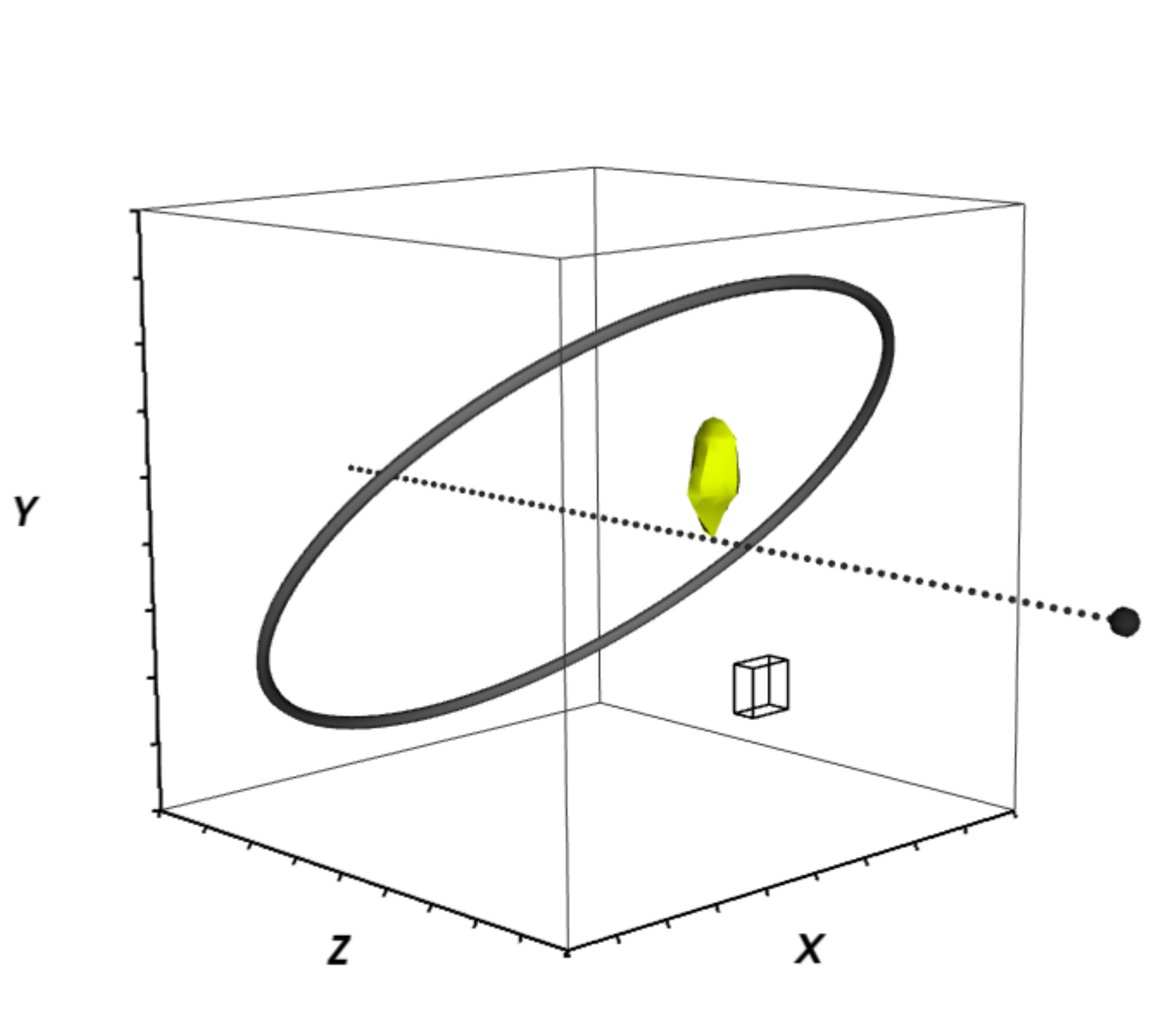}}
\caption{ 3D iso-surfaces for  [Si~I]$+$[Fe~II] (top row) and H$\alpha$ (bottom row).  Plots from left to right show the contours corresponding to $30\%$, $50\%$ and $70\%$ of the maximal intensity of each line, as indicated by the color bar. The ring shows the location of the reverse shock at the inner edge of the equatorial ring, while the dotted line and filled sphere indicate the line of sight and the position of the observer. The tick marks on the axes correspond to $1000\ \kms$. The boxes to the lower, right of each plot shows the resolution along the three axes. The [Si~I]$+$[Fe~II] plots include all data out to $4500\ \kms$ in all directions, with the central $\pm 450\ \kms$ along the line of sight removed due to contamination from the ring. The latter cut creates a gap in the centre which, however, is not seen in the top, left panel due to projection effects. The data cube was binned by a factor of six in the spectral direction. The H$\alpha$ plots were created from the ring-corrected spectra in Fig.~\ref{hamap}, which cover $[-2000,+2300]\ \kms$ in the x-direction and $[-3300,+2800]\ \kms$ in the y-direction. The main effect of these cuts is that some of the bright ejecta in the south is excluded (see lower, right panel of Fig.~\ref{slices}). Along the line of sight, all data within $\pm 4500\ \kms$ are included. Animated versions of this figure are available in the online version of the journal. The animations include different viewing angles, as well as semi-transparent versions that show multiple contour levels simultaneously.} 
\label{cont3d}
\end{center}
\end{figure*}

As a first step in constructing a 3D view of the ejecta we extracted STIS and SINFONI spectra from small spatial regions. In the case of H$\alpha$, which has the best signal, we used the  $0\farcs{1} \times 0\farcs{1}$ regions shown in the bottom panel of Fig.~\ref{hamap}. The resulting spectra are also shown in Fig.~\ref{hamap}. The original line profiles, shown as dotted lines, are contaminated by the narrow emission lines from [N~II]~$\lambda 6548$, H$\alpha$ and [N~II]~$\lambda 6583$ originating from the ring. In order to correct for this we extracted spectra from the northern and southern part of the ring, and subtracted a combination of these spectra in each panel so that the narrow lines were removed. The resulting spectra are shown as solid lines. 

We corrected the line profiles of the SINFONI data in the same way and show the results for the [Si~I]$+$[Fe~II] $1.644~\mu$m line in Fig.~\ref{simap}. As this line is somewhat weaker we used larger regions of $0\farcs{1} \times 0\farcs{25}$ to extract the spectra. The regions are shown in the bottom panel of Fig.~\ref{simap}. Fig.~\ref{hsimap} shows a comparison of H$\alpha$ and  [Si~I]$+$[Fe~II]. In the Appendix we include a comparison of these two lines with some of the weaker lines from the ejecta: [O I]~$\lambda \lambda 6300,\ 6364$, [Ca II]~$\lambda \lambda 7292,\ 7324$, Mg~II~$\lambda \lambda 9218,\ 9244$ and He~I~$2.058~\mu$m

To visualize the results for the two strongest lines, H$\alpha$ and [Si~I]$+$[Fe~II], we have created image slices through the ejecta (Fig.~\ref{slices}), integrated side views (Fig.~\ref{sideview}) and 3D iso-surfaces (Fig.~\ref{cont3d}). For a homologously expanding ejecta, the image slices in Fig.~\ref{slices} correspond to planes perpendicular to the line of sight. Each slice has a width of $1000\ \kms$. In the case of H$\alpha$, the images were produced from the ring-corrected spectra in Fig.~\ref{hamap}. Each panel in Fig.~\ref{hamap} thus corresponds to a pixel of size $0\farcs{1}\times0\farcs{1}$ in the resulting images. In order to ease comparison with the full WFC3/F625W image (shown in the bottom panel of Fig.~\ref{slices}) we also smoothed the images with a spline function\footnote{We use the ``spline16" function implemented in matplotlib, http://matplotlib.org}. In the case of  the [Si~I]$+$[Fe~II] line the velocity slices were produced from the full SINFONI H-band data cube, except for the central $\pm 500\ \kms$ bin, where we instead used the ring-corrected spectra in Fig.~\ref{simap}. This means that the pixels size in this bin is $0\farcs{1} \times 0\farcs{25}$, compared to $0\farcs{025}\times 0\farcs{025}$ in the other bins. This tradeoff in pixel size is necessary in order to have enough signal in each ``pixel" to verify that the narrow line has been subtracted. We smoothed the images in the same way as for H$\alpha$.

Figure~\ref{slices} illustrates a similar trend in both lines in the sense that the northern part of the ejecta is dominated by blueshifted emission, while the redshifted emission is stronger in the south. In addition, the blueshifted emission extends to higher velocities than the redshifted emission for both lines. The image slices also highlight several differences between the two lines. The H$\alpha$ emission is dominated by a very bright blueshifted region in the western part of the ejecta (see also central part of the fourth slit from the east in Fig.~\ref{hamap}), which has no correspondence in the [Si~I]$+$[Fe~II] emission. The latter line is instead dominated by a blueshifted region in the north and a redshifted region in the south. The two lines also differ in the centre of the ejecta, where there is significant [Si~I]$+$[Fe~II] emission around $0\ \kms$, but where H$\alpha$ is blueshifted.  

We note that the 3D geometry of  [Si~I]$+$[Fe~II] is consistent with previous results (\citealt{Kjaer2010}, L13), apart from the expansion. The lack of changes in the morphology is consistent with the nearly constant flux, discussed in section \ref{time} below. For H$\alpha$ we have no detailed recent 3D results to compare with, but we note that the lower-resolution data from day 6355 (approximately ten years earlier) showed the same trends in terms of the transition to increasingly dominant blueshifted emission when going from south to north, as well as the maximal blueshifted velocities being higher than the redshifted ones.

In Fig.~\ref{sideview} we show the ejecta as viewed from the side in the two lines.  We extracted the data from the same region as the velocity slices in Fig.~\ref{slices}, but in this case we summed over the east-west direction and used the spectral information to create an image along the line of sight to the observer.  The resulting ``images" have a significantly higher resolution along the line of sight due to the good spectral resolution. In order to create the velocity scale in the y-direction we use the fact that the ejecta are expanding freely  ($0\farcs{1} \equiv 866\ \kms$ at $10,000$ days) and chose the zero-point as the geometric centre of the ring. This position is offset by $0\farcs{03}$ (or $260\ \kms$) from the peak of the emission in the first HST image from 1994, which gives an idea of the uncertainty in the position of the centre of the explosion.  

Fig~\ref{sideview} shows all the similarities and differences between the two lines that were discussed above for the image slices. In addition, this figure provides information about the distribution of emission relative to the plane of the ring, which presumably defines the equator of the progenitor star.  While the emission is broadly located close to the ring  (as opposed to perpendicular to it) it is also clear that the emission is not symmetric around the ring plane. In particular, in both lines the brightest emission is closer to the plane of the ring in the north than in the south. In the south, the brightest emission is instead close to the plane of the sky.

Finally, we show  3D iso-surfaces for both lines in Fig.~\ref{cont3d}, obtaining the velocity information as described above. The [Si~I]$+$[Fe~II] plots include all data out to $4500\ \kms$ in all directions, with the central $\pm 450\ \kms$ along the line of sight removed due to contamination from the ring. The former cut removes the southernmost ejecta that overlap with the ring, while the latter cut creates a gap in the centre. For H$\alpha$, the plots were created from the ring-corrected spectra in Fig.~\ref{hamap} and the central region is therefore included. On the other hand, the outer edges of the extraction box (see lower, right panel of Fig.~\ref{slices}) excludes some of the highest velocities. The main effect of this is that the bright ejecta in the south is excluded, which creates a systematic difference compared to [Si~I]$+$[Fe~II].

For both lines the figure shows iso-surfaces for three different levels, corresponding to $30\%$, $50\%$ and $70\%$ of the maximal intensity of each line after subtraction of the continuum. The lower levels give a good view of the large-scale structure, while the highest level shows that the brightest emission is concentrated in a blueshifted clump in the west for H$\alpha$ and in two separate clumps in the north and south for [Si~I]$+$[Fe~II], as previously shown in Figs.~\ref{slices} and \ref{sideview}.  The space velocity of the centre of these clumps is $\sim 1800\ \kms$ (H$\alpha$ west), $\sim 2000\ \kms$ ([Si~I]$+$[Fe~II] north) and $\sim 2700\ \kms$  ([Si~I]$+$[Fe~II] south). 

Fig.~\ref{cont3d} also shows that the emission occupies a similar, relatively small fraction of the volume for the two lines. To quantify this, we determine the fractional volume ($f_{\rm{vol}}$) occupied by emission brighter than the three sigma error on the continuum level. This limit corresponds to approximately $15\%$ of the maximal (continuum-subtracted) intensity for both lines. Considering the total plotted volumes, this translates to $f_{\rm{vol}}=0.1$ for both lines. While this number gives an idea of the concentration of the emission, it should be noted that the different velocity cuts discussed above introduce a systematic uncertainty of about a factor of two. If we only consider the volume inside $2000\ \kms$, $f_{\rm{vol}}=0.5$ for H$\alpha$  and $f_{\rm{vol}}=0.7$ for  [Si~I]$+$[Fe~II].

In the weaker ejecta lines (see Appendix), the best signal is offered by [Ca~II] and He~I. In these lines we confirm the same trend as for H$\alpha$ and [Si~I]$+$[Fe~II] in the sense that northern part of the ejecta is dominated by blueshifted emission, while the redshifted parts of the lines become stronger in the south.  This is quantified in Table \ref{ratios}, where we provide the ratios of the fluxes in the blueshifted and redshifted wings of the lines in the northern, central and southern parts of the ejecta. These regions correspond to three rows shown on the images in the bottom panel of Fig.~\ref{hsimap}.

However, [Ca~II] and He~I do not clearly correlate with the brightest regions in H$\alpha$ and [Si~I]$+$[Fe~II] discussed above. In particular, neither line correlates with the H$\alpha$ western clump (corresponding to the strong peak in the line profile at  $-1500\ \kms$ in the middle, fourth panels of Figs.~\ref{camap} and \ref{hemap}), or the north clump in [Si~I]$+$[Fe~II]  (corresponding to the peak at  $-1300\ \kms$ in the north in Figs.~\ref{sicamap} and \ref{sihemap}). While He~I is similar to [Si~I]$+$[Fe~II] in the south, it should be noted that the correction for scattered light from the ring is uncertain for He~I in this region. In the case of  [O~I] and Mg~II, the only clear conclusion that can be drawn is that the locations of their brightest emission coincide with the western clump in H$\alpha$ (Figs.~\ref{omap} and \ref{mgmap}).

\begin{deluxetable}{lllll}
\tabletypesize{\scriptsize}
\tablecaption{Ratio of blueshifted and redshifted fluxes\tablenotemark{1} \label{ratios}}
\tablewidth{0pt}
\tablehead{
\colhead{Region\tablenotemark{2}} & \colhead{H$\alpha$} & \colhead{[Ca~II]} & \colhead{[Si~I]$+$[Fe~II] } & \colhead{He I} 
}
\startdata
North & $2.15\pm 0.06$ & $1.42\pm 0.09$ & $1.96\pm 0.06$  & $1.36\pm 0.08$  \\ 
Center & $2.12\pm 0.04$ & $1.16\pm 0.06$ & $1.14\pm 0.02$ & $1.08\pm 0.05$ \\
South & $1.09\pm 0.03$ & $0.69\pm 0.05$ & $0.53\pm 0.01$ & $0.73\pm 0.05$ 
\enddata
\tablenotetext{1}{Fluxes were calculated between $\pm (500-4500)\ \kms$. The errors are one sigma.}
\tablenotetext{2}{The regions correspond to the total top, middle and bottom rows shown on the images in the bottom panel of Fig.~\ref{hsimap}.}

\end{deluxetable}

\subsection{Time evolution}
\label{time}

\begin{figure}
\begin{center}
\resizebox{80mm}{!}{\includegraphics{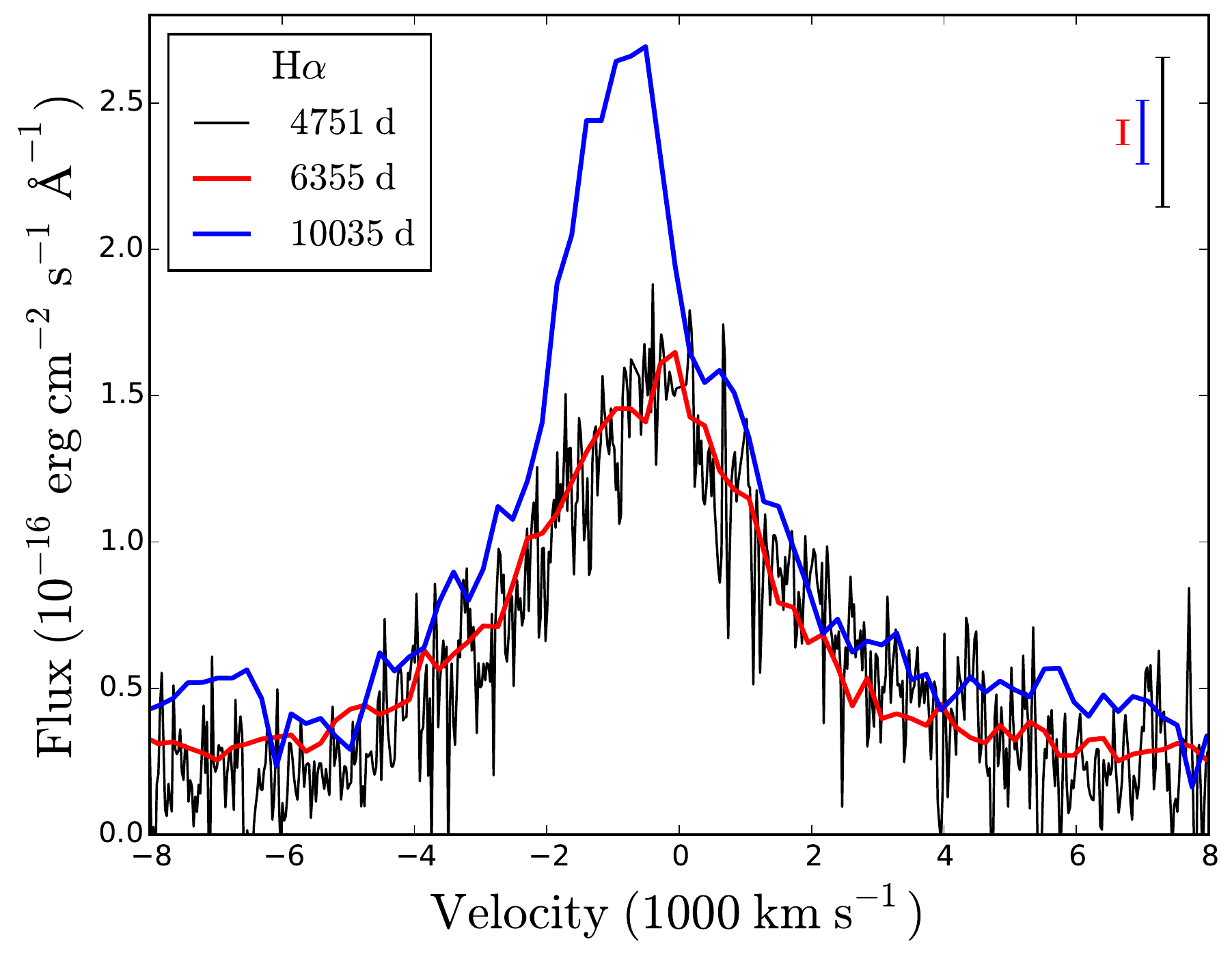}}
\caption{Time evolution of the H$\alpha$ profile from the full ejecta.  The spectrum from the most recent observation was extracted from the total $0\farcs{5}\times0\farcs{7}$ region shown in the bottom panel of Fig.~\ref{hamap}. The extraction regions for the earlier epochs are shown in L13. Error bars corresponding to the one-sigma statistical error for the three spectra are shown in the upper, right corner. All spectra have been corrected for contamination by narrow emission lines from the ring. For the spectrum at 6355 days, which was obtained using wider slits, this introduces a systematic uncertainty in the region $\pm 1500\ \kms$.} 
\label{htime}
\end{center}
\end{figure}
\begin{figure*}
\begin{center}
\resizebox{80mm}{!}{\includegraphics{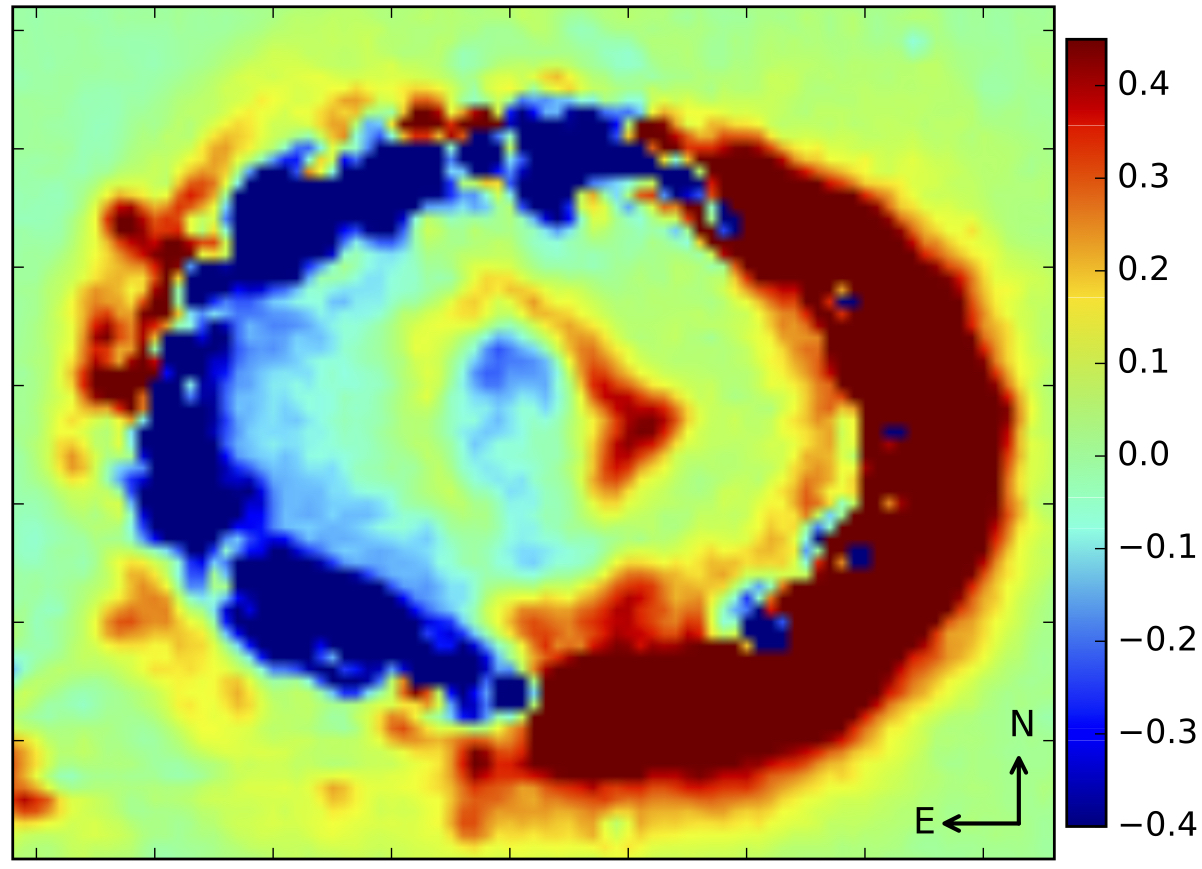}}
\resizebox{80mm}{!}{\includegraphics{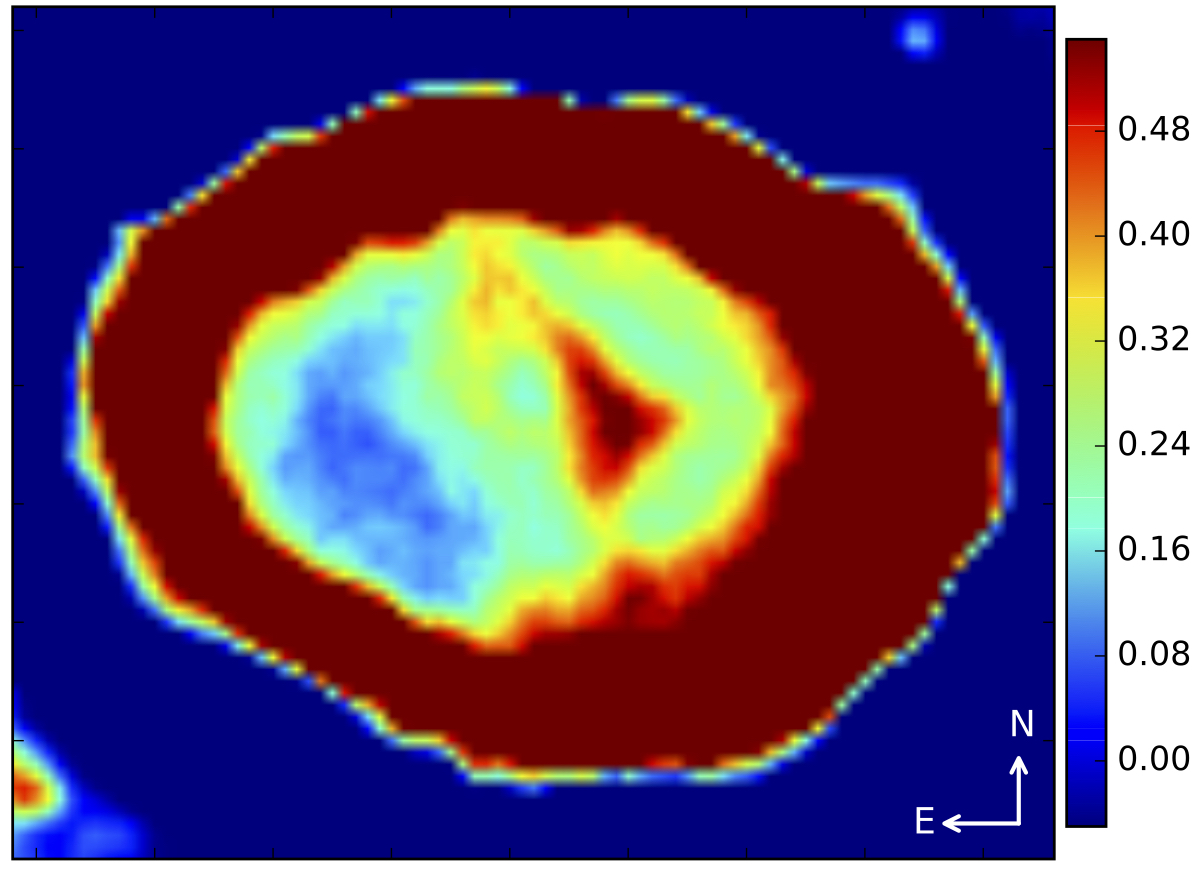}}
\caption{HST F625W difference images between days 9973 and 6790, showing the evolution of H$\alpha$.  These images are close in time to the spectra from days 10,035 and 6355 shown in Fig.~\ref{htime}. In the right panel the image from day 6790 was  expanded by a factor of 1.5 before the subtraction in order to account for the expansion of the ejecta between the two epochs. As a result, this image does not provide any information about the evolution of the ring.  The color bars show the intensity in units of $2.5 \times 10^{-16}\ \rm{erg\ cm^{-2}\ s^{-1}}$. A comparison between these images and Fig.~\ref{htime} confirms that the flux increase in the range $[-2000,0]\ \kms$ is primarily driven by the western clump. There is also a significant increase in the southern ejecta, but about half of this region is outside the spectral extraction box (see Fig.~\ref{hamap}) due to the spatial overlap with the ring. The bright region at the inner edge of the ring in the west is due to the reverse shock.}
\label{diffim}
\end{center}
\end{figure*}

Here we investigate the temporal evolution of the full line profiles from the inner ejecta, using our previous observations with both HST/STIS and SINFONI. The evolution of the H$\alpha$ profile is shown in Fig.~\ref{htime}, including observations from days 4751, 6355 and 10,035. The first two observations were previously discussed in L13. Some uncertainty is introduced in the comparison due to the fact that the observations were taken with different gratings, slit widths and position angles. Nevertheless, for each epoch the majority of the bright emission from the ejecta was included in the extraction. 

The main evolution that can be seen in the line profile is a highly significant brightening in the range $[-2000,0]\ \kms$ between the last two epochs. To investigate the spatial location of this increase we show in Fig.~\ref{diffim} difference images of the HST~F625W images that are closest in time to the spectral observations. The images were obtained on days 9973 and 6790, compared to days 10,035 and 6355 for the spectra. In the left panel the subtraction was performed without any corrections, whereas in the right panel the image from day 6790 was  expanded by a factor of 1.5 before the subtraction in order to account for the expansion of the ejecta. This provides the most accurate representation of the flux evolution of the ejecta, but, as opposed to the left panel, it gives a distorted view of the evolution of the ring.

The difference images show a non-uniform flux evolution of the ejecta, with the strongest brightening seen in the western region. There is also a significant brightening of the southernmost ejecta, but only about half of this region is included in the spectral extraction region (cf. Fig.~\ref{hamap}) due to the proximity of the ring. The conclusion that the flux increase in Fig.~\ref{htime} is primarily driven by the western clump is further supported by the spatially resolved maps in Fig.~\ref{hamap}, which show a strong emission line peaking around $-1500\ \kms$ originating from this region. We further note that the left panel of Fig.~\ref{diffim} also shows a non-uniform flux evolution of the ring, with the eastern side fading and the western side brightening.  The evolution of the ring is discussed in detail in \cite{Fransson2015}. 

\cite{Fransson2013} measured the flux evolution of the H$\alpha$ line from the ejecta inside $\pm 2500\ \kms$ using VLT/UVES, finding an increase from $0.22\pm 0.04 \times 10^{-14\ }\ \rm{erg\ s^{-1}\ cm^{-2}}$ on day 5000 to $1.2\pm 0.2 \times 10^{-14\ }\ \rm{erg\ s^{-1}\ cm^{-2}}$ on day 9035. For comparison, the observed H$\alpha$ flux inside $\pm 2500\ \kms$ in the STIS observation on day 10,035 is $1.8\pm 0.1\  \times 10^{-14\ }\ \rm{erg\ s^{-1}\ cm^{-2}}$, which implies a flux increase of $\sim 30\%$ in the last 1000 days. However, this increase is only marginally significant due to the large systematic uncertainties in correcting for contamination by narrow lines from the ring in the UVES data. 

\begin{figure*}
\begin{center}
\resizebox{80mm}{!}{\includegraphics{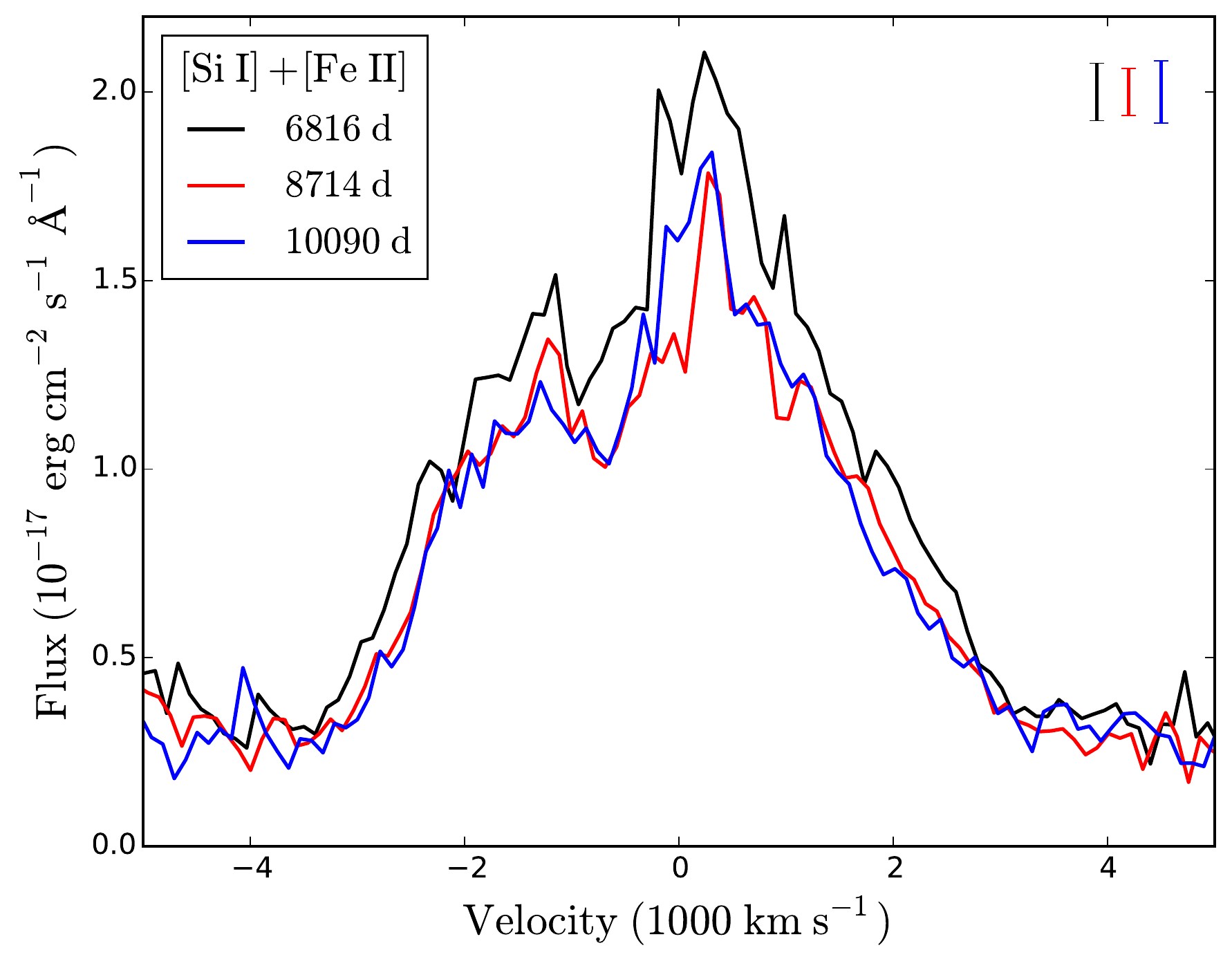}}
\resizebox{80mm}{!}{\includegraphics{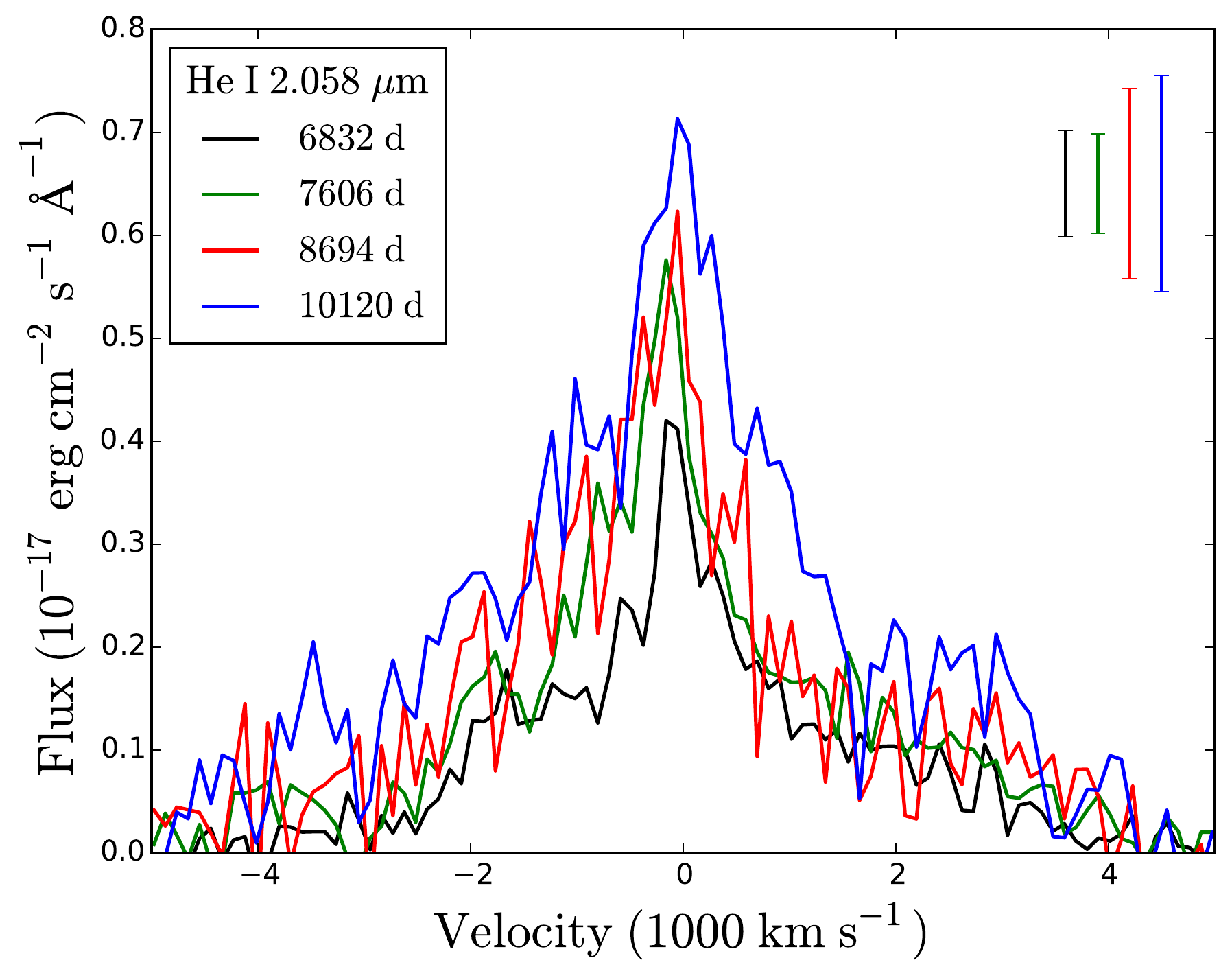}}
\caption{Comparison of line profiles from the full ejecta. Left: evolution of the [Si~I]$+$[Fe~II] $1.644\ \mu$m line. Right: Evolution of the He~I~$2.058\ \mu$m line. All spectra have been binned by a factor of three and corrected for scattered light from the ring. For the He~I line the continuum level has also been subtracted (see text for details). Error bars corresponding to the one-sigma statistical error for the different spectra are shown in the upper, right corners.}
\label{hsitime}
\end{center}
\end{figure*}

In Fig.~\ref{hsitime} we show the time evolution of the [Si~I]$+$[Fe~II] $1.644\ \mu$m line (three observations) and the He~I  $2.058\ \mu$m line (four observations), spanning the time interval from $\sim 6800 - 10,100$ days. As in the case of H$\alpha$, the spectra were extracted from regions covering the full inner ejecta. The previous observations of the [Si~I]$+$[Fe~II] line were discussed in L13, while the K-band observations of the He~I line were discussed in Fransson et al. (2016).  Since the K-band has a strong and increasing continuum we show the He~I line after subtracting the flux level between $2.09-2.11\ \mu$m, i.e. the region between the He I and H$_2$ lines, see Fig.~\ref{totspec}. 

The two lines clearly evolve in different ways. The  [Si~I]$+$[Fe~II] line has remained remarkably constant, with the only change being a $10\%$  decrease in flux between 6816 and 8714 days. Given the $10\%$ systematic uncertainties in the SINFONI flux calibration, this change is not significant. The He~I line, on the other hand, has increased by approximately a factor of 2.3 between the first and last observations (Fransson et al. 2016). Neither of the lines show any evidence for changes in the line profile, although it should be noted that such changes would be hard to detect in the case of He~I due to the high noise level.  

A comparison with our STIS observations from day 6355 also reveals a flux increase in [O~I], [Ca~II]  and Mg~II. For the latter two lines this is consistent with the flux evolution measured from VLT/UVES spectra (\citealt{Fransson2013}), while the evolution of [O I] was not reported in that work due to problems with contamination from the reverse shock. The fact that these lines are relatively weak, together with the systematic uncertainties in correcting for contamination from the ring in the previous observations, prevents us from carrying out a meaningful comparison with the increase in H$\alpha$ and He~I.

\section{Discussion}
\label{discussion}

One-dimensional models for core collapse SNe result in a stratified structure with an Fe core surrounded by layers of Si/S, O/Ne, C/O, He and H. Multidimensional calculations of the explosion show, however, that Rayleigh-Taylor instabilities cause a strong mixing of these layers.  Except in thin boundary layers, this mixing is likely to be macroscopic, rather than microscopic. 
The resulting morphology will therefore carry information about the late stellar evolution as well as the explosion mechanism. Below we discuss the interpretation of the morphology of the ejecta in \8 in this context. We first discuss in which zones the different lines are thought to arise and how the observed morphology relates to the energy sources, as well as the properties of dust. We then discuss the geometry of the ejecta, compare this geometry with numerical models, and discuss other observations of asymmetries in SNe.

\subsection{Energy sources and their effect on the morphology}
\label{einput}

The main energy sources that currently power the ejecta are radioactive decay of $^{44}$Ti and X-ray emission from the equatorial ring. The details of the energy deposition and the effect on the morphology and spectrum have previously been discussed in \cite{Kozma1998,Jerkstrand2011,Larsson2011,Fransson2013} and L13. Here we briefly summarize the main points and discuss how the new results fit into the picture. 

In the optical, there was a clear sign of a change in the dominant energy source after about 5000 days, when the ejecta started re-brightening after initially decaying as expected from radioactive decay (\citealt{Larsson2011}). The brightening can be explained in terms of energy input from the X-ray emission from the ring.  As noted in L13, the change of the energy source is accompanied by a change in the morphology of the ejecta, from a centrally peaked elliptical shape before $\sim 5000$ days to an edge-brightened morphology thereafter. The brightening and change in morphology is seen in all the HST filters that were used, which together cover the wavelength interval $\sim 2000 - 10000\ \rm{\AA}$.  \cite{Fransson2013} showed that the change in morphology is a natural consequence of the X-rays being absorbed at the boundary between the H envelope and the core, where the density gradient is steep and where the abundances of metals increase, resulting in a dramatic increase in the X-ray absorption.

\subsubsection{H$\alpha$}
\label{ha}

In this work we have shown that the H$\alpha$ line profile has developed a significantly stronger blue wing compared to the last spectral observations that covered the full ejecta on day 6355 (see Fig.~\ref{htime}). This is primarily driven by the bright clump in the western part of the ejecta, as shown in Figs.~\ref{hamap} and \ref{diffim}. This region is close to the brightest part of the ring in soft X-rays (\citealt{Frank2016}), in agreement with the scenario of energy input from the X-rays powering H$\alpha$ emission. 

\subsubsection{[Si~I]$+$[Fe~II]  $1.644\ \mu$m}
In contrast to H$\alpha$, the flux in the [Si~I]$+$[Fe~II]  $1.644\ \mu$m line is consistent with staying constant since the first observation at $\sim 6800$ days. The shape of the line profile has also remained remarkably constant over the last 10 years (Fig.~\ref{hsitime}). The part of the ejecta probed by this emission line, i.e. the metal core, is thus most likely still dominated by radioactive input. This conclusion is further supported by the centrally peaked morphology (Fig.~\ref{conts}), which has also remained constant with time apart from the expansion. 

\begin{figure}
\begin{center}
\resizebox{80mm}{!}{\includegraphics{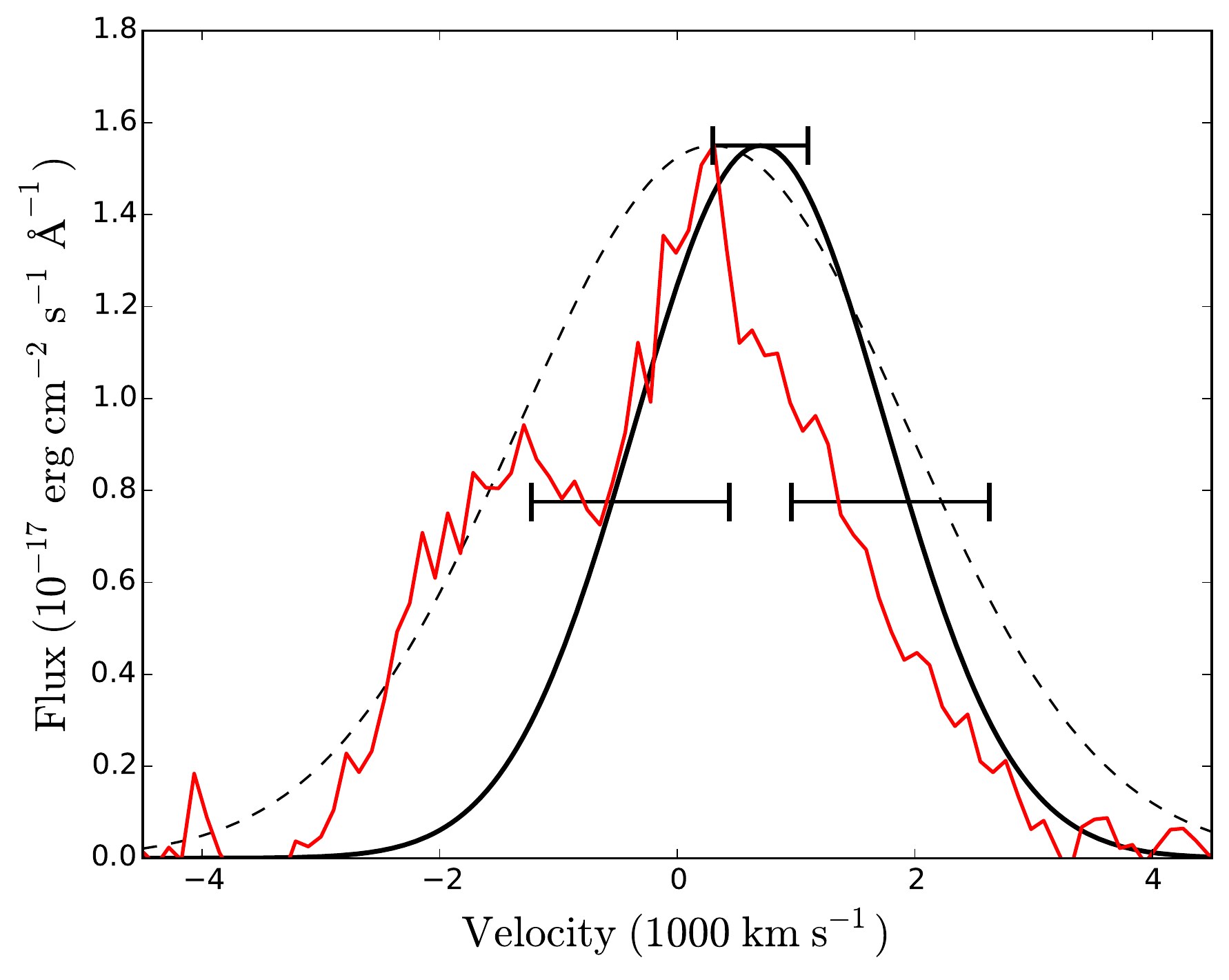}}
\caption{Comparison of  the total [Si~I]$+$[Fe~II] $1.644\ \mu$m line profile from 10,090 days (red) with the best-fit Gaussian for the $\rm{^{44}Ti}$ lines detected by NuSTAR (black, from \citealt{Boggs2015}). The error bars indicate the $90\%$ confidence region for the peak position (show at the top of the line) and the width of the Gaussian. The dashed line represents the case of the lowest peak energy and largest width consistent with these confidence intervals.}
\label{nustar}
\end{center}
\end{figure}

In order to investigate how well the [Si~I]$+$[Fe~II]~$1.644\ \mu$m emission traces the $^{44}$Ti, we can compare with the  $^{44}$Ti decay lines at  $67.87$~keV and $78.32$~keV detected by NuSTAR (\citealt{Boggs2015}). A Gaussian fit to these lines gives a redshift of $700\pm 400\ \kms$ and a width of $\sigma = 0.24^{+0.13}_{-0.19}$~keV. This best-fit model is shown together with the [Si~I]$+$[Fe~II]  line in Fig.~\ref{nustar}. A comparison is not straight forward given the clearly non-Gaussian shape of the  [Si~I]$+$[Fe~II]  line, which has much better statistics and resolution than the NuSTAR data. However, we note that the peak of the [Si~I]$+$[Fe~II] line at $\sim 300\ \kms$ coincides with the lower boundary of the $90\%$ confidence interval from the NuSTAR fit. The width of the NuSTAR line is also consistent with our observations, given the large uncertainties. 

\cite{Boggs2015} suggest that the NuSTAR observations can be explained with a single-lobe explosion model, where the lobe is at an angle pointing away from the observer. Assuming that the [Si~I]$+$[Fe~II]~$1.644\ \mu$m emission is a good tracer of $^{44}$Ti, this is clearly inconsistent with our observations (e.g, Fig.~\ref{cont3d}). 

The only other SNR where the spatial distribution of $^{44}$Ti, Fe and Si has been investigated is Cas~A. In this case $^{44}$Ti does not  correlate with Fe or Si (\citealt{Grefenstette2014}). However, this can likely be attributed to the Fe and Si in the interior of the remnant being in ionization states that are difficult to observe. The geometry of Cas~A as compared to  \8 is discussed in more detail in section \ref{comp}.

\subsubsection{He~I $2.058\ \mu$m}
The He~I $2.058\ \mu$m line has a similar morphology to the [Si~I]$+$[Fe~II] line in the images, but has nevertheless increased by more than a factor of two between $\sim 6800$ and $\sim 10,100$ days, indicating that X-rays from the ring play a significant part in powering the emission. As already noted by Fransson et al.~(2016), this seemingly surprising result may be explained by the fact that the He~I emission has contributions from the Fe/He, He and H zones. The inner zone, where the He results from $\alpha$-rich freeze-out during the explosion, would thus explain the centrally peaked morphology, while the outer zones are responsible for the flux increase. The contributions from several different zones also agrees with the fact that the spatially resolved maps show some differences compared to both H$\alpha$ and [Si~I]$+$[Fe~II] (Figs.~\ref{hemap} and \ref{sihemap}).

The fact that the whole line profile increases and not only the wings is explained by the fact that the flux at a given wavelength, or velocity, is an integral of the emission in a plane perpendicular to the line of sight for a $V \propto r$ velocity law. Even at the line center there is therefore a contribution from large radii, which may be affected  by the X-rays.

\subsubsection{[O I] $\lambda \lambda 6300, 6364$, [Ca II] $\lambda \lambda 7292,\ 7324$,  and Mg~II~$\lambda \lambda 9218,\ 9244$}

The observed increase in the fluxes of the [O I], [Ca II] and Mg II lines (Sect. \ref{time}), indicate that these lines are affected by the X-ray illumination. In line with this, the velocity maps of both [O I] and Mg II show that their brightest emission coincides with the bright H$\alpha$ clump in the west (middle, forth panel of Figs.~\ref{omap} and \ref{mgmap}). Given the low temperature in the ejecta, this implies that there is a high abundance of O nearly co-spatial with the H in the region that is most affected by the X-rays. The Mg~II~$\lambda \lambda 9218,\ 9244$ lines are likely powered by Ly~$\alpha$ fluorescence (\citealt{Fransson2013}), and it is therefore expected that these lines should have similar line profiles as H$\alpha$. Taking into account the low S/N of the Mg~II lines, this is consistent with the observations (Fig.~\ref{mgmap}). 

In contrast to this, the velocity maps of [Ca~II] do not show a very high concentration of flux in the western clump (Fig.~\ref{camap}). There is no clear correlation with the brightest clumps in the [Si~I]$+$[Fe~II] emission either (Fig.~\ref{sicamap}). The [Ca~II] lines, like the Mg II lines, are mainly excited by fluorescence \citep{Jerkstrand2011}. In this case it is by radiative excitation in the H and K lines in the near UV. The main requirement for this is that the lines should be optically thick, which is the case in the regions with synthesized Ca as well as the core regions with primordial Ca. In the model of the spectrum at 8 years, the H and K lines were also optically thick in the H envelope out to $\sim 8000 \ \kms$ (\citealt{Jerkstrand2011}), although this should be considerably reduced at the present time. The scattered UV photons  may originate either in the core or from emission from the ring. The detailed emissivity distribution of [Ca~II] may therefore be different from the other lines, as is indicated by our velocity maps. 

\subsection{Dust in the ejecta}
\label{dust}

Observations by {\it Herschel} and ALMA have revealed a large mass ($\sim 0.5-0.8 \ \msun$) of cold dust in the ejecta of \8 (\citealt{Matsuura2011,Indebetouw2014,Matsuura2015}). This large dust mass is in contrast to early observations, where  the mass was estimated to be $10^{-4} - 10^{-3}\ \msun$ at $\sim 500$ days (\citealt{Lucy1989,Wooden1993}), as well as the low dust masses reported in a number of other type II SN (e.g.~\citealt{Meikle2007}). This finding has prompted theoretical work aimed at understanding the composition and formation of the dust. Both \cite{Sarangi2015} and \cite{Dwek2015} find that the dust is dominated by silicates and that most of the dust formed within the first $500-1200$ days. On the other hand, \cite{Wesson2015} find that the dust is dominated by carbon and that the overwhelming majority formed after $1500$ days (with $90\%$ of the dust forming after 3000 days). From modeling of line profiles, \cite{Bevan2015} also favor a scenario where the majority of the dust formed late. However, their assumption that the line asymmetry is solely due to dust rather than intrinsic asymmetries in the ejecta is likely to lead to large uncertainties. 

While dust will inevitably affect our observations, it is unlikely that it is responsible for the main differences that we see in the different emission lines, such as the central low-surface brightness region (or ``hole") seen in H$\alpha$ but not in [Si~I]$+$[Fe~II]. As discussed in L13, one of the main arguments for this is the observed time evolution in H$\alpha$. If the ``hole"  were due to dust, we would expect it to become less prominent with time as the optical depth of the dust is expected to decrease as $t^{-2}$, where $t$ is the time since explosion, contrary to what we observe. Instead, the ``hole" appears at the same time as the ejecta start brightening due to the X-ray illumination, and is a natural consequence of the absorption in the outer parts of the ejecta (L13 and \citealt{Fransson2013}). In addition, as also discussed in L13, considerable fine tuning would be required for the dust to be optically thick at $\sim 0.85\ \mu m$ (where the ``hole" is seen in the HST/F814W filter) but not in the H-band at $\sim 1.6\ \mu m$. A likely explanation for the observations is that the dust is in optically thick clumps that are affecting the optical and NIR in the same way. 

From the work presented here we have added additional arguments in favor of this interpretation. In particular, we have shown that  Br$\gamma$ also has a central, low-surface brightness region (Fig.~\ref{conts}). To estimate the extinction in the different observed lines we use the extinction curve for the Galactic ISM by \cite{Cardelli1989}, which gives extinction ratios, A($\lambda$)/A(V), of 0.82, 0.09 and 0.19, for H$\alpha$, [Si~I]$+$[Fe~II] and Br$\gamma$, respectively. We stress that these numbers are only meant to illustrate typical values for the extinction, and will not be precise if the dust composition of \8 differs from the ISM, as one expects. If the differences in morphology were solely due to extinction, [Si~I]$+$[Fe~II] and Br$\gamma$ should therefore be similarly affected at a much lower level than H$\alpha$, which is not what we observe. We therefore suggest that the central  ``hole" seen in both H$\alpha$ and Br$\gamma$, reflects the actual shape of the H emitting region. The fact that the morphology in these two lines is not identical may indicate some effects of dust extinction, although it should also be noted that the Br$\gamma$ image is affected by the wider wings of the SINFONI PSF (which could reduce the contrast between the hole and the western clump), a contribution from the ``continuum" emission from the ejecta, as well as brighter emission from the ring in the south. 

One of the first arguments for dust formation in \8 was the shift of the peak of the [O I] $\lambda \lambda 6300, 6364$ lines from $0\ \kms$ to $-600\ \kms$ around 500 days (\citealt{Lucy1989}). \cite{Fransson2013} showed that the line profile was still consistent with this blueshift at around 4400 days. This was the last optical spectrum obtained during the epoch when the energy input to the whole ejecta was dominated by radioactivity. Here we have shown that this line, as well as all the other optical lines, are now affected by the X-rays (sections \ref{time} and \ref{einput}). The line profiles are therefore no longer tracers of dust. As expected from this, as well as the preceding discussion, there is no clear wavelength-dependence that can be associated with dust in the ratios of blueshifted and redshifted fluxes in Table \ref{ratios}.

The only line that is not significantly affected by the X-rays is the  $1.644\ \mu$m [Si~I]$+$[Fe~II]. While there are no observations of this line from early epochs, we have shown that the line profile has not changed in the last ten years (since day $\sim 6800$, left panel of Fig.~\ref{hsitime}), indicating that any dust affecting the line is optically thick during the whole period. This places strong constraints on models that invoke significant dust formation in the last decade (\citealt{Wesson2015,Bevan2015}).

It should also be emphasized that it is clear that dust alone cannot explain many of the observed asymmetries. In particular, this holds for the lack of highly blueshifted emission in the southern ejecta (Fig.~\ref{sideview}) as well as the elongation of the ejecta along the northwest-southeast direction (Fig.~\ref{conts}). However, an interesting question is whether the dust could be responsible for the north-south asymmetry seen in Fig.~\ref{sideview}. For example, a large amount of dust near the plane of the ring could enhance the observed asymmetry with respect to the ring plane. To fully explain the asymmetry the dust would then have to extend to the southernmost parts of the ejecta, which is not indicated by the current ALMA observations (Cigan et al., in prep.).

\subsection{Geometry of the ejecta}
\label{geometry}

All emission lines that have sufficient signal-to-noise (i.e H$\alpha$, [Ca II] $\lambda \lambda 7292,\ 7324$, [Si~I]$+$[Fe~II]~$1.644\ \mu$m and He~I $2.058\ \mu$m) reveal a similar large-scale structure. As discussed above, these lines originate from the metal core, the H envelope and the zones in between. On the plane of the sky, the ejecta are elongated along the north-east - south-west direction (Fig.~\ref{conts}). In 3D, the large-scale structure has a clear north-south asymmetry, resembling a broken dipole. In particular, as shown in Figs.~\ref{sideview} and \ref{cont3d}, the brightest ejecta in the north are distributed between the plane of the ring and the line of sight, while the brightest regions in the south are closer to the plane of the sky. In the case of [Si~I]$+$[Fe~II] and He I, this structure has been seen in previous observations (\citealt{Kjaer2010} and L13). This structure was also indicated by the lower-resolution map of H$\alpha$ obtained at day 6355 (L13), but our new STIS observations have provided a high-resolution confirmation of this, as well as evidence that [Ca II] has a similar structure (Fig.~\ref{camap} and Table \ref{ratios}). 

The fact that there is less ejecta in the plane of the ring on the south (far) side also agrees qualitatively with the fact that the reverse shock emission, which is produced as the ejecta interact with the CSM, is fainter on this side (e.g., \citealt{Fransson2013}). In addition, several of the lines in the reverse shock also show that the maximal blueshifted velocities are higher than the redshifted ones (\citealt{France2011}), just as we find for the ejecta. There is thus some evidence that the reverse shock reflects the overall ejecta structure. However, it should be noted that the reverse shock emission also depends on the properties of the CSM in a wide region around the ring plane ($\sim \pm30^{\circ}$, \citealt{Michael2003}). 

It is interesting to compare the large-scale structure to early observations, which probe the geometry of the outermost ejecta. In particular, the so-called ``mystery" spots revealed by speckle imaging at 30-50 days were located in the north-east and south-west, respectively (\citealt{Nisenson1987,Meikle1987,Nisenson1999}), consistent with the symmetry axis in the current images,  as well as early polarimetric measurements (e.g., \citealt{Schwarz1987,Jeffery1991}). In addition, recent observations of light echoes dominated by the flux of the first few 100 days reveal an H$\alpha$ emission varying with direction (\citealt{Sinnott2013}, cf. their Fig.~20 to our Fig.~\ref{sideview}). This is reminiscent of the ``Bochum" event \citep{Hanuschik1990}, which has been interpreted as a result of an asymmetric  $^{56}$Ni distribution \citep{Lucy1988,Chugai1991}. All these observations are consistent with an explosion geometry that is similar from the inner metal core to the H envelope.

When investigating the structure of the H$\alpha$ and [Si~I]$+$[Fe~II] emission in more detail, it is clear that the morphology is irregular and that there are significant differences between the two lines (Figs.~\ref{slices}, \ref{sideview} and \ref{cont3d}). Substructure is seen close to the limit of the resolution in both lines. For the H$\alpha$ 3D maps this is $\sim 900\ \kms$ in the plane of the sky and $\sim 500\ \kms$ along the line of sight. For the [Si~I]$+$[Fe~II]  3D maps, the resolution is similar to H$\alpha$ in the sky plane, but significantly better along the line of sight, where substructure can be seen down to $\sim 200\ \kms$. In line with this, the $1000\ \kms$ wide images slices in Fig.~\ref{slices} show significant changes between consecutive slices in both lines.

The main differences between the two lines are well illustrated by the iso-surfaces in Fig.~\ref{cont3d}. The [Si~I]$+$[Fe~II] emission is concentrated in two main regions (artificially completely separated due to problems with contamination from the ring) that simply get smaller when increasing the contour levels. On the other hand, the structure of the H$\alpha$ emission changes more with the contour levels, going from an irregular shape that shows the ``hole", to a broken-dipole shape, and finally a single region. This edge-brightened morphology and dominance of the western part most likely reflects the external X-ray illumination, as discussed above.

Despite these differences, it is interesting to note that the fractional volume occupied by significant emission is similar for the two lines. For the total volumes plotted in Fig.~\ref{cont3d}, this is $f_{\rm{vol}}=0.1$ for both lines, albeit with a systematic uncertainty of about a factor of two due to different velocity cuts. In addition, the space velocity of the centre of the very brightest regions are similar for both lines. For H$\alpha$, the centre of the western clump is at $\sim 1800\ \kms$, while the brightest regions in the north and south for  [Si~I]$+$[Fe~II] are located at $\sim 2000\ \kms$ and $\sim 2700\ \kms$, respectively. Even though the energy sources are different, these positions are likely to correspond to regions of high density in both cases.

As discussed in section \ref{time}, the western clump dominates the integrated H$\alpha$ profile. Similarly, the two bright regions for [Si~I]$+$[Fe~II] can be identified with the two main peaks in the integrated line profile (Fig.~\ref{hsitime}). As this line profile has been corrected for scattered light from the ring, this confirms that there is a true reduction of intensity in between these regions (which, however, is enhanced by the cut in the 3D contours in Fig.~\ref{cont3d}).  

The 3D maps also allow us to investigate the minimum velocity of the two lines. From Figs.~\ref{hamap} and \ref{slices} we see that there is virtually no H$\alpha$ emission at zero Doppler shifts close to the center. The central panel of Fig.~\ref{hamap} (i.e. zero velocity in the x,y plane) has its peak at $\sim -1500\ \kms$ with very little emission extending to $0\ \kms$. On the other hand, going out to $450\ \kms$ in the plane of the sky (corresponding to the boundary of the central extraction region in Fig.~\ref{hamap}) there is significant emission at $0\ \kms$ in all directions except the east. The image of Br$\gamma$ (Fig.~\ref{conts}) is also consistent with this. We thus conclude that H is mixed in to about $450\ \kms$, in agreement with previous findings (\citealt{Kozma1998} find that H is mixed in to $<700\ \kms$). We further note that H is present in the centre of the ejecta in the form of H$_2$, where it is shielded from the X-ray emission from the ring (\citealt{Fransson2016}). However, a study of the 3D emissivity of H$_2$ will have to await future deep observations in the K-band.  

In the case of [Si~I]$+$[Fe~II], the best view of the lowest velocities is provided in Fig.~\ref{simap}. This figure shows that there is significant emission at zero Doppler shift in the middle panel even after the scattered light from the ring has been subtracted.  However, because of the size of this region (covering the central $\pm 450 \times 1100\ \kms$), this only provides a loose constraint on the minimum velocity. 

Regarding the highest velocities, we note that the integrated H$\alpha$ profile shows weak emission out to at least $8000\ \kms$ (Fig.~\ref{htime}) and that emission at even higher velocities of up to $11,000\ \kms$ is seen in the reverse shock (\citealt{Fransson2013}). For [Si~I]$+$[Fe~II], we see emission at least out to $4500\ \kms$ in the plane of the sky in the southern ejecta.

\subsection{Comparison with explosion models}

As noted above, the morphology of the ejecta in the homologous phase is one of few observational probes of the explosion mechanism. In our previous work (L13), we compared the morphology of the [Si~I]$+$[Fe~II] emission with the 3D explosion model based on a $15\ \msun$ Blue Super Giant (BSG) by \cite{Hammer2010}. This progenitor mass is likely somewhat low for \8, with \cite{Fransson2002} reporting a range of $18-20\ \msun$ and \cite{Smartt2009} a range $14-20\ \msun$. In addition, there is an uncertainty in the comparison in that we observe emissivities, while the models show mass distributions. With these caveats in mind, our main conclusions from the comparison were that although there are some qualitative similarities, the observations show a higher degree of asymmetry as well as higher velocities than the models. As our new observations of [Si~I]$+$[Fe~II] are consistent with the previous results, these conclusions still hold. 

However, new 3D models have been published since the previous comparison. In particular, \cite{Wongwathanarat2015} have performed self-consistent full 3D calculations of the explosion from the time of core-collapse until shock breakout for several progenitors with 15 and $20 \msun$. While the explosion of the $15 \msun$ BSG progenitor results in a highly mixed ejecta, the $20\ \msun$ BSG, which is most relevant for \8, is considerably less mixed. This progenitor instead results in a fragmented, rather spherical ejecta structure at shock breakout, as well as limited radial mixing. Our previous finding that the observations show a higher degree of large-scale asymmetry than the models is therefore even stronger when comparing with the $20\ \msun$ model. 

\cite{Wongwathanarat2015} find that the dramatic difference in  the ejecta structure between the 15 and $20 \msun$ models comes from the relative velocities between the ejecta and the shock at the time of the crossing of the O-C/He interface. Because of the lower relative velocities in the  $20 \msun$ model, the growth of the Rayleigh-Taylor fingers is slow. The fingers are also further compressed and 'flattened' by the reverse shock.  

However, there is an important caveat in that the simulations are only carried out until 0.66 days, which is not enough to reach the homologous phase (H-Th Janka priv. comm.). This can be seen in the velocity contours (their fig.~10), which show a factor of $\sim 3$ variation in the velocities at the same radii for the $20 \msun$ progenitor. As the ejecta evolve, these velocity variations will amplify the fluctuations in radii by a large factor. 
In addition, the distributions shown are for an abundance of only 3\% ${}^{56}$Ni, and are therefore not showing where most of the  ${}^{56}$Ni resides. Because of these complications it is difficult to draw any firm conclusions about which (if any) of these models are compatible with our observations. 

A more representative illustration of the bulk of the ${}^{56}$Ni mass may be that shown in Fig. 15 in \cite{Wongwathanarat2013} for the $15 \msun$ models. Although these snapshots are taken at only a few hundred seconds, they show that most of the mass resides in two main clumps moving in opposite directions. This major asymmetry should hardly change during the subsequent evolution and may therefore indicate a structure similar to what we observe. 

Previous to these models, there are also SPH-simulations of the post-explosion dynamics of a SN1987A progenitor by \cite{Herant1991,Herant1992}, and of a $15 \msun$ progenitor with a radius of $3\times 10^{13}$ cm by \citet{Ellinger2012}. Having a low mass and a radius corresponding to a red supergiant, the latter simulations are more directed to a comparison with Cas A, and a quantitative comparison with SN 1987A is difficult. 

In contrast to the simulation by \cite{Wongwathanarat2015}, both sets of simulations introduce the seed perturbations artificially, either as fluctuations in the smoothing length or as large scale ``jets''. Nevertheless, they contain several results which are of general interest. The most important aspects from the point of view of our observations are the inclusion of the ${}^{56}$Ni heating (not included in the simulations by  \citealt{Wongwathanarat2015}) and that their simulations are run to very late stages. The \cite{Herant1991,Herant1992} simulations extend to 90 days and the \citet{Ellinger2012} simulations run to several decades in some cases. 

Both groups find that the Rayleigh-Taylor instability results in fingers with dense clumps at the ends, mainly containing O, C and He, but no Fe-nuclei from the explosion. This is in agreement with the simulations by \cite{Wongwathanarat2015}. Besides mixing of these elements to higher velocities, there is also inward mixing of H, which reaches $\la 1000 \ \kms$ in the simulations by  \cite{Herant1992}. 

The inclusion of ${}^{56}$Ni heating results in a ``Ni-bubble'', filling most of the central volume. While the  ${}^{56}$Ni heating does not significantly change the general morphology of the rest of the ejecta, it leads to a further compression of the regions containing O, He and H clumps mixed into the bubble. It also results in a minor boosting of the velocities of the non-Fe clumps. From these results we expect that the general morphology in the simulations by \cite{Wongwathanarat2015} is not dramatically affected by the neglect of the ${}^{56}$Ni heating, although the density contrasts of the clumps will probably be increased. 
 
It is clear that the observations of \8 presented here can provide a unique test of explosion models, but that more work remains to be done in terms of calculating  the models to later epochs, as well as in modeling the connection between the observed intensity distribution and the underlying mass distribution for the relevant energy sources. This is especially relevant for H$\alpha$, where the observed morphology is strongly affected by the X-ray illumination.

In terms of using \8 to draw conclusions about SN explosions in general, we note that \8-like SNe (i.e.~SNe with compact blue progenitors) account for only 1-3 \% of all core collapse SNe \citep{Pastorello2012}. However, while the blue  progenitor of SN 1987A makes it different from most other Type II progenitors, the core of the SN is typical of  Type II SNe, both in terms of nucleosynthesis and core velocity \citep[e.g.,][and references therein]{McCray2016}. The main difference is in the compact H-envelope. Although the morphology of the ejecta is affected by this, as discussed above, the core collapse and explosion mechanism  were probably normal for a massive star.

\subsection{Other observations of asymmetries in SNe}
\label{comp}

Evidence for asymmetric explosions is seen in many other SNRs.  Among these, Cas~A provides the best probe of the hydrodynamical structure of the ejecta, since it is young enough not to have been severely affected by the environment.  Another important example is the Crab SNR, but in this case the structure is affected by the pulsar wind nebula. This is also a complication for the LMC remnant SNR~0540-69.3 (\citealt{Sandin2013}). In comparing our results with other SNRs we therefore limit ourselves to Cas~A.

Information about the structure of the Cas A ejecta comes from optical, IR and X-ray observations. The observations reveal a highly inhomogeneous structure, including bubbles seen in both the main shell and the interior as [S III] emission \citep{Isensee2010,Milisavljevic2015}. In addition, there are high-velocity Si and S rich ``jets" , moving in opposite directions with velocities of up to $15,000 \ \kms$ \citep{Fesen2001,Hwang2004}. 

A complication with Cas A is that the observations, with the exception of the near- and mid-IR range, mainly reflect the emission between the reverse shock and the forward blast wave. The recent NuSTAR observations of ${}^{44}$Ti \citep{Grefenstette2014} are therefore very valuable. This emission reveals a highly inhomogeneous distribution with a few large ${}^{44}$Ti emitting blobs. Most importantly, at least 80\% of the ${}^{44}$Ti emission is contained within the reverse shock radius, as projected on the plane of the sky, and shows little correlation with the Fe K emission. Much of the Fe, which is expected to be at roughly the same location  as the ${}^{44}$Ti, may therefore be in the interior, and in ionization stages difficult to observe in the optical or IR. 

When comparing these observations with the ejecta emission in SN 1987A, the differences in the excitation have to be considered. The [S III] in the interior of  Cas A is probably photoionized by the emission from the hot gas behind the reverse shock. In SN 1987A, the  [Si~I]$+$[Fe~II] emission is powered by the positrons from the ${}^{44}$Ti decay, and hence reflects this distribution, as well as the distribution of Si and Fe. It is therefore most relevant to compare the [Si~I]$+$[Fe~II] emission in SN 1987A with the ${}^{44}$Ti emission in Cas A, rather than the [S III] emission. While the 3D distribution of ${}^{44}$Ti in Cas~A is not yet known, there is at least qualitatively some resemblance between the Cas~A ${}^{44}$Ti  images and the \8 [Si~I]$+$[Fe~II] distribution, with a dominance of large structures with a small total filling factor. 

Although this similarity is interesting, one has to keep in mind that Cas~A and \8 are different SN types. Cas A was a Type IIb SN (\citealt{Krause2008}), with an H envelope of less than $1 \msun$, while the progenitor of SN 1987A had a massive envelope of  $\sim 10 \msun$ \citep[e.g.,][]{Sukhbold2016}.  The ejecta structure is therefore expected to be very different, with Cas~A having a considerably higher core velocity, $\sim 6000\ \kms$, excluding the ``jets", compared to $\sim 2000-3000 \ \kms$ for \8. A low envelope mass also leads to a weak reverse shock during the first hours. The convective instabilities generated by the explosion are therefore expected to be less affected by the reverse shock in the case of Cas~A, in contrast to the case with a  massive H envelope in \8 \citep{Kifonidis2003}. The two SNe therefore provide complementary information about the 3D structure of core-collapse SNe. 

There is also information about asymmetric explosions from distant SNe. Polarization provides some knowledge about the structure of the star \citep[e.g.,][]{Wang2008,Maund2009,Reilly2016}, while line profiles provide important information regarding both velocities and asymmetries of different elements. For example, the line profiles of the Type Ib SN 1985F showed that there had to be macroscopic mixing,  which erased the stratified structure of the progenitor, resulting in nearly identical line profiles of different elements \citep[][]{Fransson1989}. There have also been extensive discussions about highly asymmetric explosions (``jets") for Type Ic~SNe \citep[e.g.,][]{Modjaz2008,Taubenberger2009,Milisavljevic2010}. In connection with this, we note that \8 serves as a cautionary example of how very non-spherical structures can result in nearly symmetric line profiles when integrated over the spatial dimensions. This is most clearly seen in the integrated H$\alpha$ profile, which was nearly symmetric before the western clump started brightening (Fig.~\ref{htime}).

SN~1987A is not unique in still being observable at 30 years after the explosion. Other, even older cases observed in the optical include SN~1957D, SN~1970G, SN~1978K, SN~1979C and SN~1980K (\citealt{Milisavljevic2012} and references therein; \citealt{Kuncarayakti2016}). These objects are, however, all powered by circumstellar interaction, which makes it difficult to compare to \8. While the interaction results in external illumination by the X-rays, similar to what is now happening with SN 1987A, the X-rays only reach part of the core and the spatial information is therefore limited. 

There are also several examples of SNe where VLBI observations provide direct spatial information, such as SN~1993J \citep{Bietenholz2003,Marti2011} and SN~2008iz \citep{Kimani2016}. However, the radio emission only  provides information about the mass loss history of the progenitor, and not the structure of the core.

\section{Conclusions and future prospects}
\label{conclusions}

We have analyzed HST/STIS observations of \8 obtained $10,000$ days after the explosions. From this we present the most detailed 3D map of the H$\alpha$ emissivity to date, as well as the first 3D information for  [O I] $\lambda \lambda 6300, 6364$, [Ca II] $\lambda \lambda 7292,\ 7324$ and  Mg~II~$\lambda \lambda 9218,\ 9244$. We have also analyzed SINFONI observations of the 3D emissivity of [Si~I]$+$[Fe~II]~$1.644\ \mu$m and He~I $2.058\ \mu$m from the same epoch, which we compare with the STIS results, as well as previous SINFONI observations. 

The interpretation of the 3D emissivities depend on the energy sources powering the different lines, as well as the properties of dust in the ejecta. We therefore first summarize our conclusions  regarding these issues, followed by the conclusions for the morphology.

\begin{itemize}

\item The integrated H$\alpha$ profile has changed dramatically since the last spectral observations covering the full ejecta from day 6355, brightening significantly in the range $[-2000,0]\ \kms$. We show that this is primarily due to the brightening of a region in the western ejecta. This is in agreement with our previous finding that H$\alpha$ is primarily powered by the X-ray emission from the ring, which also gives rise to an edge-brightened morphology. We also observe flux increases for [O I], [Ca II],  Mg~II and He~I, showing that these lines are also affected by the X-rays. However, the centrally peaked morphology of the He~I  line indicates that it is powered by both $^{44}$Ti and X-rays. 

\item The [Si~I]$+$[Fe~II] emission does not show any significant changes in flux, line profile or morphology compared to previous observations obtained during the past ten years (days 8714 and 6816). This shows that it is powered by the radioactive decay of $^{44}$Ti. In agreement with this, the integrated line profile is consistent with the NuSTAR profile of the $^{44}$Ti decay lines, albeit given large uncertainties.  

\item The Br$\gamma$ line shows a similar low-surface brightness region (or ``hole") at the centre as H$\alpha$, supporting our previous arguments that the dust is likely to be in in optically thick clumps that affect the optical and NIR in the same way. Furthermore, the constant [Si~I]$+$[Fe~II] line profile argues against dust becoming optically thin during the last ten years.

\item A similar large-scale geometry is seen in all lines that have sufficient signal to noise (i.e H$\alpha$, [Ca II], [Si~I]$+$[Fe~II] and He~I). As projected on the sky, the ejecta extend in the north-east to south-west direction. In 3D, the ejecta emission is concentrated between the line of sight and the plane of the ring in the north, but closer to the plane of the sky in the south, resembling a broken dipole. This is consistent with the structure seen in previous SINFONI observations and indicated by previous low-resolution H$\alpha$ observations (\citealt{Kjaer2010}, L13). Remarkably, the large-scale structure correlates with very early observations of  asymmetries. We therefore conclude that there is a global asymmetry that extends from the inner metal core to the outer H envelope. 

\item In [Si~I]$+$[Fe~II] and H$\alpha$, which offer the best signal, substructure is seen on a scale of $\sim 200 - 1000\ \kms$, close to the level of resolution in the different directions. As seen in previous observations, the [Si~I]$+$[Fe~II] is concentrated to two main regions. The H$\alpha$ 3D map differs from this, with, for example, an arc being seen around the ``hole",  and the brightest emission being concentrated to the clump in the west. These differences are due to a combination of the true distribution of the different elements and the energy sources powering the two lines. 

\item The centre of the bright, western clump in H$\alpha$ has a space velocity of $\sim 1800\ \kms$. Both [O I] and Mg~II also have their brightest emission concentrated to this region. This shows that there is high-abundance of O in the H-region that is most affected by X-rays. For Mg~II, a correlation with H$\alpha$ is expected since it is powered by Ly~$\alpha$ fluorescence. The H$\alpha$ map also shows that H is mixed in to about $450\ \kms$. For [Si~I]$+$[Fe~II], the space velocities of the centers of the brightest regions are $\sim 2000\ \kms$ and $\sim 2700\ \kms$ in the north and south, respectively.

\end{itemize}

\8 is unique among SNe and allows us to probe the explosion physics during the first seconds through its radioactivity, nucleosynthesis and morphology. Our observations have provided additional information about the latter. As has been shown by \cite{Wongwathanarat2015} and other simulations, there is a direct connection between the large-scale instabilities created in the explosion and those in the homologous phase for at least for some progenitors. However, this needs to be probed in more detail for different progenitor models and different  explosion physics. In addition, rotational effects and magnetic fields remain to be included in the models. 

On the observational side, there will be additional important information from different molecules and dust from ALMA. A taste of this has already been seen \citep{Kamenetzky2013, Indebetouw2014,Matsuura2015}, but the full ALMA will have a resolution comparable or higher than that of HST. We also hope to get 3D maps of both the He I $\lambda 2.058 \ \mu$m line and the H$_2$ lines in the NIR, using deeper SINFONI observations. The connection between these different molecules and atomic lines with the mass distribution, however, require a detailed modeling of the emission, including the excitation by the positrons \citep{Jerkstrand2011} and the X-rays from the ring collision. 

Finally, the compact object in the centre of \8 remains to be revealed. HST, ALMA, NuStar and JWST may provide new opportunities for this.

\acknowledgments
We are grateful to Hans-Thomas Janka  and Michael Gabler for discussions about their explosion models.  
This work was supported by the Swedish National Space Board and Swedish Research Council.  
Support for HST GO program numbers 13401 and 13405 was provided by NASA through grants from the Space Telescope Science Institute, which is operated by the Association of Universities for Research in Astronomy, Inc., under NASA contract NAS5-26555. 
The ground-based observations were collected at the European Organization for Astronomical Research in the Southern Hemisphere, Chile (ESO Program 094.D-0505(C)).

{\it Facilities:} \facility{HST (STIS, WFC3)}, \facility{VLT (SINFONI)}.

\bibliographystyle{apj}
\bibliography{87a_ejecta_2016_p.bbl}

\appendix

\section{Maps of weaker ejecta lines}

\begin{figure*}
\begin{center}
\resizebox{160mm}{!}{\includegraphics{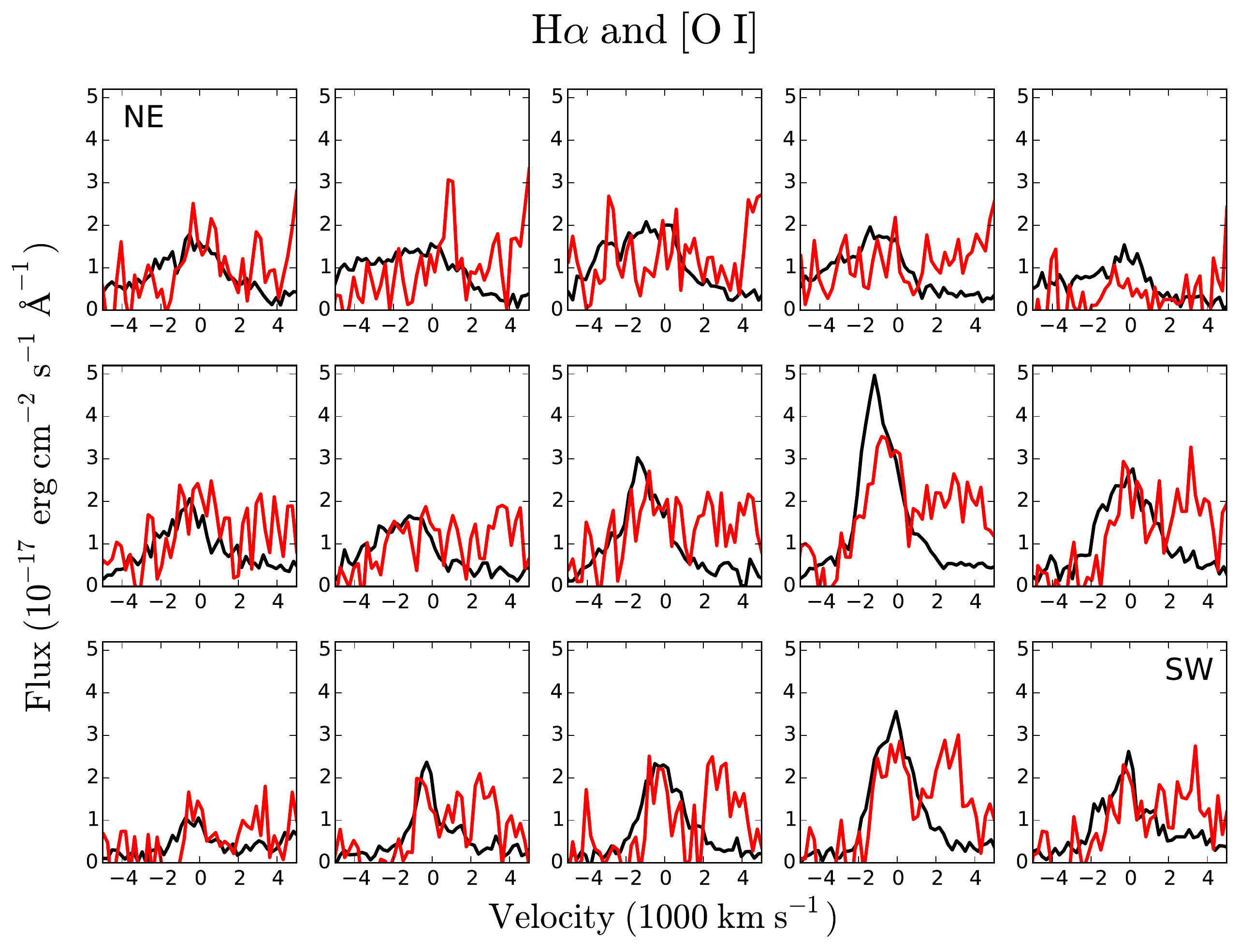}}
\caption{Comparison of the H$\alpha$ (black) and [O I]~$\lambda \lambda 6300,\ 6364$ (red) lines extracted from the regions shown in Fig.~\ref{hsimap}. For [O I] the spectrum has been multiplied by a factor of four and centered on zero velocity for the $6300\ \rm{\AA}$ component. The two components are separated by about $3000\ \kms$}
\label{omap}
\end{center}
\end{figure*}

\begin{figure*}
\begin{center}
\resizebox{160mm}{!}{\includegraphics{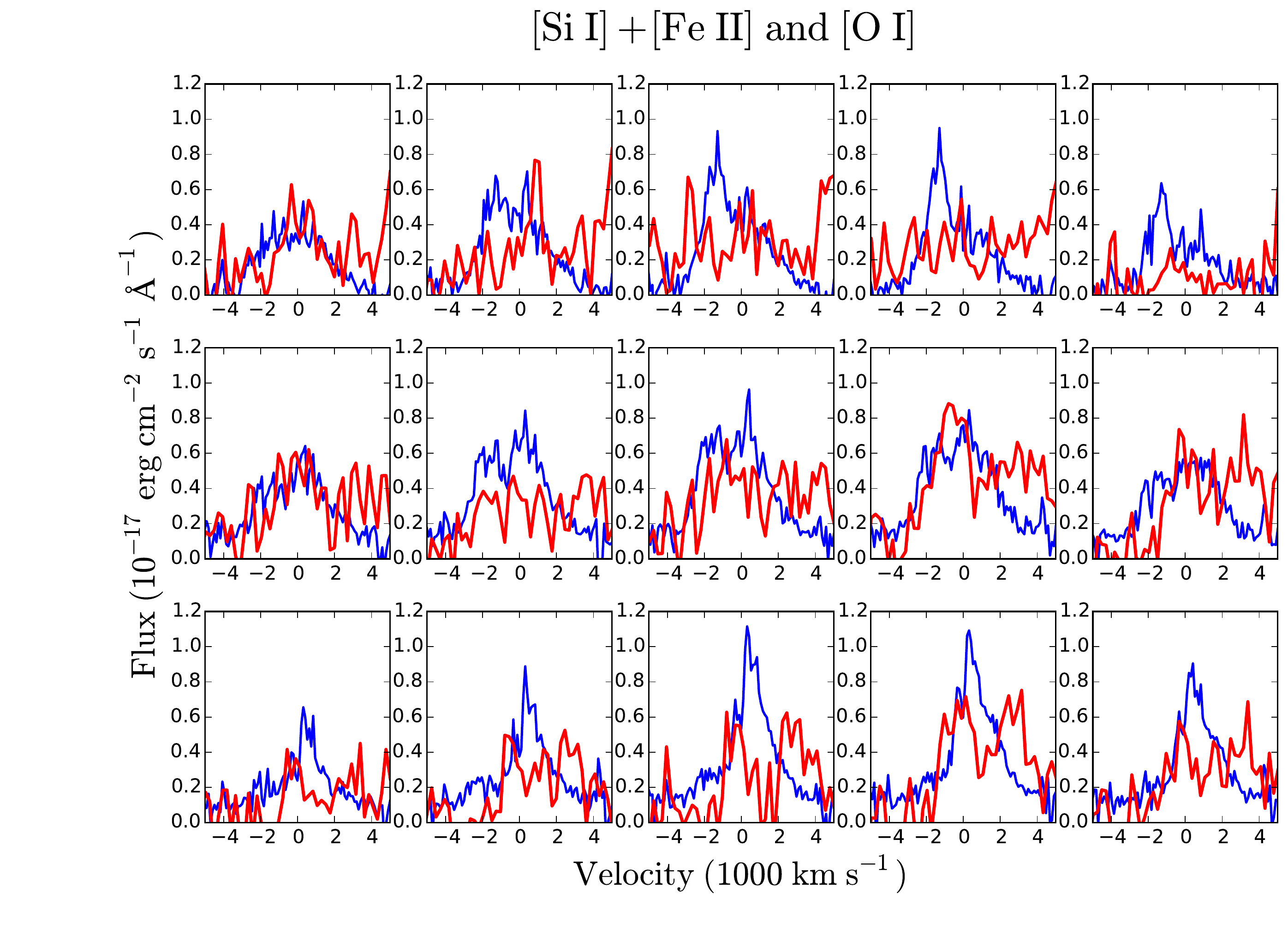}}
\caption{Comparison of the [Si~I]$+$[Fe~II] $1.644\ \mu$m  (blue) and [[O I]~$\lambda \lambda 6300,\ 6364$ (red) lines extracted from the regions shown in Fig.~\ref{hsimap}. The spectrum has been multiplied by a factor of four for [Si~I]$+$[Fe~II]. For [O I] the spectrum is centered on zero velocity for the $6300\ \rm{\AA}$ component.}
\label{siomap}
\end{center}
\end{figure*}

\begin{figure*}
\begin{center}
\resizebox{160mm}{!}{\includegraphics{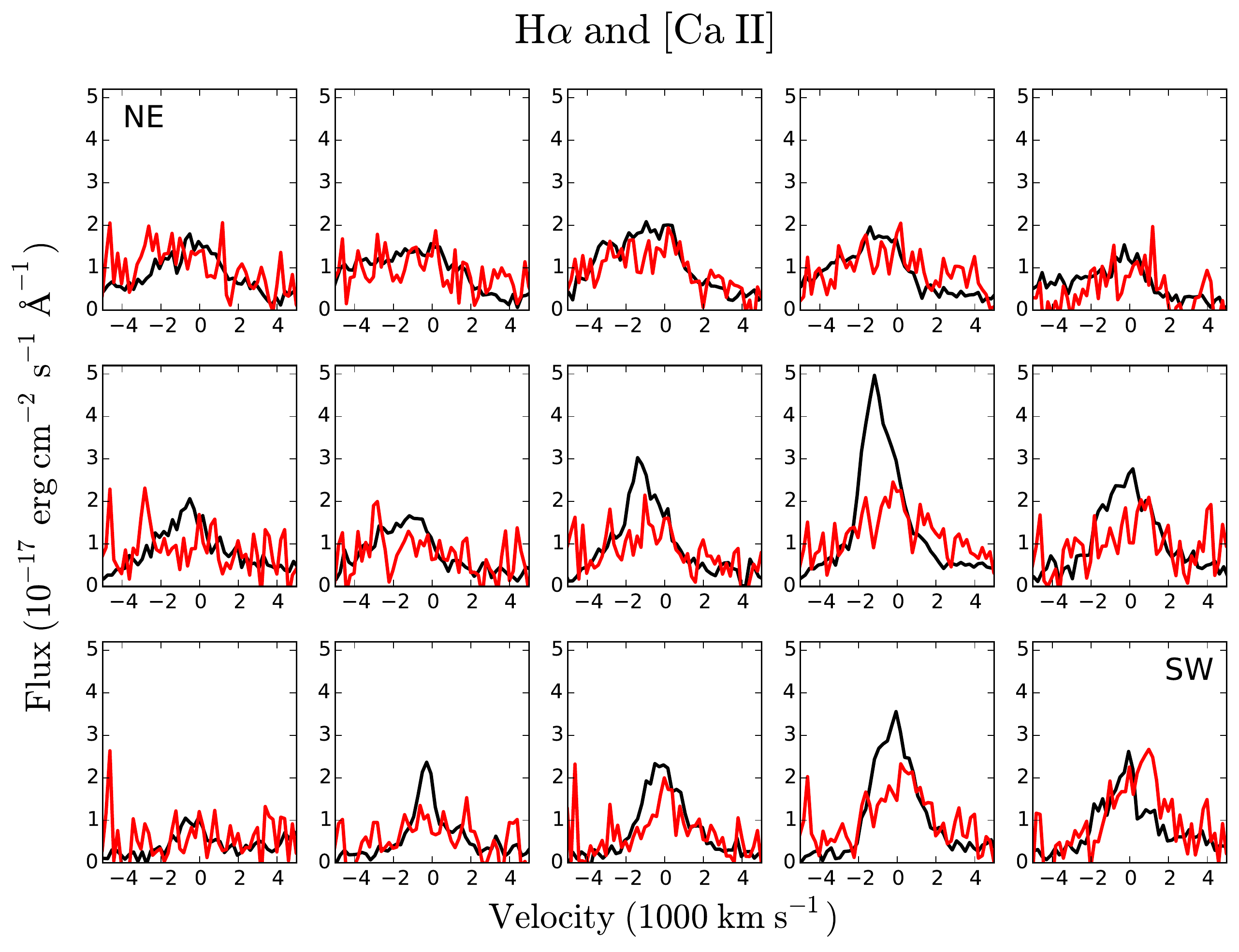}}
\caption{Comparison of the H$\alpha$ (black) and [Ca II] ~$\lambda \lambda 7292,\ 7324$ (red) lines extracted from the regions shown in Fig.~\ref{hsimap}. For [Ca II] the spectrum has been multiplied by a factor of three and centered on zero velocity for the average wavelength of the two components. The velocity difference between the two components is about $1300\ \kms$.}
\label{camap}
\end{center}
\end{figure*}

\begin{figure*}
\begin{center}
\resizebox{160mm}{!}{\includegraphics{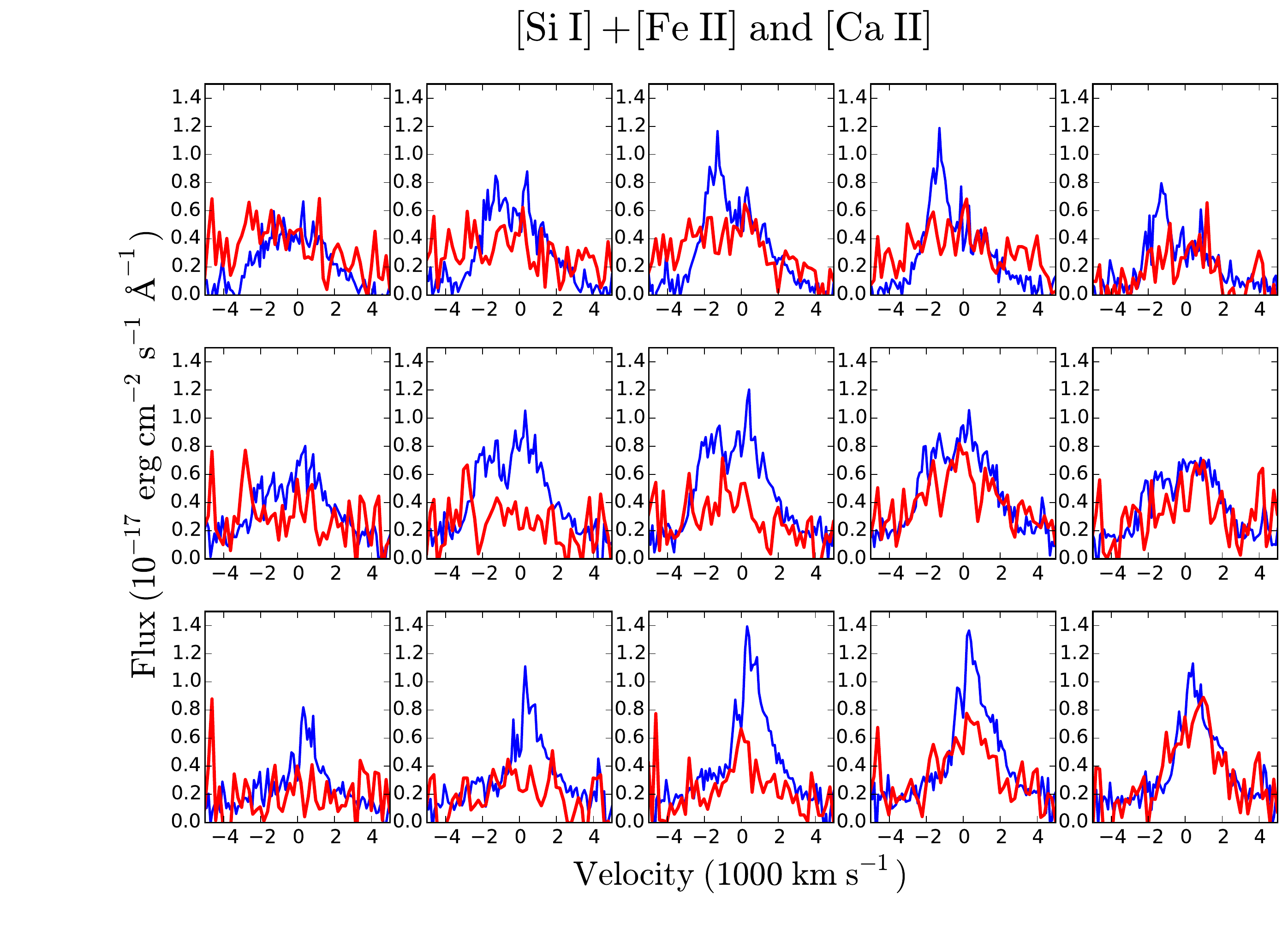}}
\caption{Comparison of the [Si~I]$+$[Fe~II] $1.644\ \mu$m  (blue) and [Ca II] ~$\lambda \lambda 7292,\ 7324$ (red) lines extracted from the regions shown in Fig.~\ref{hsimap}. The spectrum has been multiplied by a factor of five for [Si~I]$+$[Fe~II]. For [Ca II] the spectrum is centered on zero velocity for the average wavelength of the two components.}
\label{sicamap}
\end{center}
\end{figure*}

\begin{figure*}
\begin{center}
\resizebox{160mm}{!}{\includegraphics{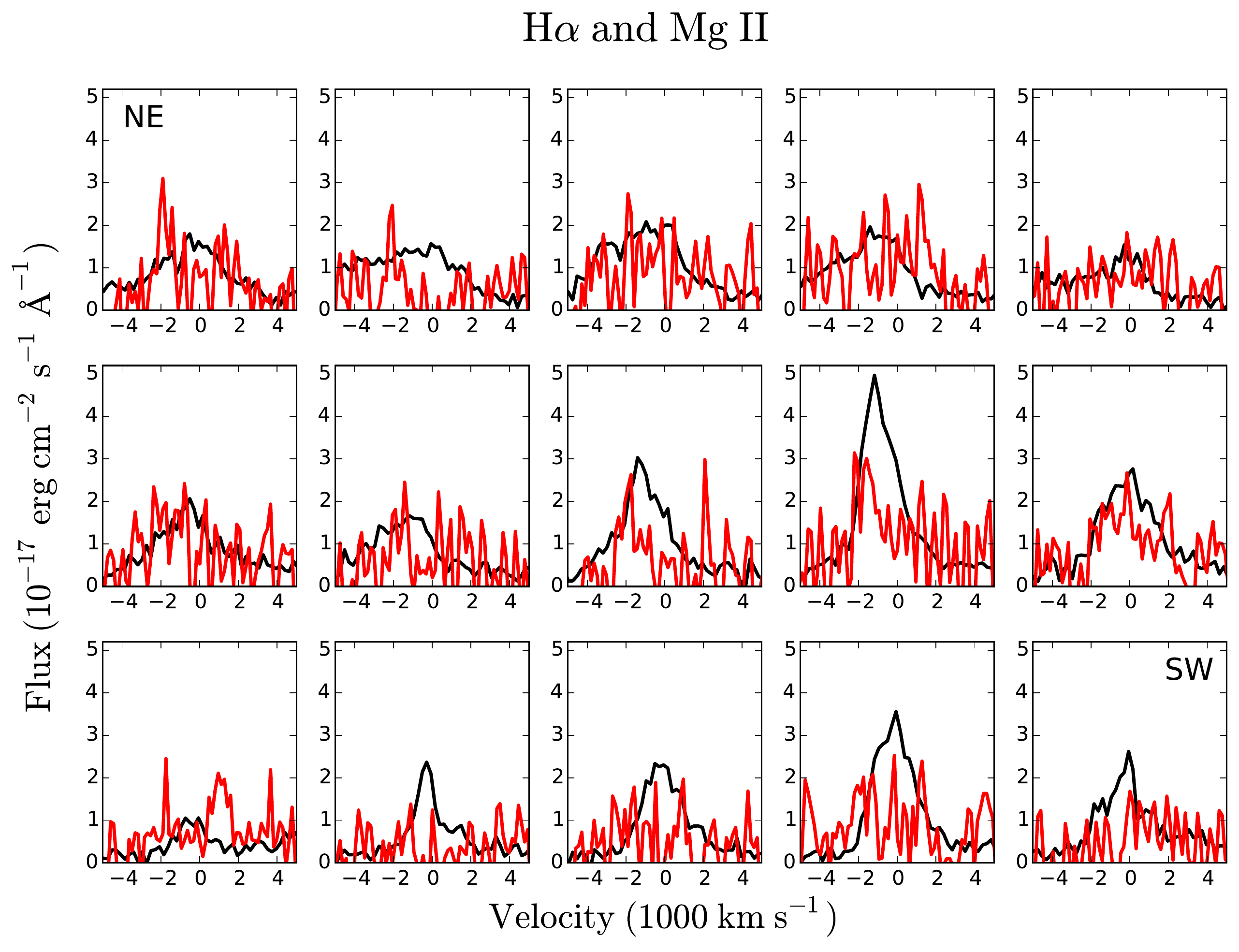}}
\caption{Comparison of the H$\alpha$ (black) and Mg~II~$\lambda \lambda 9218,\ 9244$ (red) lines extracted from the regions shown in Fig.~\ref{hsimap}. For Mg~II the spectrum has been multiplied by a factor of four and centered at $9226 \rm{\AA}$ based on the statistical weight of the two components.}
\label{mgmap}
\end{center}
\end{figure*}

\begin{figure*}
\begin{center}
\resizebox{160mm}{!}{\includegraphics{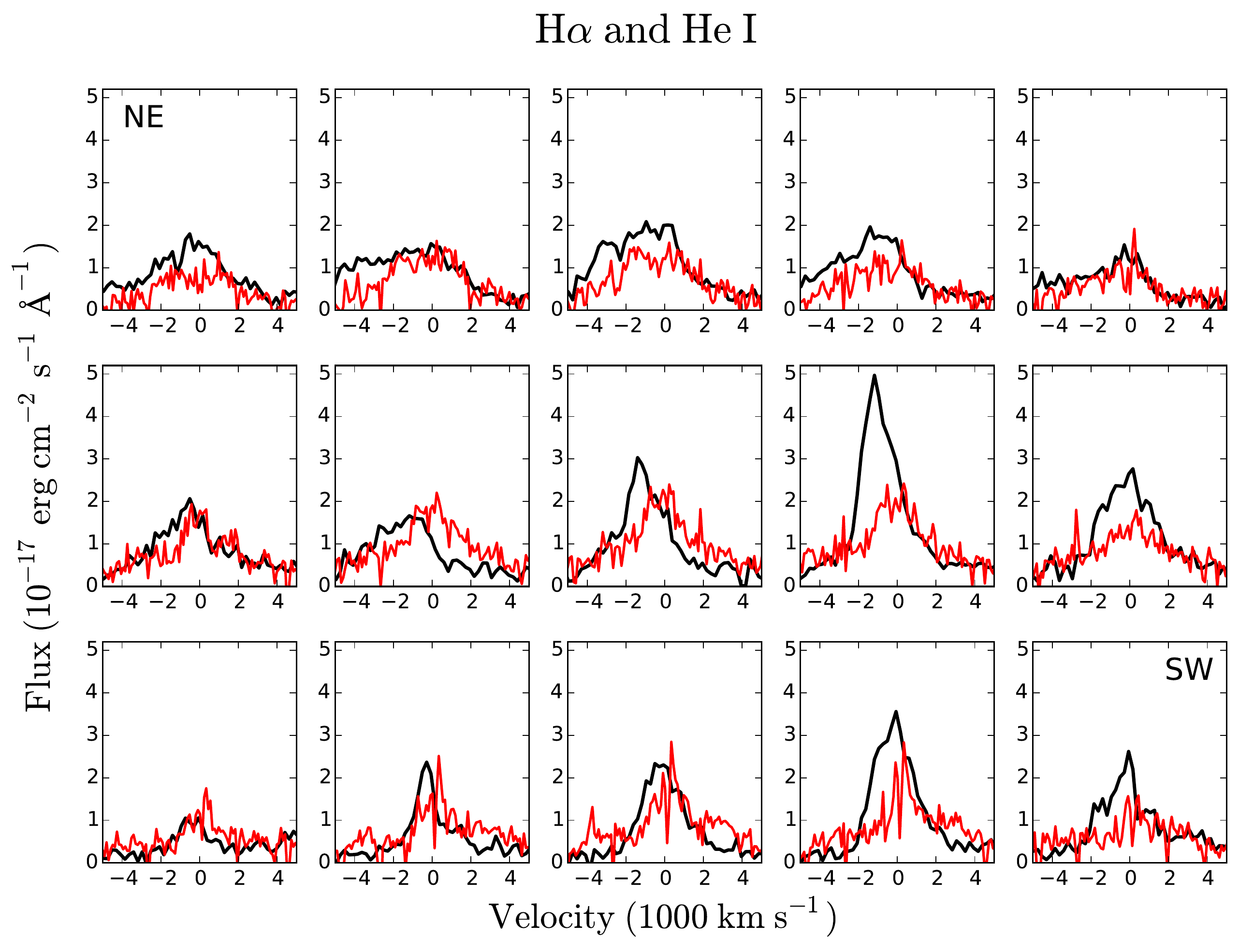}}
\caption{Comparison of the H$\alpha$ (black) and He I~$2.058\ \mu$m (red) lines extracted from the regions shown in Fig.~\ref{hsimap}. The SINFONI spectrum of He~I has been multiplied by a factor of 20. }
\label{hemap}
\end{center}
\end{figure*}

\begin{figure*}
\begin{center}
\resizebox{160mm}{!}{\includegraphics{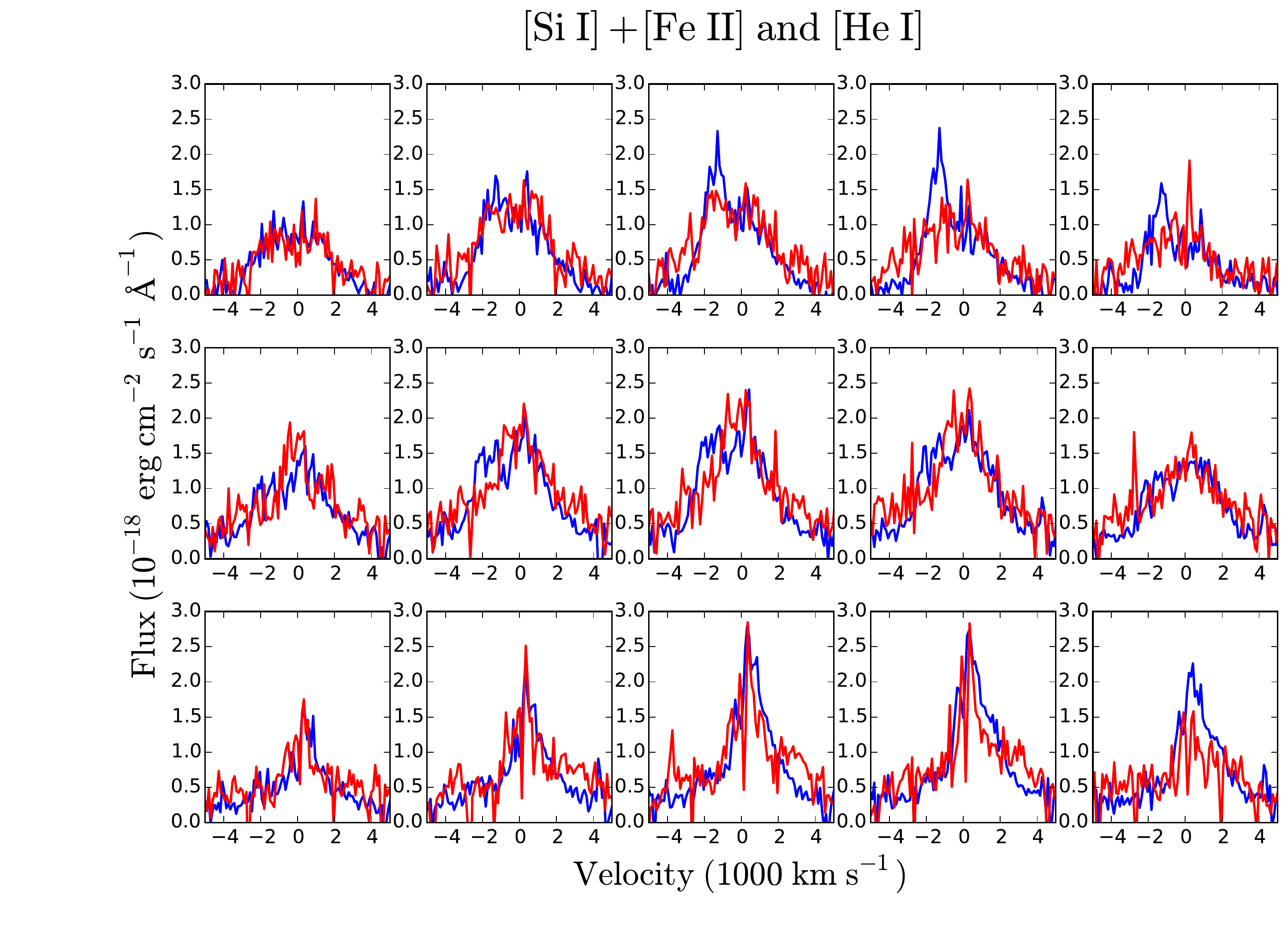}}
\caption{Comparison of the  [Si~I]$+$[Fe~II] $1.644\ \mu$m (blue) and  He I~$2.058\ \mu$m (red) lines extracted from the regions shown in Fig.~\ref{hsimap}.  The He I lines have been multiplied by a factor of two.}
\label{sihemap}
\end{center}
\end{figure*}

Here we show spectral maps of several of the weaker lines from the ejecta compared to H$\alpha$ and [Si~I]$+$[Fe~II] $1.644~\mu$m, which are the two strongest lines. The spectra were extracted from the regions shown in Fig.~\ref{hsimap} and corrected for scattered light from the ring as described in section \ref{3ddist}. Figs.~\ref{omap} and \ref{siomap} show the [O I]~$\lambda \lambda 6300,\ 6364$ lines, Figs.~\ref{camap} and \ref{sicamap} show the [Ca II]~$\lambda \lambda 7292,\ 7324$ lines, Fig.~\ref{mgmap} shows the Mg~II~$\lambda \lambda 9218,\ 9244$ lines, and Figs.~\ref{hemap} and \ref{sihemap} show the He~I~$2.058~\mu$m line. 

For [O I] the spectrum is centered on zero velocity for the $6300\ \rm{\AA}$ component. As the redshifted part of this line is blended with the other component of the doublet as well at the reverse shock, only the blueshifted side should be used for comparison with the other lines. For [Ca II] on the other hand, the two components are closer together and more similar in strength ($\lambda 7292/\lambda 7324 = 3/2$ assuming the upper levels are in statistical equilibrium and that the lines are optically thin). The spectrum is therefore centered on zero velocity for the weighted average wavelength of the two components. For Mg~II we centre the line at $9226 \rm{\AA}$ based on the statistical weight of the two components.  The ``spikes" seen in the central part of the He~I line is due to problems with the correction for scattered light from the ring.  This especially affects the south-western region, which is close to the brightest part of the ring. The ratio of ring to ejecta emission is extremely high in this line, and an improvement in the correction would require a detailed model for contamination from different parts of the ring in different parts of the ejecta, which is beyond the scope of this work. 

\clearpage


\clearpage

\clearpage

\end{document}